\documentclass[12pt,twoside]{article}

\usepackage{graphicx, epsfig}
\usepackage{amsfonts, amssymb, amsmath, amsthm}
\usepackage{latexsym}
\usepackage{cite}
\newcommand{\partdif}[2]{\ensuremath{ \frac{\partial #1}{\partial #2}}}

\begin{document}

\title{A Hybrid Godunov Method for Radiation Hydrodynamics}
\author{Michael D Sekora$^{\dagger}$ \\ 
James M Stone$^{\dagger,~\ddagger}$ \\ 
\textit{{\small Program in Applied and Computational Mathematics$^{\dagger}$}} \\ 
\textit{{\small Department of Astrophysical Sciences$^{\ddagger}$}} \\ 
\textit{{\small Princeton University, Princeton, NJ 08544, USA}}}
\date{24 March 2010}
\maketitle

\begin{abstract}
\noindent
From a mathematical perspective, radiation hydrodynamics can be thought of as a system of hyperbolic balance laws with dual multiscale behavior (multiscale behavior associated with the hyperbolic wave speeds as well as multiscale behavior associated with source term relaxation). With this outlook in mind, this paper presents a hybrid Godunov method for one-dimensional radiation hydrodynamics that is uniformly well behaved from the photon free streaming (hyperbolic) limit through the weak equilibrium diffusion (parabolic) limit and to the strong equilibrium diffusion (hyperbolic) limit. Moreover, one finds that the technique preserves certain asymptotic limits. The method incorporates a backward Euler upwinding scheme for the radiation energy density $E_{r}$ and flux $F_{r}$ as well as a modified Godunov scheme for the material density $\rho$, momentum density $m$, and energy density $E$. \\

\noindent
The backward Euler upwinding scheme is first-order accurate and uses an implicit HLLE flux function to temporally advance the radiation components according to the material flow scale. The modified Godunov scheme is second-order accurate and directly couples stiff source term effects to the hyperbolic structure of the system of balance laws. This Godunov technique is composed of a predictor step that is based on Duhamel's principle and a corrector step that is based on Picard iteration. The Godunov scheme is explicit on the material flow scale but is unsplit and fully couples matter and radiation without invoking a diffusion-type approximation for radiation hydrodynamics. This technique derives from earlier work by Miniati \& Colella 2007. Numerical tests demonstrate that the method is stable, robust, and accurate across various parameter regimes.
\end{abstract}

\pagestyle{myheadings}
\markboth{M D Sekora \& J M Stone}{A Hybrid Godunov Method for Radiation Hydrodynamics}


\section{Introduction}
\noindent
Radiation hydrodynamics is a dynamical description of fluid material interacting with electromagnetic radiation and is appropriate whenever radiation governs the transport of energy and momentum in the fluid. Many phenomena in plasma physics and astrophysics are governed by radiation hydrodynamics, some examples include: star formation, supernovae, accretion disks, radiatively driven outflows, stellar convection, and inertial confinement fusion \cite{mmbook, mBib2002, castorbook}. In these applications, the radiation field heterogeneously couples to the material dynamics such that radiative effects are strong in some parts of the system and weak in other parts. These variations give rise to characteristically different dynamical properties (i.e., advection versus diffusion behavior). The primary objective for developing the numerical technique presented in this paper is to have a computational tool that accurately solves radiation hydrodynamical problems across a range of asymptotic limits. The new algorithmic ideas are cast in such a way that they seem familiar with respect to classical Godunov schemes and can be implemented in existing codes with minimal computational overhead. A future research endeavor is to combine the hybrid Godunov method for radiation hydrodynamics with an existing code for MHD (magnetohydrodynamics) such as Athena \cite{athena} and investigate full radiation MHD in multiple spatial dimensions. \\

\noindent
Some of the initial research in developing numerical methods to solve radiation hydrodynamical problems was carried out by Castor 1972, Pomraning 1973, Levermore \& Pomraning 1981, Mihalas \& Klein 1982, and Mihalas \& Mihalas 1984. One of the simplest and most successful approaches used in astrophysics was the Zeus code of Stone et al 1992, which relied on operator splitting and a Crank-Nicholson temporal finite difference scheme. Since the introduction of that code, finite volume schemes (e.g., Godunov-type methods) have emerged as a powerful technique for solving hyperbolic conservation laws (i.e., the mathematical framework describing radiation hydrodynamics) \cite{leveque1, leveque2}. Moreover, this integral formulation allows one to more naturally treat boundary conditions, capture shock waves and other discontinuous behavior, investigate complex geometries and multi-dimensions, and implement adaptive mesh refinement \cite{yip2005}. Despite these advantages, there have been significant difficulties in developing a Godunov method to accurately represent radiation hydrodynamical behavior across a range of asymptotic limits. Earlier attempts to construct a Godunov-type technique either $(i)$ neglected the heterogeneity of the matter-radiation coupling and solved the system of equations in a specific limit \cite{dai1, dai2}, $(ii)$ were based on a model system \cite{jin, buet}, $(iii)$ invoked artificial coupling terms that were based on the reference frame in which the problem was solved \cite{balsara1999} as pointed out by Lowrie \& Morel 2001, or $(iv)$ used a variation of flux limited diffusion \cite{levermore1981, turner2001, heracles2007, krumholtz2007}. Lowrie \& Morel 2001 was even critical of the likelihood of developing such a method for full radiation hydrodynamics. \\

\noindent
Developing a Godunov method for radiation hydrodynamics has been difficult because numerical difficulties arise from $(i)$ the reference frame one chooses for taking moments of the photon transport equation, $(ii)$ multiscale waves, $(iii)$ stiff source terms, and $(iv)$ solving a hierarchy of radiation transport moment equations to compute the variable tensor Eddington factor $\mathsf{f}$. The first difficulty occurs because one takes moments of the photon transport equation in order to define the radiation quantities that interact with the material components of the system. One encounters problems with the radiation field's specific intensity function $I(\nu,\mathbf{n})$ which is governed by frame-dependent quantities, the radiation frequency $\nu$ and directional vector $\mathbf{n}$. Here, one either casts the photon transport equation into the comoving frame (at rest with respect to the local fluid velocity) and contends with complicated transport operators but simple interaction terms $S(\nu,\mathbf{n})$; or one casts the photon transport equation into the Eulerian frame (at rest with respect to the system as a whole) and contends with simple transport operators but complicated interaction terms. The specific intensity is written below in the Eulerian frame, where $c$ is the speed of light and $t$ is time:
\begin{equation}
\left( \frac{1}{c} \partdif{~}{t} + \mathbf{n} \cdot \nabla \right) I(\nu,\mathbf{n}) = S(\nu,\mathbf{n}) .
\end{equation}

\noindent
The Eulerian frame approach is also referred to as the mixed frame approach if the radiation intensity is measured in the Eulerian frame while the opacities $\sigma$ (embedded in the interaction terms) are measured in the comoving frame \cite{lowrie2001, castorTalk}. The second difficulty arises because although the transport operators have a simple formulation in the mixed/Eulerian frame, the radiation quantities are still characterized by waves propagating at the speed of light $c$. This dynamical scale is significantly larger than the speed of sound $a_{\infty}$ which characterizes the material quantities in the absence of radiation, such variation in propagation speeds defines the nature of multiscale waves associated with radiation hydrodynamics. It is important to note that the reference frame one chooses to take moments of the photon transport equation does not affect how one defines the material quantities. A mixed frame approach was adopted because the resulting equations most closely resemble a system of hyperbolic balance laws which is advantageous for constructing a Godunov-type method. The third difficulty occurs because in addition to defining balance laws for the radiation quantities, one must also add relativistic stiff source terms that are correct to $\mathcal{O}(u/c)$ to the right hand sides of the non-relativistic conservation laws for the material quantities (i.e.,  Euler equations). These source terms are stiff because of the variation in time and length scales associated with radiation hydrodynamical problems \cite{mk1982}. Having to contend with stiffness arising from some waves propagating at the speed of light as well as source terms having large magnitudes, it is obvious to see how such numerical difficulties make conventional techniques like operator splitting and the method of lines breakdown \cite{leveque1, leveque2, torobook}. \\

\noindent
This paper presents a hybrid Godunov method that addresses the above numerical difficulties. The technique adopts a mixed frame approach, includes the appropriate frame-dependent terms to $\mathcal{O}(u/c)$, is implicit with respect to the fastest hyperbolic wave speeds, and semi-implicitly updates the stiff source terms. The paper proceeds in the following manner. After defining the full system of equations for radiation hydrodynamics, the paper discusses what dynamics characterize the various asymptotic limits. Then, the paper gives an overview of the hybrid Godunov method and explains certain numerical properties that the algorithm possesses. The next three sections present the main algorithmic ideas behind the hybrid Godunov method. Lastly, the paper describes numerical tests that demonstrate the technique to be stable, robust, and accurate across various parameter regimes.


\section{Radiation Hydrodynamics}
\noindent
As presented in Lowrie et al 1999 and Lowrie \& Morel 2001, the system of equations for radiation hydrodynamics can be non-dimensionalized with respect to the material flow scale so that one can compare hydrodynamical and radiation effects as well as identify terms that are $\mathcal{O}(u/c)$ \cite{lowrie1999, lowrie2001}. This scaling gives two important parameters: $\mathbb{C} = c / a_{\infty}$ which measures relativistic effects and $\mathbb{P} = a_r T^4_{\infty} / \rho_{\infty}a^2_{\infty} $ which measures how radiation affects material dynamics. Additionally, $a_r = 8 \pi^5 k^4 / 15 c^3 h^3 = 7.57 \times 10^{-15} ~ erg ~ cm^{-3} ~ K^{-4}$ is a radiation constant, $T_{\infty}$ is a reference material temperature in the absence of radiation, and $\rho_{\infty}$ is a reference material density in the absence of radiation. The full system of equations for radiation hydrodynamics in the mixed frame that is correct to $\mathcal{O}(1/\mathbb{C})$ is:
\begin{equation}
\partdif{\rho}{t} + \nabla \cdot \left( \mathbf{m} \right) = 0 \label{eq:rh1} ,
\end{equation}
\begin{equation}
\partdif{\mathbf{m}}{t} + \nabla \cdot \left( \frac{\mathbf{m} \otimes \mathbf{m}}{\rho} \right) + \nabla p = -\mathbb{P} \left [ -\sigma_t \left( \mathbf{F_r} - \frac{ \mathbf{u}E_r + \mathbf{u} \cdot \mathsf{P_r} }{\mathbb{C}} \right) + \sigma_a \frac{\mathbf{u}}{\mathbb{C}} (T^4 - E_r) \right ] \label{eq:rh2} , 
\end{equation}
\begin{equation}
\partdif{E}{t} + \nabla \cdot \left( (E+p) \frac{\mathbf{m}}{\rho} \right) = -\mathbb{P} \mathbb{C} \left [ \sigma_a(T^4 - E_r) + (\sigma_a - \sigma_s) \frac{\mathbf{u}}{\mathbb{C}} \cdot \left( \mathbf{F_r} - \frac{ \mathbf{u}E_r + \mathbf{u} \cdot \mathsf{P_r} }{\mathbb{C}} \right) \right ] \label{eq:rh3} , 
\end{equation}
\begin{equation}
\partdif{E_r}{t} + \mathbb{C} \nabla \cdot \mathbf{F_r} = \mathbb{C} \left [ \sigma_a(T^4 - E_r) + (\sigma_a - \sigma_s) \frac{\mathbf{u}}{\mathbb{C}} \cdot \left( \mathbf{F_r} - \frac{ \mathbf{u}E_r + \mathbf{u} \cdot \mathsf{P_r} }{\mathbb{C}} \right) \right ] \label{eq:rh4} , 
\end{equation}
\begin{equation}
\partdif{ \mathbf{F_r} }{t} + \mathbb{C} \nabla \cdot \mathsf{P_r} = \mathbb{C} \left [ -\sigma_t \left( \mathbf{F_r} - \frac{ \mathbf{u}E_r + \mathbf{u} \cdot \mathsf{P_r} }{\mathbb{C}} \right) + \sigma_a \frac{\mathbf{u}}{\mathbb{C}} (T^4 - E_r) \right ] \label{eq:rh5} ,
\end{equation}
\begin{equation}
\mathsf{P_r} = \mathsf{f} E_r ~~ \textrm{(closure relation)} \label{eq:rad_closure} .
\end{equation}

\noindent
For the material quantities, $\rho$ is density, $\mathbf{m}$ is momentum density, $p = (\gamma-1)e $ is pressure, $E$ is energy density, and $T$ is temperature. For the radiation quantities, $E_r$ is energy density, $\mathbf{F_r}$ is flux, $\mathsf{P_r}$ is pressure, and $\mathsf{f}$ is the variable tensor Eddington factor. In the source terms, $\sigma_a$ is the absorption cross section, $\sigma_s$ is the scattering cross section, and $\sigma_t = \sigma_a + \sigma_s$ is the total cross section. $\mathsf{f}$ is used to close the hierarchy of radiation transport moment equations such that:
\begin{equation}
E_r = \int_{0}^{\infty} \oint_{\mathbf{n}} I d \mathbf{n} d \nu , ~~~~ \mathbf{F_r} = \int_{0}^{\infty} \oint_{\mathbf{n}} \mathbf{n} I d \mathbf{n} d \nu , ~~~~ \mathsf{P_r} = \int_{0}^{\infty} \oint_{\mathbf{n}} \mathbf{n} \otimes \mathbf{n} I d \mathbf{n} d \nu .
\end{equation}

\noindent
The above non-dimensionalization admits the equation of state $p = \rho T \left( R T_{\infty} / a^{2}_{\infty} \right)$. If one assumes physically relevant reference quantities (e.g., gas constant $R = 8.31 \times 10^{7} ~ erg ~ K^{-1}$, $T_{\infty} \approx 10 ~ K$, and $a_{\infty} \approx 3 \times 10^{4} ~ cm ~ s^{-1}$), then one finds that $\mathbb{C} \approx 10^{5}$, $R T_{\infty} / a^{2}_{\infty} \approx 1$, and $p = \rho T$. If one further assumes that $\rho_{\infty} \approx 10^{-9} ~ g ~ cm^{-3}$, then $\mathbb{P} \approx 10^{-10}$. However, the choice of these values is arbitrary. These quantities were chosen because they are similar to the physical conditions associated with well known test problems (e.g., Su-Olson non-equilibrium diffusion as well as subcritical and supercritical radiative shocks). It is important to distinguish that while $\mathbb{C}$ and $\mathbb{P}$ set the physical scale for the overall system, $\mathbb{P}$ is different from the ratio $P_{r} / p$ which can be greater than unity in certain regions of a physical system. \\

\noindent
For the above system of equations, one has assumed that scattering is isotropic and coherent in the comoving frame, emission is defined by local thermodynamic equilibrium (LTE), and that spectral averages for the cross-sections can be employed (gray approximation). The source terms are given by the modified Mihalas-Klein description \cite{lowrie2001, lowrie1999} which improves upon the original Mihalas-Klein description \cite{mk1982} because it maintains important $\mathcal{O}(1 / \mathbb{C}^2)$ terms that ensure the correct relaxation rate to thermal equilibrium. See Section 4 of Lowrie et al 1999 for a physical description of what these source terms mean. \\

\noindent
Lastly, because the focus of this paper is exploring the dynamical properties associated with resolving a system of stiff hyperbolic balance laws, one has assumed a gray approximation. Under this assumption and given the algorithmic machinery built in this paper, a mixed frame approach is straightforward to implement. However, this formulation becomes increasingly complicated for problems defined by multigroup physics where spectral averages for the cross-sections cannot be employed. In this context, a comoving frame approach is attractive.

\subsection{System of Equations in One Spatial Dimension}
\noindent
This paper is concerned with formulating the algorithmic ideas for a hybrid Godunov method for radiation hydrodynamical problems and develops the method in only one spatial dimension. However, the technique should generalize to multidimensions and such work will be the subject of a future paper. If on neglects transverse flow, then in one spatial dimension the equations can be written as:
\begin{equation}
\partdif{U}{t} + \partdif{F(U)}{x} = S(U) \label{eq:cons_law_1d} ,
\end{equation}
\begin{equation}
U = 
\left( \begin{array}{c} \rho \\
                        m \\
                        E \\
                        E_r \\
                        F_r \end{array} \right) , ~~~~
F(U) = 
\left( \begin{array}{c} m \\
                        \frac{{m}^2}{\rho} + p \\
                        (E + p)\frac{m}{\rho} \\
                        \mathbb{C} F_r \\
                        \mathbb{C} f E_r \end{array} \right) \label{eq:cons_law_u_and_f_1d} , 
\end{equation}
\begin{equation}
S(U) = 
\left( \begin{array}{c} 0 \\
                        -\mathbb{P} S^F \\
                        -\mathbb{P} \mathbb{C} S^E \\
                        \mathbb{C} S^E \\
                        \mathbb{C} S^F \end{array} \right) = 
\left( \begin{array}{c} 
0 \\
- \mathbb{P} \left [ -\sigma_t \left( F_r - \frac{ (1+f) m E_r }{\rho \mathbb{C}} \right) + \sigma_a \frac{m}{\rho \mathbb{C}} (T^4 - E_r) \right ] \\
- \mathbb{P} \mathbb{C} \left [ \sigma_a(T^4 - E_r) + (\sigma_a - \sigma_s) \frac{m}{\rho \mathbb{C}} \left( F_r - \frac{ (1+f) m E_r }{\rho \mathbb{C}} \right) \right ] \\
\mathbb{C} \left [ \sigma_a(T^4 - E_r) + (\sigma_a - \sigma_s) \frac{m}{\rho \mathbb{C}} \left( F_r - \frac{ (1+f) m E_r }{\rho \mathbb{C}} \right) \right ] \\
\mathbb{C} \left [ -\sigma_t \left( F_r - \frac{ (1+f) m E_r }{\rho \mathbb{C}} \right) + \sigma_a \frac{m}{\rho \mathbb{C}} (T^4 - E_r) \right ] 
\end{array} \right) \label{eq:cons_law_source_1d} .
\end{equation}

\noindent
The quasi-linear form of this system of balance laws is:
\begin{equation}
\partdif{U}{t} + A \partdif{U}{x} = S(U) \label{eq:cons_law_quasi_1d} ,
\end{equation}
\begin{equation}
A = \left( \begin{array}{ccccc} 
0                                                 & 1                       & 0          & 0            & 0          \\
\frac{\gamma-3}{2}u^2                             & -(\gamma-3)u            & (\gamma-1) & 0            & 0          \\
u\left( \frac{\gamma-1}{2}u^2-\tilde{H} \right)   & \tilde{H}-(\gamma-1)u^2 & \gamma u   & 0            & 0          \\ 
0                                                 & 0                       & 0          & 0            & \mathbb{C} \\
0                                                 & 0                       & 0          & \mathbb{C} f & 0          \end{array} \right) \label{eq:cons_law_jacobian_1d} .
\end{equation}

\noindent
Here, $u = m/\rho$ is velocity and $\tilde{H} = \frac{\gamma E}{\rho} - \frac{\gamma-1}{2}u^2$ is specific enthalpy. The Jacobian $A$ has eigenvalues: $\lambda = u, u \pm a, \pm f^{1/2} \mathbb{C}$. However, one must account for how the stiff momentum and energy source terms affect the hyperbolic structure.

\subsection{Asymptotics}
\noindent
Before describing the method, it is instructive to consider the properties of Equations \ref{eq:cons_law_1d}-\ref{eq:cons_law_jacobian_1d} in various limits. Following the asymptotic analysis of Lowrie et al 1999, one considers the hyperbolic-parabolic behavior associated with the above system of equations. For non-relativistic flows, $1/\mathbb{C} = \mathcal{O}(\epsilon)$ where $\epsilon \ll 1$. Assume that there is a moderate amount of radiation in the flow and that scattering effects are small where $\sigma_s / \sigma_t = \mathcal{O}(\epsilon)$. \\

\noindent
The optical depth $\mathcal{L}$ is a useful quantity for classifying the limiting behavior of a system that is driven by radiation hydrodynamics:
\begin{equation}
\mathcal{L} = \int_{x_{min}}^{x_{max}} \sigma_t dx = \sigma_t (x_{\max}-x_{\min}) .
\end{equation}

\noindent
Optically thin regimes are characterized by $\mathcal{L} < \mathcal{O}(1)$, whereas optically thick regimes are characterized by $\mathcal{L} > \mathcal{O}(1)$. In optically thin regimes (free streaming limit), radiation and hydrodynamics decouple such that the resulting dynamics resemble an advection process. In optically thick regimes (weak/strong equilibrium diffusion limit), radiation and hydrodynamics are strongly coupled and the resulting dynamics resemble a diffusion process. Given these definitions, one makes the following assumption $\mathcal{L} = \ell_{\textrm{mat}} / \lambda_{t} = \ell_{\textrm{mat}} ~ \sigma_{t}$, where $\lambda_{t}$ is the total mean free path of the photos and $\ell_{\textrm{mat}} = \mathcal{O}(1)$ is the material flow length scale \cite{lowrie1999}.

\subsubsection{Free Streaming Limit: $\sigma_a, \sigma_t \sim \mathcal{O}(\epsilon)$}
\noindent
In this regime, the right-hand-side of Equation \ref{eq:cons_law_source_1d} is negligible such that Equation \ref{eq:cons_law_1d} is strictly hyperbolic. Moreover, $f \rightarrow 1$ and the Jacobian of the quasi-linear conservation law in Equation \ref{eq:cons_law_quasi_1d} has eigenvalues $\lambda = u, u \pm a, \pm \mathbb{C}$.
\begin{equation}
\partdif{\rho}{t} + \partdif{}{x} \left( m \right) = 0 , \label{eq:streamingD} 
\end{equation}
\begin{equation}
\partdif{m}{t} + \partdif{}{x} \left( \frac{m^2}{\rho} + p \right) = 0 , \label{eq:streamingM}
\end{equation}
\begin{equation}
\partdif{E}{t} + \partdif{}{x} \left( \left( E+p \right) \frac{m}{\rho} \right) = 0 , \label{eq:streamingE}
\end{equation}
\begin{equation}
\partdif{E_r}{t} + \partdif{}{x} \left( \mathbb{C} F_r \right) = 0 , \label{eq:streamingER}
\end{equation}
\begin{equation}
\partdif{F_r}{t} + \partdif{}{x} \left( \mathbb{C} E_r \right) = 0 . \label{eq:streamingFR}
\end{equation}

\subsubsection{Weak Equilibrium Diffusion Limit: $\sigma_a, \sigma_t \sim \mathcal{O}(1)$}
\noindent
One obtains this limit by defining $\sigma_a$ and $\sigma_t$ to be of order unity in Equation \ref{eq:cons_law_source_1d}, matching terms of like order, and combining the resulting expressions. From the definition of the equilibrium state, $E_r = T^4 + \mathcal{O}(\epsilon)$ and $F_r = \epsilon u (E_r+P_r) - \frac{1}{\sigma_t} \partdif{P_r}{x}$. Therefore, the system is parabolic and resembles a diffusion equation, where $f \rightarrow 1/3$.
\begin{equation}
\partdif{\rho}{t} + \partdif{}{x} \left( m \right) = 0 \label{eq:weak_diff_D} ,
\end{equation}
\begin{equation}
\partdif{m}{t} + \partdif{}{x} \left( \frac{m^2}{\rho} + p^{*} \right) = 0  \label{eq:weak_diff_M} , 
\end{equation}
\begin{equation}
\partdif{E^*}{t} + \partdif{}{x} \left( \left( E^{*} + p^{*} \right) \frac{m}{\rho} \right) = \partdif{^2}{x^2} \left( \frac{\mathbb{P} \mathbb{C} E_r}{3 \sigma_a} \right) \label{eq:weak_diff_E} ,
\end{equation}

\noindent
where:
\begin{equation}
p^* = p + \mathbb{P} P_r = p + \frac{1}{3} \mathbb{P} T^4 , ~~~~
E^* = E + \mathbb{P} E_r = E + \mathbb{P} T^4 , ~~~~
e^* = e + \mathbb{P} T^4 .
\end{equation}

\noindent
If one only considers the left-hand-side of Equations \ref{eq:weak_diff_D}-\ref{eq:weak_diff_E}, the Jacobian $A_{\textrm{diff}}$ has eigenvalues $\lambda = u,u \pm a^*$, where $a^*$ is the total radiation modified sound speed. $a^*$ represents the combined influence of material and radiation (i.e., effects associated with source terms as well as multiscale wave speeds) and is the propagation speed which characterizes the above reduced dynamical system. Therefore, this quantity is different than the sound speed associated with just the material components of radiation hydrodynamics.
\begin{equation}
(a^*)^2 = p^*_{\rho} + \frac{p^* p^*_{e^*}}{\rho}, ~~~~~ 
p^*_{\rho} \equiv \partdif{p^*}{\rho}, ~~~~~ 
p^*_{e^*} \equiv \partdif{p^*}{e^*}.
\end{equation}

\noindent
Additionally, the radiation modifies the material such that \cite{lowrie1999} $(i)$ if the equation of state is thermally stable, then $a$ and $a^*$ are real-valued quantities, $(ii)$ if the hydrodynamic system without radiation is hyperbolic, then the reduced system is also hyperbolic, and $(iii)$ for $\gamma$-law gases, $a^* \geq a$. If one assumes a $\gamma$-law gas, then one can evaluate $a^*$ according to the following relationship \cite{mmbook}:
\begin{equation}
(a^*)^2 = \frac{\Gamma p*}{\rho} =  \frac{\Gamma (p + \mathbb{P} f E_r)}{\rho}, ~~~~
\Gamma = \frac{(\frac{\gamma}{\gamma-1} + 20\xi + 16\xi^2)}{(\frac{1}{\gamma-1} + 12\xi)(1+\xi)} , ~~~~
\xi = \frac{P_r}{p} ,
\end{equation}

\noindent
where $\Gamma$ is the Gruneisen coefficient and $\xi$ admits the following limits \cite{mmbook, lowrie1999, zeldovich1966}:
\begin{eqnarray}  
\xi \rightarrow 0 & \Rightarrow & \Gamma \rightarrow \gamma ~~ \textrm{(no radiation)} , \\
\xi \rightarrow \infty & \Rightarrow & \Gamma \rightarrow \frac{4}{3} ~~ \textrm{(pure radiation)} .
\end{eqnarray}

\subsubsection{Strong Equilibrium Diffusion Limit: $\sigma_a, \sigma_t \sim \mathcal{O}(1/\epsilon)$}
\noindent
One obtains this limit by defining $\sigma_a$ and $\sigma_t$ to be much larger than order unity in Equation \ref{eq:cons_law_source_1d} and following the steps outlined for the weak equilibrium diffusion limit. From the definition of the equilibrium state, $E_r = T^4 + \mathcal{O}(\epsilon)$ and $F_r = \epsilon u (E_r+P_r)$. Therefore, the right-hand-side of Equations \ref{eq:weak_diff_D}-\ref{eq:weak_diff_E} is negligible such that the system can be considered hyperbolic. The Jacobian of the quasi-linear conservation law has eigenvalues $\lambda = u, u \pm a^*$, where $f \rightarrow 1/3$.
\begin{equation}
\partdif{\rho}{t} + \partdif{}{x} \left( m \right) = 0 , 
\end{equation}
\begin{equation}
\partdif{m}{t} + \partdif{}{x} \left( \frac{m^2}{\rho} + p^{*} \right) = 0 , 
\end{equation}
\begin{equation}
\partdif{E^*}{t} + \partdif{}{x} \left( \left( E^{*} + p^{*} \right) \frac{m}{\rho} \right) = 0 ,
\end{equation}

\subsubsection{Isothermal Limit}
\noindent
Lastly, Lowrie et al 1999 evidenced an additional limit which comprises a large portion of parameter space: $\mathcal{O}(\epsilon) \leq \mathcal{L} \leq \mathcal{O}(1/\epsilon)$. This isothermal limit has some dynamical properties in common with the weak equilibrium diffusion limit, but its defining characteristic is that the material temperature $T(x,t)$ is constant. Additionally, acoustic waves associated with this parameter regime propagate according to the isothermal sound speed $(a^*)^2 = p / \rho$. This limiting behavior results from temperature deviations on the slow material flow scale quickly reaching equilibrium on the fast radiation flow scale \cite{lowrie1999}. This paper will later show how the hybrid Godunov method reproduces this isothermal limit.


\section{Overview of Algorithm and the Nike Code}
\noindent
In radiation hydrodynamics, there are three important dynamical scales: the speed of sound (material flow), speed of light (radiation flow), and speed at which the source terms interact. When the matter-radiation coupling is weak (free streaming limit), the source terms define the slowest scale and the speed of light defines the fastest scale. However, when the matter-radiation coupling is strong (equilibrium diffusion limit), the speed of sound defines the slowest scale and the source terms as well as the speed of light define the fastest scale. Therefore, one must contend with stiffness arising from some waves propagating at the speed of light as well as source terms that can have large magnitudes. \\

\noindent
From a mathematical perspective, radiation hydrodynamics can be thought of as a system of hyperbolic balance laws with the following dual multiscale behavior: $(i)$ multiscale behavior associated with the hyperbolic wave speeds, such that $c / a_{\infty} \sim 10^{6}$, which causes breakdowns in monotonicity and $(ii)$ multiscale behavior associated with source term relaxation, such that $S / a_{\infty} \sim [0, 10^{6}]$, which causes breakdowns in stability. Despite these dual behaviors being different, they both are sources of stiffness that influence the temporal resolution of the problem. Rigorous definitions for monotonicity and stability are presented in Matus \& Lemeshevsky 2009. Given these variations, one desires a numerical technique that treats the material flow (slowest hyperbolic scale) explicitly, radiation flow (fastest hyperbolic scale) implicitly, and source terms (slow and fast relaxation terms) semi-implicitly. \\

\noindent
The hybrid Godunov method was chosen over two other algorithmic ideas (fully implicit methods and $P_N P_M$ schemes) because of the method's reliance on familiar notions from Godunov-type techniques, ability to be easily implemented with minimal computational overhead, and accuracy across a wide range of physical behavior. One could have built a fully implicit method that advanced time according to the material flow scale, thereby circumventing any stiffness related to some hyperbolic waves propagating at the speed of light as well as matter-radiation coupling terms. Such an approach was not pursued because these methods often have difficulties associated with conditioning, invoke root finding techniques when nonlinearities are present, are computationally expensive because of matrix inversion, are usually built into central difference schemes rather than higher-order Godunov methods, and have trouble sharply resolving discontinuous solutions associated with advection and shock phenomena. $P_N P_M$ schemes are a new family of arbitrary higher-order accurate numerical methods for hyperbolic conservation laws. The $N$ designates the degree of the polynomials used for test functions in a quasi-WENO (Weighted Essentially Non-Oscillatory) technique to spatially reconstruct cell-centered quantities \cite{det2008, dbtm2008}. This WENO-like reconstruction is found in ADER (Advection-Diffusion-Reaction) schemes \cite{det2008, titarev2002}. The $M$ designates the degree of the polynomials used for flux and source term computation in a local space-time DG (discontinuous Galerkin) finite element procedure for temporal evolution \cite{det2008, dbtm2008}. A general feature of the $P_N P_M$ family of schemes is that they contain classical higher-order finite volume methods $(N=0)$ as well as DG methods $(M=N)$. Due to the inherent stiffness of radiation hydrodynamical problems, some authors have suggested abandoning classical Godunov methods in favor of DG techniques \cite{lowrie2001, lowrie2002}. However, such methods would be computationally more expensive than the hybrid Godunov method discussed here and more difficult to implement in existing codes. Additionally, $P_N P_M$ schemes would not be able to contend with stiffness arising from some waves propagating at the speed of light and some modification is required. Nevertheless, $P_N P_M$ schemes have been successfully applied to problems in resistive relativistic MHD \cite{dz2009} and could be applicable to other characteristically stiff problems.

\subsection{Effective CFL Condition}
\noindent
Resolving the evolution of material quantities is the primary interest when solving problems in radiation hydrodynamics. The hybrid method computes $(\rho,m,E)$ to second-order accuracy as the entire algorithm is advanced according to an effective CFL condition:
\begin{equation}
\Delta t = \frac{\nu \Delta x}{\max_{i} ( |u_i| + a_{\textrm{eff},i} )} ,
\end{equation}

\noindent 
where $\Delta t$ is the time step, $\nu \in [0,1]$ is the CFL number, $\Delta x = (x_{\max}-x_{\min}) / N_{\textrm{cell}}$ is the grid spacing or spatial resolution for a given number of computational grid cells $N_{\textrm{cell}}$, and $\max_{i} ( |u_i| + a_{\textrm{eff},i} )$ is the maximum material wave speed over all grid cells. Furthermore, $a_{\textrm{eff}}$ is an estimate of the radiation modified sound speed that is obtained by carrying out an effective eigen-analysis of the material Jacobian. This analysis is presented in a later section.

\subsection{Algorithmic Steps}
\noindent
After defining $\Delta t$, the algorithm loops over the following steps:
\begin{enumerate}
\item Backward Euler Upwinding Scheme - implicitly advances the radiation quantities from time $t_{n}$ to time $t_{n+1}$: $(E_r^{n}, F_r^{n}) \rightarrow (E_r^{n+1},F_r^{n+1})$ and handles breakdowns in monotonicity related to the multiscale hyperbolic wave speeds
\item Modified Godunov Predictor Scheme - couples stiff source term effects to the hyperbolic structure of the balance laws for the material quantities and uses effective piecewise linear extrapolation to spatially reconstruct material quantities at the left/right sides of cell interfaces $i \pm 1/2$: $U_{L/R,i+1/2}^{m,n+1/2}$
\item Flux Function - evaluates the passage of material across cell interfaces using left/right material states $U_{L/R,i+1/2}^{m,n+1/2}$ and an approximate Riemann solver
\item Modified Godunov Corrector Scheme - semi-implicitly advances the material quantities from time $t_{n}$ to time $t_{n+1}$: $(\rho^{n},m^{n},E^{n}) \rightarrow (\rho^{n+1},m^{n+1},E^{n+1})$ and handles breakdowns in stability related to multiscale source term relaxation
\item Apply boundary conditions
\item Compute next time step $\Delta t$
\end{enumerate}

\noindent
In the above expressions, $U$ represents all of the conserved quantities, $U^{r}$ represents the conserved radiation quantities $(E_{r},F_{r})$, and $U^{m}$ represents the conserved material quantities $(\rho,m,E)$. $i$ represents the location of a cell center, $i \pm 1/2$ represents the location of a cell interface to the right/left of $i$, and $n$ represents the time discretization. Details about each step are explained in later sections.

\subsection{Numerical Properties}
\noindent
When solving a system of hyperbolic balance laws with stiff source terms, there are six properties that a method must satisfy in order to produce confident numerical solutions. The first three properties pertain to solving hyperbolic conservation laws and are: \textbf{consistency} (truncation error $T_{\Delta} \rightarrow 0$ as $\Delta t , \Delta x \rightarrow 0$), \textbf{stability} (numerical solution remains bounded even in the limit of $n \rightarrow \infty$ and $\Delta t \rightarrow 0$, where time $t_{n} = n \Delta t$ is finite), and \textbf{accuracy} (solution to a finite difference equation $\psi_{\Delta}$ converges to the true solution $\psi$ at some rate as $\Delta t , \Delta x \rightarrow 0$). These properties are guaranteed because the hybrid method is based on the classical Godunov scheme \cite{mc2007, det2008, treb2005, leveque1}. \\

\noindent
If there are source terms, stiffness, and multiple scales associated with the system of equations, then a method must also be: \textbf{coarsely gridded} (grid size does not need to resolve small scale features attributed to the source terms, therefore $\Delta t$, $\Delta x$ are based on the associated homogeneous hyperbolic conservation law), \textbf{well-balanced} (method preserves steady states), and \textbf{asymptotic preserving} (method gives the correct asymptotic behavior during the relaxation of the system by the stiff source terms) \cite{det2008}. The first property is satisfied because the radiation quantities are implicitly advanced and the material quantities are semi-implicitly advanced according to a time step $\Delta t$ which is defined by a conventional CFL condition that is governed by the material flow scale. The second property pertains to problems where there is a balance between flux and source terms and the system exhibits no temporal change. This property is satisfied in the modified Godunov corrector scheme when the error estimate $\epsilon(\Delta t) = 0$, such that $\frac{1}{2} \left( S^{m}(\hat{U},U^{r,n+1}) + S^{m}(U^{m,n},U^{r,n+1}) \right) = \nabla \cdot F^{m,n+1/2}$. Lastly, Miniati \& Colella 2007 showed their modified Godunov method to be asymptotic preserving by looking at the truncation error for a general system of stiff balance laws and examining the behavior of the eigenvalues $\lambda_{\textrm{eff}}$ for the effective Jacobian $A_{\textrm{eff}}$. The last property is satisfied by $a_{\textrm{eff}}$ obtaining the correct asymptotic behavior as one changes the parameters $\{ \mathbb{C},\mathbb{P},\sigma_a, \sigma_t \}$. This effect is examined in a later section.


\section{Backward Euler Upwinding Scheme}
\noindent
This section presents the implicit scheme that advances the radiation quantities $(E_{r}, F_{r})$ according to the material flow scale. The stability of explicit schemes (e.g., Godunov-type methods) is governed by the CFL condition which restricts the allowable time step according to the fastest characteristic speed. However, if the overall hyperbolic system consists of multiscale wave speeds, then explicit schemes can become inefficient. Radiation hydrodynamics (characterized by the speed of light $c$ and material sound speed $a_{\infty}$ as well as the speed at which source terms interact) is an example of a system that exhibits dual multiscale behavior. This section addresses the multiscale behavior associated with the hyperbolic wave speeds.

\subsection{Implicit Schemes and TVD Conditions}
\noindent
It is well known that numerical difficulties arise when applying implicit schemes to systems of hyperbolic conservation laws, particularly when discontinuities are present. Moreover, in order to ensure non-oscillatory solutions when using linear implicit higher-order time integration methods, one must impose time step restrictions because of the total variation diminishing (TVD) condition \cite{duraisamy2007, gottlieb2001, harten1983}. Following the presentation in \cite{harten1984}, the total variation of a mesh function $v$ is defined by the following equation:
\begin{equation}
TV(v) = \sum_{i = - \infty}^{\infty} | v_{i+1} - v_{i} | , 
\end{equation}

\noindent
where the computational grid is indexed by $i$. One uses this relation to refer to a numerical scheme $( v^{n+1} = \mathcal{S} \circ v^{n} )$ as being TVD, if $TV(v^{n+1}) \leq TV(v^{n})$ where $v^{n}$ denotes an approximation to the true solution at time $t_{n}$. This idea is important to numerical methods because Harten 1983 proved that:
$(i)$ a monotone scheme is TVD and $(ii)$ a TVD scheme is monotonicity preserving. In this context, a numerical method is monotonicity preserving if $(i)$ no new local extrema are created within the spatial domain of the numerical solution and $(ii)$ the value of a local minimum is non-decreasing which is to say, the value of a local maximum is non-increasing. \\

\noindent
From a TVD perspective, Duraisamy \& Baeder 2007 presents ratios of the maximum allowable time step for a few implicit schemes with respect to the maximum allowable time step for the explicit forward Euler scheme. From these ratios, one sees that linear implicit higher-order schemes are impractical when solving flow fields with discontinuities such that $(\Delta t)_{\textrm{imp}}^{\textrm{HO}} / (\Delta t)_{\textrm{exp}} \sim \mathcal{O}(1)$. Only the implicit backward Euler method allows for $(\Delta t)_{\textrm{imp}}^{\textrm{BE}} / (\Delta t)_{\textrm{exp}} \rightarrow \infty$ \cite{duraisamy2007}. This condition is important when numerically solving radiation hydrodynamical problems because the backward Euler scheme, which is unconditionally TVD stable, can handle the high degree of stiffness associated with the following time scales:
\begin{equation}
\frac{ (\Delta t)_{\textrm{imp}}^{\textrm{BE}} }{ (\Delta t)_{\textrm{exp}} } > \frac{ (\Delta t)_{\textrm{mat}} }{ (\Delta t)_{\textrm{rad}} } \approx \frac{c}{a_{\infty}} \approx 10^{6}. 
\end{equation}

\noindent
Furthermore, the above statement is consistent with the \textbf{Godunov order barrier theorem}, which states: linear numerical schemes for solving partial differential equations that have the property of not generating new extrema (i.e., a monotone scheme) can be at most first-order accurate \cite{godunov1959}.

\subsection{Linearity and Choice of Algorithm}
\noindent
If one thinks about these numerical difficulties from a physical perspective, then one can intuitively understand why implicit higher-order schemes are not viable approaches for a coupled hyperbolic system where some waves move at the speed of sound and other waves move at the speed of light. Clearly, there are going to be issues with higher-order schemes that do not enforce limiting conditions because the fast waves will have traversed a large distance in the computational domain, altered the reconstruction, and caused significant oscillations. Therefore, one needs to ensure monotonicity in a higher-order implicit scheme that attempts to solve a stiff problem. However, higher-order implicit updates for the radiation quantities are unlikely because in order to preserve monotonicity one must invoke TVD limiting or artificial viscosity conditions. When applying limiting techniques to the implicitly defined linear radiation subsystem , one forces the subsystem to be highly nonlinear. This approach makes the problem difficult to solve, especially in multiple spatial dimensions where convergence is not guaranteed. When using artificial viscosity, one adds a competing parabolic/diffusion process. This is a dangerous approach and will likely give the incorrect solution in certain parameter regimes because radiation hydrodynamics already exhibits parabolic relaxation. Therefore, one is forced to use the first-order backward Euler technique. Yet, one way to obtain better resolution while maintaining monotonicity is to use adaptive mesh refinement (AMR). Such a technique would be invaluable for accurately depicting discontinuous radiative phenomena and is an future area of algorithmic research.  \\

\noindent
Since backward Euler-type schemes naturally handle stiff problems and since a primary goal of this work is to resolve the material components of radiation hydrodynamics, a backward Euler-type scheme provides the suitable framework for defining the implicit update step. The backward Euler framework can be cast in the following form:
\begin{equation}
U^{n+1} = U^{n} + \Delta t \mathbb{L} \left( U^{n+1} \right)
\end{equation}

\noindent
where $\mathbb{L}$ is a linear operator that defines spatial differences and numerical fluxes. The linearity of this operator is important because the radiation subsystem has no nonlinearities with respect to $E_{r}$ or $F_{r}$. Nonlinearities only arise in the material quantities $(\rho, m, E)$ which are held at time $t_{n}$. Lastly, Duraisamy \& Baeder 2007 describe how to approximate the flux integral at cell interfaces which takes the form:
\begin{equation}
F_{i+1/2} = \frac{1}{\Delta t} \int_{t_{n}}^{t_{n+1}} F \left( U (x_{i+1/2},t) \right) dt , 
\end{equation}

\noindent 
such that one constructs an implicit flux function. With this idea in mind, an HLLE framework was chosen for this algorithm because the HLLE flux function is a simple approximate Riemann solver with numerical properties that guarantee conservation and good behavior for isolated discontinuities as well as smooth solutions. Choosing to implicitly advance the radiation quantities following the backward Euler HLLE scheme ultimately splits the radiation and material components even though the modified Godunov scheme that semi-implicitly advances the material quantities fully maintains the coupling between the material components and source terms. However, this procedural choice does not decrease the accuracy or stability of the respective schemes. Moreover, the sum of the total energy $(E+E_r)$ across all space $x$ is conserved over time. These results are evidenced in the numerical tests presented later in Sections 8-10.

\subsection{HLLE Framework}
\noindent
The HLLE scheme is based on estimating the minimum and maximum wave speeds $(s_{\min}, s_{\max})$ that arise in the Riemann problem: $U_{i+1/2} = \mathcal{R}(U_{L,i+1/2},U_{R,i+1/2})$. The numerical flux is calculated from:
\begin{equation}
F_{i+1/2}^{\textrm{HLLE}} (\mathcal{R}(U_{L},U_{R})) = \frac{1}{2} \left( \mathcal{F}_{L} + \mathcal{F}_{R} \right) + \frac{1}{2} \left( \frac{s_{\max} + s_{\min}}{s_{\max} - s_{\min}} \right) \left( \mathcal{F}_{L} - \mathcal{F}_{R} \right) , 
\end{equation}
\begin{equation}
\mathcal{F}_{L} = F(U_{L}) - s_{\min} U_{L} , ~~~~ \mathcal{F}_{R} = F(U_{R}) - s_{\max} U_{R} ,
\end{equation}

\noindent
By combining these relations, one arrives at:
\begin{equation}
F_{i+1/2}^{\textrm{HLLE}} (\mathcal{R}(U_{L},U_{R})) = \frac{1}{2} \left( \left( 1 + C^{s} \right) \left( F(U_{L}) - s_{\min} U_{L} \right) + \left( 1 - C^{s} \right) \left( F(U_{R}) - s_{\max} U_{R} \right) \right) . \label{eq:HLLE1} 
\end{equation}

\noindent 
where $C^{s} = (s_{\max} + s_{\min}) / (s_{\max} - s_{\min})$. Inserting $C^{s}$ into Equation \ref{eq:HLLE1} results in the more familiar relation:
\begin{equation}
F_{i+1/2}^{\textrm{HLLE}} (\mathcal{R}(U_{L},U_{R})) = \frac{s_{\max} F(U_{L}) - s_{\min} F(U_{R}) + s_{\min} s_{\max} (U_{R} - U_{L})}{s_{\max} - s_{\min}} \label{eq:HLLE2} .
\end{equation}

\noindent
However, Equation \ref{eq:HLLE1} is the form used by the algorithm's backward Euler update step. Defining the left/right states of the Riemann problem according to a first-order accurate (piecewise constant) reconstruction, forces the HLLE flux function to become:
\begin{equation}
F_{i+1/2}^{\textrm{HLLE}} (\mathcal{R}(U_{L},U_{R})) = F_{i+1/2}^{\textrm{HLLE}} (\mathcal{R}(U_{i},U_{i+1})) ,
\end{equation}

\noindent
such that the exact integral formulation of the conservative differencing is:
\begin{equation}
U_{i}^{n+1} = U_{i}^{n} - \frac{\Delta t}{\Delta x} \left( F_{i+1/2}(\mathcal{R}(U_{i},U_{i+1})) - F_{i-1/2}(\mathcal{R}(U_{i-1},U_{i})) \right) + \Delta t S(U_{i}^{n}) .
\end{equation}

\noindent
The choice of using a first-order accurate (piecewise constant) reconstruction was made to simplify the calculation and because higher order accuracy is not required since backward Euler-type schemes are already first-order accurate. One makes the above explicit HLLE scheme implicit by defining the variables in the flux and source terms to be at time $t_{n+1}$:

{\footnotesize {\begin{equation}
U_{i}^{n+1} = U_{i}^{n} - \frac{\Delta t}{\Delta x} \left( F_{i+1/2}^{\textrm{HLLE}}(\mathcal{R}(U_{i}^{n+1},U_{i+1}^{n+1})) - F_{i-1/2}^{\textrm{HLLE}}(\mathcal{R}(U_{i-1}^{n+1},U_{i}^{n+1})) \right) + \Delta t S(U_{i}^{n+1}) \label{eq:con_diff_imp} ,
\end{equation}
\begin{equation}
F_{i+1/2}^{\textrm{HLLE}} = \frac{1}{2} \left( \left( 1 + C_{i+1/2}^{s} \right) \left( F(U_{i}^{n+1}) - s_{\min} U_{i}^{n+1} \right) + \left( 1 - C_{i+1/2}^{s} \right) \left( F(U_{i+1}^{n+1}) - s_{\max} U_{i+1}^{n+1} \right) \right) \label{eq:hlle_imp} .
\end{equation} }}

\subsection{Applying the Backward Euler HLLE Scheme}
\noindent
Consider only the radiation part of the equations for radiation hydrodynamics (i.e., Equations \ref{eq:rh4} and \ref{eq:rh5}) \cite{sekora2009}. This simpler system is termed the radiation subsystem and the variables, fluxes, and source terms are:
\begin{equation}
U^{r} = 
\left( \begin{array}{c} E_r \\
                        F_r \end{array} \right) , ~~~~
F^{r}(U) = 
\left( \begin{array}{c} \mathbb{C} F_r \\
                        \mathbb{C} f E_r \end{array} \right) , \label{eq:radsub1}
\end{equation}
\begin{equation}
S^{r}(U) = 
\left( \begin{array}{c} \mathbb{C} S^E \\
                        \mathbb{C} S^F \end{array} \right) = 
\left( \begin{array}{c} 
\mathbb{C} \left [ \sigma_a(T^4 - E_r) + (\sigma_a - \sigma_s) \frac{m}{\rho \mathbb{C}} \left( F_r - \frac{ (1+f) m E_r }{\rho \mathbb{C}} \right) \right ] \\
\mathbb{C} \left [ -\sigma_t \left( F_r - \frac{ (1+f) m E_r }{\rho \mathbb{C}} \right) + \sigma_a \frac{m}{\rho \mathbb{C}} (T^4 - E_r) \right ] 
                        \end{array} \right) , \label{eq:radsub2}
\end{equation}

\noindent
where the eigenvalues of the radiation subsystem in the free streaming limit $( \sigma_a, \sigma_t \sim \mathcal{O}(\epsilon) )$ are $\lambda^{\pm} = \pm f^{1/2} \mathbb{C}$. Given that the HLLE scheme uses minimum and maximum wave speeds to compute fluxes at cell interfaces $i+1/2$, one defines the following equations:
\begin{equation}
s_{\min} = \lambda_{L,i  }^{-} = - f_{i  }^{1/2} \mathbb{C} , ~~~~
s_{\max} = \lambda_{R,i+1}^{+} =   f_{i+1}^{1/2} \mathbb{C} , ~~~~
C_{i+1/2}^{s} = \frac{f_{i+1}^{1/2} - f_{i}^{1/2}}{f_{i+1}^{1/2} + f_{i}^{1/2}} ,
\end{equation}

\noindent
where $f$ arises from the closure relation $P_{r} = f E_{r}$ and is either a user defined quantity or obtained by solving the radiation transport equation. If $f$ varies spatially, then $C_{i+1/2}^{s}$ is non-zero. Defining or computing $f(x,t)$ precedes the backward Euler update of $U^{r}$. For all of the numerical tests presented in this paper, $f$ is assumed to be spatially and temporally constant, thereby setting $C_{i+1/2}^{s} = 0$. Future work will update $f(x,t)$ at each iteration by solving the radiation transport equation.

\subsection{Matrix Equation for the Radiation Components}
\noindent
Inputting $U^{r,n+1}$, $F^{r}(U^{r,n+1})$, and $S^{r}(U^{m,n},U^{r,n+1})$ into Equations \ref{eq:con_diff_imp} and \ref{eq:hlle_imp} gives the following difference equations:

{\footnotesize {\begin{eqnarray}
& ~ & E_{r,i-1}^{n+1} \left[ -d_{1} \left( 1 + C_{i-1/2}^{s} \right) f_{i-1}^{1/2} \right] \\
& + & F_{r,i-1}^{n+1} \left[ -d_{1} \left( 1 + C_{i-1/2}^{s} \right) \right] \nonumber \\
& + & E_{r,i}^{n+1} \left[ 1 + d_{1} \left( 1 + C_{i+1/2}^{s} \right) f_{i}^{1/2} + d_{1} \left( 1 - C_{i-1/2}^{s} \right) f_{i}^{1/2} + d_{2} \sigma_{a} + \frac{ d_{2} \left( \sigma_{a} - \sigma_{s} \right) \left( 1 + f_{i} \right) \left( m_{i}^{n} \right)^2 }{ \left( \rho_{i}^{n} \right)^2 \mathbb{C}^2 } \right] \nonumber \\
& + & F_{r,i}^{n+1} \left[ d_{1} \left( 1 + C_{i+1/2}^{s} \right) - d_{1} \left( 1 - C_{i-1/2}^{s} \right) - \frac{ d_{2} \left( \sigma_{a} - \sigma_{s} \right) m_{i}^{n} }{ \rho_{i}^{n}\mathbb{C} } \right] \nonumber \\
& + & E_{r,i+1}^{n+1} \left[ -d_{1} \left( 1 - C_{i+1/2}^{s} \right) f_{i+1}^{1/2} \right] \nonumber \\
& + & F_{r,i+1}^{n+1} \left[  d_{1} \left( 1 - C_{i+1/2}^{s} \right) \right] \nonumber \\
& = & E_{r,i}^{n} + d_{2} \sigma_{a} \left( T_{i}^{n} \right)^4 \nonumber , 
\end{eqnarray} 
\begin{eqnarray}
& ~ & E_{r,i-1}^{n+1} \left[ -d_{1} \left( 1 + C_{i-1/2}^{s} \right) f_{i-1} \right] \\
& + & F_{r,i-1}^{n+1} \left[ -d_{1} \left( 1 + C_{i-1/2}^{s} \right) f_{i-1}^{1/2} \right] \nonumber \\
& + & E_{r,i}^{n+1} \left[ d_{1} \left( 1 + C_{i+1/2}^{s} \right) f_{i} - d_{1} \left( 1 - C_{i-1/2}^{s} \right) f_{i} - \frac{ d_{2} \sigma_{t} \left( 1 + f_{i} \right) m_{i}^{n} }{ \rho_{i}^{n} \mathbb{C} } + \frac{ d_{2} \sigma_{a} m_{i}^{n} }{ \rho_{i}^{n} \mathbb{C} } \right] \nonumber \\
& + & F_{r,i}^{n+1} \left[ 1 + d_{1} \left( 1 + C_{i+1/2}^{s} \right) f_{i}^{1/2} + d_{1} \left( 1 - C_{i-1/2}^{s} \right) f_{i}^{1/2} + d_{2} \sigma_{t} \right] \nonumber \\
& + & E_{r,i+1}^{n+1} \left[  d_{1} \left( 1 - C_{i+1/2}^{s} \right) f_{i+1} \right] \nonumber \\
& + & F_{r,i+1}^{n+1} \left[ -d_{1} \left( 1 - C_{i+1/2}^{s} \right) f_{i+1}^{1/2} \right] \nonumber \\
& = & F_{r,i}^{n} + \frac{ d_{2} \sigma_{a} \left( T_{i}^{n} \right)^4 m_{i}^{n} }{ \rho_{i}^{n} \mathbb{C} } \nonumber ,
\end{eqnarray} }}

\noindent
where $d_{1} = \Delta t \mathbb{C} / 2 \Delta x$, $d_{2} = \Delta t \mathbb{C}$, and $T_{i}^{n} = p_{i}^{n} / \rho_{i}^{n} = \left( \gamma - 1 \right) \left( \frac{E_{i}^{n}}{\rho_{i}^{n}} - \frac{1}{2} \frac{\left( m_{i}^{n} \right)^2}{\left( \rho_{i}^{n} \right)^2} \right)$. If one attributes variables $\theta$ and $\phi$ to the bracketed quantities in the above relations, then the difference equations can be written as:
\begin{equation}
\theta_{1} E_{r,i-1}^{n+1} + \theta_{2} F_{r,i-1}^{n+1} + \theta_{3} E_{r,i}^{n+1} + \theta_{4} F_{r,i}^{n+1} + \theta_{5} E_{r,i+1}^{n+1} + \theta_{6} F_{r,i+1}^{n+1} = E_{r,i}^{n} + \theta_{7} ,
\end{equation}
\begin{equation}
\phi_{1} E_{r,i-1}^{n+1} + \phi_{2} F_{r,i-1}^{n+1} + \phi_{3} E_{r,i}^{n+1} + \phi_{4} F_{r,i}^{n+1} + \phi_{5} E_{r,i+1}^{n+1} + \phi_{6} F_{r,i+1}^{n+1} = F_{r,i}^{n} + \phi_{7} .
\end{equation}

\noindent
Since there are no nonlinearities in the radiation quantities for which root finding (e.g., Newton's method) must be implemented, one casts these equations into a sparse matrix format that can be solved exactly with basic linear algebra techniques (i.e., Gaussian elimination and back substitution). The formulation $Ax = b$ has dimensions $\dim(A) = 2N \times 2N$, where elements $\{ E_{r,1},F_{r,1},E_{r,N},F_{r,N} \}$ account for the boundary conditions. Efficient matrix operations can be performed across the relevant radiation hydrodynamical parameter space because the matrix $A$ is diagonally dominant. Four types of boundary conditions (periodic, outflow, reflecting, and inflow) were applied to the numerical tests presented in this paper. The ways in which these boundary conditions modify the matrix equations are shown in Appendix 1. \\

\noindent
It is important to note that the backward Euler upwinding scheme naturally accounts for radiation dominated problems $(P_{r} > p$ or $E_{r} > T^{4})$. As was discussed earlier and will be shown in the following section, the modified Godunov method preserves the isothermal limit. When $P_{r}$ increases with respect to $p$, material effects as well as updates from the modified Godunov scheme become less important. For such physical phenomena, the backward Euler upwinding scheme becomes the dominant numerical method in the overall algorithm. If one examines Equations \ref{eq:rh1}-\ref{eq:rh5} as well as Equations \ref{eq:radsub1} and \ref{eq:radsub2}, then one notices that $\mathbb{P}$ is only associated with the source terms of the material components. To this end, radiation dominated systems are naturally handled by the backward Euler upwinding scheme and is evidenced in the numerical tests presented in this paper.


\section{Modified Godunov Predictor Scheme}
\noindent
Given that the radiation quantities $(E_{r},F_{r})$ are at time $t_{n+1}$, one computes the flux divergence $(\nabla \cdot F^{m})^{n+1/2}$ for the material quantities $(\rho, m, E)$ that are at time $t_{n}$. Following the analysis of Trebatich et al 2005, Miniati \& Colella 2007, and Sekora \& Stone 2009, one applies Duhamel's principle to the quasi-linear system of balance laws in Equations \ref{eq:cons_law_quasi_1d} and \ref{eq:cons_law_jacobian_1d} for only the material components. This technique defines the following system that locally includes in space-time the effects of the stiff source terms on the hyperbolic structure:
\begin{equation}
\frac{D U^{m}_{\textrm{eff}}}{Dt} = \mathcal{I}(\eta) \left( - A^{m}_{L} \partdif{U^{m}}{x} + S^{m}(U^{m,n},U^{r,n+1}) \right) ,
\end{equation}

\noindent 
where $\frac{DU^{m}}{Dt} = \partdif{U^{m}}{t} + u \partdif{U^{m}}{x}$ is the total derivative, $\mathcal{I}$ is a propagation operator that projects the dynamics of the stiff source terms onto the hyperbolic structure, and $A^{m}_{L} = A^{m} - uI$ is the Jacobian for a Lagrangian trajectory with $I$ being the identity matrix. Since the predictor scheme is a first-order accurate step in a second-order accurate predictor-corrector method, one chooses $\eta = \Delta t / 2$ and the effective balance law becomes:
\begin{equation} 
\partdif{U^{m}}{t} + A^{m}_{\textrm{eff}} \partdif{U^{m}}{x} = \mathcal{I}(\Delta t / 2) S^{m}(U^{m,n},U^{r,n+1}) ,
\end{equation}

\noindent 
where $A^{m}_{\textrm{eff}} = \mathcal{I}(\Delta t / 2) A^{m}_{L} + u I$. In order to compute $\mathcal{I}$, one first computes $\nabla_{U^{m}} S^{m}(U)$.

\subsection{Applying the Modified Godunov Predictor Scheme}
\noindent
If one only considers the material component in Equations \ref{eq:cons_law_1d}-\ref{eq:cons_law_source_1d}, then the variables, fluxes, and source terms are:
\begin{equation}
U^{m} = 
\left( \begin{array}{c} 
\rho \\
m \\
E \end{array} \right) , ~~~~
F^{m}(U) = 
\left( \begin{array}{c} 
m \\
\frac{m^2}{\rho} + p \\
\left( E + p \right) \frac{m}{\rho} \end{array} \right) , 
\end{equation}
\begin{equation}
S^{m}(U) = 
\left( \begin{array}{c} 
0 \\
- \mathbb{P} S^{F} \\
- \mathbb{P} \mathbb{C} S^{E} \end{array} \right) = 
\left( \begin{array}{c} 
0 \\
- \mathbb{P} \left [ -\sigma_t \left( F_r - \frac{ (1+f) m E_r }{\rho \mathbb{C}} \right) + \sigma_a \frac{m}{\rho \mathbb{C}} (T^4 - E_r) \right ] \\
- \mathbb{P} \mathbb{C} \left [ \sigma_a(T^4 - E_r) + (\sigma_a - \sigma_s) \frac{m}{\rho \mathbb{C}} \left( F_r - \frac{ (1+f) m E_r }{\rho \mathbb{C}} \right) \right ]
\end{array} \right)  \label{eq:Sm} .
\end{equation}

\noindent
Therefore:
\begin{equation}
\nabla_{U^{m}} S^{m}(U) = \left( \begin{array}{ccc}
0 & 0 & 0 \\
-\mathbb{P} S^{F}_{\rho} & -\mathbb{P} S^{F}_{m} & -\mathbb{P} S^{F}_{E} \\
-\mathbb{P} \mathbb{C} S^{E}_{\rho} & -\mathbb{P} \mathbb{C} S^{E}_{m} & -\mathbb{P} \mathbb{C} S^{E}_{E} 
\end{array} \right) \label{eq:dSodU} ,
\end{equation}

\noindent
where the partial derivatives are:
\begin{eqnarray}
S^{F}_{\rho} & = & \frac{-\sigma_{t} \left( 1 + f \right) m E_{r}}{\rho^2 \mathbb{C}} - \frac{\sigma_{a} m \left( T^{4} - E_{r} \right) }{\rho^{2} \mathbb{C}} + \frac{4 \sigma_{a} m T^{3}}{\rho \mathbb{C}} \left( \gamma - 1 \right) \left( \frac{-E}{\rho^{2}} + \frac{m^{2}}{\rho^{3}} \right) , \label{eq:part1} \\
S^{F}_{m} & = & \frac{\sigma_{t} \left( 1 + f \right) E_{r}}{\rho \mathbb{C}} + \frac{\sigma_{a} \left( T^{4} - E_{r} \right)}{\rho \mathbb{C}} + \frac{4 \sigma_{a} m T^{3}}{\rho \mathbb{C}} \left( \gamma - 1 \right) \left( \frac{-m}{\rho^{2}} \right) , \label{eq:part2} \\
S^{F}_{E} & = & \frac{4 \sigma_{a} m T^{3}}{\rho \mathbb{C}} \left( \gamma - 1 \right) \left( \frac{1}{\rho} \right) , \label{eq:part3} \\
S^{E}_{\rho} & = & 4 \sigma_{a} T^{3} \left( \gamma - 1 \right) \left( \frac{-E}{\rho^{2}} + \frac{m^{2}}{\rho^{3}} \right) - \frac{m \left( \sigma_{a} - \sigma_{s} \right) F_{r}}{\rho^{2} \mathbb{C}} + \frac{2 \left( \sigma_{a} - \sigma_{s} \right) \left( 1 + f \right) m^{2} E_{r}}{\rho^{3} \mathbb{C}^{2}} , \label{eq:part4} \\ 
S^{E}_{m} & = & 4 \sigma_{a} T^{3} \left( \gamma - 1 \right) \left( \frac{-m}{\rho^{2}} \right) + \frac{\left( \sigma_{a} - \sigma_{s} \right) F_{r}}{\rho \mathbb{C}} - \frac{2 \left( \sigma_{a} - \sigma_{s} \right) \left( 1 + f \right) m E_{r}}{\rho^{2} \mathbb{C}^{2}} , \label{eq:part5} \\
S^{E}_{E} & = & 4 \sigma_{a} T^{3} \left( \gamma - 1 \right) \left( \frac{1}{\rho} \right) . \label{eq:part6}
\end{eqnarray}

\subsubsection{Simplifying $\nabla_{U^{m}} S^{m}(U)$}
\noindent
In its current form, $\nabla_{U^{m}} S^{m}(U)$ in Equation \ref{eq:dSodU} leads to a propagation operator $\mathcal{I}$ that is difficult to work with algebraically. One should focus on the fact that the material momentum source term $-\mathbb{P} S^{F}$ is not the dominant factor defining the stiffness associated with the matter-radiation coupling. By inspection, $-\mathbb{P} S^{F} < \mathcal{O}(1)$ even in the strong equilibrium diffusion limit, where the contributing terms have the following magnitudes $\mathbb{P} \sim \mathcal{O}(1 / \mathbb{C})$, $\sigma_{a,s,t} \sim \mathcal{O}(\mathbb{C})$, $F_{r} < \mathcal{O}(1)$, $(1+f) m E_{r} / \rho \mathbb{C} < \mathcal{O}(1)$, $m / \rho \mathbb{C} \sim \mathcal{O}(1 / \mathbb{C})$, and $(T^{4}-E_{r}) < \mathcal{O}(1)$. Additionally, one finds that the derivative of the material momentum source term with respect to the conserved variables has the following magnitude $-\mathbb{P} S^{F}_{ \{ \rho,m,E \} } < \mathcal{O}(1)$. Therefore, $-\mathbb{P} S^{F}$ can be included like a body force (i.e., gravity).  \\

\noindent
It is the material energy source term $-\mathbb{P} \mathbb{C} S^{E}$ that defines the stiffness associated with the problem. By inspection of the contributing terms, $-\mathbb{P} \mathbb{C} S^{E} < \mathcal{O}(\mathbb{C})$ in the strong equilibrium diffusion limit. Additionally, one finds that the derivative of the material energy source term with respect to the conserved variables has the following magnitude $-\mathbb{P} \mathbb{C} S^{E}_{ \{ \rho,m,E \} } < \mathcal{O}(\mathbb{C}^{2})$. Therefore, one only needs to use $-\mathbb{P} \mathbb{C} S^{E}$ to define $\nabla_{U^{m}} S^{m}(U)$, such that:
\begin{equation}
\nabla_{U^{m}} S^{m}(U) = \left( \begin{array}{ccc}
0 & 0 & 0 \\
0 & 0 & 0 \\
-\mathbb{P} \mathbb{C} S^{E}_{\rho} & -\mathbb{P} \mathbb{C} S^{E}_{m} & -\mathbb{P} \mathbb{C} S^{E}_{E} 
\end{array} \right) \label{eq:dSodU_simp} .
\end{equation}

\noindent
$\nabla_{U^{m}} S^{m}(U)$ is further simplified by examining $S^{E}_{ \{ \rho,m,E \} }$ in the equilibrium diffusion limit and neglecting terms that have magnitudes of or less than $\mathcal{O}(\mathbb{C})$. Therefore:
\begin{eqnarray}
S^{E}_{\rho} & \rightarrow & 4 \sigma_{a} T^{3} \left( \gamma - 1 \right) \left( \frac{-E}{\rho^{2}} + \frac{m^{2}}{\rho^{3}} \right) , \label{simp:part4} \\ 
S^{E}_{m} & \rightarrow & 4 \sigma_{a} T^{3} \left( \gamma - 1 \right) \left( \frac{-m}{\rho^{2}} \right) , \label{simp:part5} \\
S^{E}_{E} & \rightarrow & 4 \sigma_{a} T^{3} \left( \gamma - 1 \right) \left( \frac{1}{\rho} \right) . \label{simp:part6}
\end{eqnarray}

\noindent
It is important to note that these partial derivatives have the same stiff magnitude $4 \sigma_{a} T^{3} \left( \gamma - 1 \right)$. This insight simplifies algebraic manipulation. \\

\noindent
If $\nabla_{U^{m}} S^{m}(U)$ is diagonalizable, then $\nabla_{U^{m}} S^{m}(U) = R D R^{-1}$. Here, $D = \textrm{diag}(0,0,-\mathbb{P} \mathbb{C} S^{E}_{E})$ and $R$ is a matrix whose columns are the right eigenvectors. Below, one sees how the stiff magnitudes cancel out:
\begin{equation}
R = \left( \begin{array}{ccc}
\frac{-S^{E}_{E}}{S^{E}_{\rho}} & \frac{-S^{E}_{m}}{S^{E}_{\rho}} & 0 \\
0 & 1 & 0 \\
1 & 0 & 1 
\end{array} \right) , ~~~~
R^{-1} = \left( \begin{array}{ccc}
\frac{-S^{E}_{\rho}}{S^{E}_{E}} & \frac{-S^{E}_{m}}{S^{E}_{E}} & 0 \\
0 & 1 & 0 \\
\frac{S^{E}_{\rho}}{S^{E}_{E}} & \frac{S^{E}_{m}}{S^{E}_{E}} & 1
\end{array} \right) .
\end{equation}

\subsubsection{Propagation Operator $\mathcal{I}$}
\noindent
The propagation operator $\mathcal{I}$ is defined in Miniati \& Colella 2007. Since one is considering a modified Godunov scheme with a predictor step of $\Delta t/2$:
\begin{eqnarray}
\mathcal{I} \left( \frac{\Delta t}{2} \right) & = & \frac{1}{\Delta t/2} \int^{\Delta t/2}_{0} e^{\tau \nabla_{U^{m}} S^{m}(U)} d\tau \\
& = & \left( \begin{array}{ccc}
            1 & 0 & 0 \\
            0 & 1 & 0 \\
            (\alpha-1) \frac{S^{E}_{\rho}}{S^{E}_{E}} & (\alpha-1) \frac{S^{E}_{m}}{S^{E}_{E}} & \alpha  
            \end{array} \right) \\
& = & \left( \begin{array}{ccc}
            1 & 0 & 0 \\
            0 & 1 & 0 \\
            (1-\alpha) \left( \frac{E}{\rho} - \frac{m^2}{\rho^2} \right) & (1-\alpha) \frac{m}{\rho} & \alpha  
            \end{array} \right) ,
\end{eqnarray} 

\noindent 
where $\alpha = \left( 1 - \exp ( -\mathbb{P} \mathbb{C} S^{E}_{E} \Delta t / 2 ) \right) / \left( \mathbb{P} \mathbb{C} S^{E}_{E} \Delta t / 2 \right)$. Since $S^{E}_{E} \geq 0$ across all relevant parameter space, $0 \leq \alpha \leq 1$. This property is important when considering stability and the subcharacteristic condition which is discussed later in the paper.

\subsection{Effective Material Jacobian $A^{m}_{\textrm{eff}}$}
\noindent
Before applying $\mathcal{I}$ to $A^{m}_{L}$, it is useful to understand that moving-mesh methods can be accommodated in non-relativistic descriptions of radiation hydrodynamics whenever an Eulerian frame treatment is employed. These methods do not require transformation to the comoving frame \cite{lowrie2001}. Since the non-dimensionalization is associated with the hydrodynamical scale, one can use $u_{\textrm{mesh}} = u$ from Lagrangian hydrodynamical methods. \\

\noindent
The effects of the stiff source terms on the hyperbolic structure are accounted for by transforming to a moving-mesh (Lagrangian) frame $A^{m}_{L} = A^{m} - uI$, applying the propagation operator $\mathcal{I}$ to $A^{m}_{L}$, and transforming back to an Eulerian frame gives the effective material Jacobian $A^{m}_{\textrm{eff}} = \mathcal{I} A^{m}_{L} + uI$ \cite{mc2007}:

{\scriptsize {\begin{equation}
A^{m}_{\textrm{eff}} = \left( \begin{array}{ccc} 
0 & 1 & 0 \\
\frac{\gamma-3}{2} u^2 & -(\gamma-3) u & (\gamma-1) \\
u \left( \frac{\gamma-1}{2} u^2 - \alpha \tilde{H} - (1-\alpha) \left( \frac{T}{\gamma-1} + \frac{1}{2} u^2 \right) \right) & - (\gamma-1) u^2 + \alpha \tilde{H} + (1-\alpha) \left( \frac{T}{\gamma-1} + \frac{1}{2} u^2 \right) & \gamma u 
\end{array} \right) ,
\end{equation} }}

\noindent 
which has eigenvalues $\lambda^{m}_{\textrm{eff},\{-,0,+\}} = u-a_{\textrm{eff}}, u, u + a_{\textrm{eff}}$. Here, the effective sound speed $a_{\textrm{eff}}$ (i.e., the radiation modified sound speed) is:
\begin{eqnarray}
a^{2}_{\textrm{eff}} & = & -\frac{\gamma-1}{2} u^2 + \alpha (\gamma-1) \tilde{H} + (1-\alpha) \left( T + \frac{\gamma-1}{2} u^2 \right)  \\
~                    & = & \alpha \frac{\gamma p}{\rho} + (1-\alpha) T  \\
~                    & = & \left( \alpha (\gamma-1) + 1  \right) \frac{p}{\rho} , 
\end{eqnarray}

\noindent
where $T = p / \rho$ because of the non-dimensionalization in Section 2 admitting the equation of state $p = \rho T (R T_{\infty} / a^{2}_{\infty})$. Here, one notices that $\tilde{H}, \left( T + (\gamma-1) u^2/2 \right) \geq 0$ across all relevant parameter space such that the effective sound speed $a_{\textrm{eff}}$ admits the following limits:
\begin{eqnarray}  
-\mathbb{P} \mathbb{C} S^{E}_{E} \rightarrow 0 \Rightarrow \alpha \rightarrow 1 & \Rightarrow & a^{2}_{\textrm{eff}} \rightarrow -\frac{\gamma-1}{2} u^2 + (\gamma-1) \tilde{H} = \frac{\gamma p}{\rho} ~~~~ \textrm{(adiabatic)}  \label{eq:aeff.ad} \\
-\mathbb{P} \mathbb{C} S^{E}_{E} \rightarrow -\infty \Rightarrow \alpha \rightarrow 0 & \Rightarrow & a^{2}_{\textrm{eff}} \rightarrow T = \frac{p}{\rho} ~~~~ \textrm{(isothermal)} . \label{eq:aeff.iso}
\end{eqnarray}

\noindent
When examining Equations \ref{eq:aeff.ad} and \ref{eq:aeff.iso}, one sees that the subcharacteristic condition for material wave speeds is satisfied, such that \cite{mc2007}:
{\small {\begin{equation}
\lambda^{m}_{-} = u-a_{\textrm{ad}} \leq \lambda^{m}_{\textrm{eff},-} = u-a_{\textrm{eff}} \leq \lambda^{m}_{0} = \lambda^{m}_{\textrm{eff},0} = u \leq \lambda^{m}_{\textrm{eff},+} = u+a_{\textrm{eff}} \leq \lambda^{m}_{-} = u+a_{\textrm{ad}} .
\end{equation} }}

\noindent
This condition is necessary for the stability of the system and guarantees that the numerical solution tends to the solution of the equilibrium equation as the relaxation time tends to zero. Additionally, the structure of the equations remains consistent with respect to classical Godunov methods. Therefore, the CFL condition $\rm{max}(|\lambda^{\textit{m}}_{\textrm{eff},*}|) \frac{\Delta t}{\Delta x} \leq 1$ applies, where $* = \{ -,0,+ \}$. \\

\noindent
It is important to note that $a_{\textrm{eff}}$ is different from $a^{*}$ which was defined earlier as the propagation speed associated with a reduced dynamical system that captures the combined influence of material and radiation (i.e., effects associated with source terms as well as multiscale wave speeds). $a_{\textrm{eff}}$ defines the effective material sound speed that estimates the influence of the source terms on the hyperbolic structure of the split material subsystem. Although the modified Godunov method is an unsplit finite volume technique that directly couples advection-reaction processes, there is an obvious splitting between the material (modified Godunov scheme) and radiation (backward Euler upwinding scheme) because of the multiscale nature of the hyperbolic wave speeds. $a_{\textrm{eff}}$ is designed to handle the former piece of this splitting. In the dispersion analysis of Lowrie et al 1999, $a^{*}$ is plotted for various parameters. As the optical depth is increased, $a^{*}$ takes values equal to the adiabatic sound speed and then values equal to the isothermal sound speed. At very large optical depths, $a^{*}$ takes values exceeding the isothermal sound speed. Because $a_{\textrm{eff}}$ is only defined for the split material subsystem, not the global radiation hydrodynamical system, $a_{\textrm{eff}}$ should only take values ranging between the adiabatic and isothermal sound speeds which is what is seen in Equations \ref{eq:aeff.ad} and \ref{eq:aeff.iso}. In situations where large optical depths are involved, radiation is the dominant species governing the dynamical behavior and is evolved using the backward Euler upwinding scheme for the split radiation subsystem. \\

\noindent
Lastly, the right material eigenvectors $R^{m}_{\textrm{eff}}$ (stored as columns) and left material eigenvectors $L^{m}_{\textrm{eff}}$ (stored as rows) are shown below. Clearly, there are many structural similarities between the effective eigen-quantities $A^{m}_{\textrm{eff}}$, $\lambda^{m}_{\textrm{eff}}$, $R^{m}_{\textrm{eff}}$, and $L^{m}_{\textrm{eff}}$ and those from adiabatic/isothermal hydrodynamics.
\begin{equation}
R^{m}_{\textrm{eff}} = \left( \begin{array}{ccc} 
1 & 1 & 1 \\
u-a_{\textrm{eff}} & u & u+a_{\textrm{eff}} \\
\frac{u^2}{2} - u a_{\textrm{eff}} + \frac{a^{2}_{\textrm{eff}}}{\gamma-1} & \frac{u^2}{2} & \frac{u^2}{2} + u a_{\textrm{eff}} + \frac{a^{2}_{\textrm{eff}}}{\gamma-1}
\end{array} \right) \label{eq:right_eff_evec} ,
\end{equation}
\begin{equation}
L^{m}_{\textrm{eff}} = \left( \begin{array}{ccc} 
\frac{u}{2 a_{\textrm{eff}}} \left( 1 + \frac{(\gamma-1) u}{2 a_{\textrm{eff}}} \right) & \frac{-1}{2 a_{\textrm{eff}}} \left( 1 + \frac{(\gamma-1) u}{a_{\textrm{eff}}} \right) & \frac{(\gamma-1)}{2 a^{2}_{\textrm{eff}}} \\
1 - \frac{(\gamma-1) u^2}{2 a^{2}_{\textrm{eff}}} & \frac{(\gamma-1) u}{a^{2}_{\textrm{eff}}} & \frac{-(\gamma-1)}{a^{2}_{\textrm{eff}}} \\
\frac{-u}{2 a_{\textrm{eff}}} \left( 1 - \frac{(\gamma-1) u}{2 a_{\textrm{eff}}} \right) & \frac{1}{2 a_{\textrm{eff}}} \left( 1 - \frac{(\gamma-1) u}{a_{\textrm{eff}}} \right) & \frac{(\gamma-1)}{2 a^{2}_{\textrm{eff}}}
\end{array} \right) \label{eq:left_eff_evec} .
\end{equation}

\subsection{Computing Left/Right States}
\noindent
In the modified Godunov predictor scheme one uses effective piecewise linear extrapolation to spatially reconstruct material quantities at the left/right sides of cell interfaces. This technique is given by the following relations:
\begin{eqnarray}
U_{L,i+1/2}^{m,n+1/2} = U_{i}^{m,n}   & + & \frac{1}{2} \left( I - \frac{\Delta t}{\Delta x} A_{\textrm{eff}}^{m}(U_{i}^{m,n}) \right) P_{\Delta}(\Delta U_{i}^{m,n}) \\
                                      & + & \frac{\Delta t}{2} \mathcal{I} \left( \frac{\Delta t}{2} \right) S^{m}(U_{i}^{m,n},U_{i}^{r,n+1}) \nonumber , \\
U_{R,i+1/2}^{m,n+1/2} = U_{i+1}^{m,n} & - & \frac{1}{2} \left( I + \frac{\Delta t}{\Delta x} A_{\textrm{eff}}^{m}(U_{i+1}^{m,n}) \right) P_{\Delta}(\Delta U_{i+1}^{m,n}) \\
                                      & + & \frac{\Delta t}{2} \mathcal{I} \left( \frac{\Delta t}{2} \right) S^{m}(U_{i+1}^{m,n},U_{i+1}^{r,n+1}) \nonumber ,
\end{eqnarray}

\noindent
where $I$ is the identity matrix, $A_{\textrm{eff}}^{m}$ is the effective material Jacobian, $P_{\Delta}$ is a slope limiting function used to eliminate spurious oscillations, $\mathcal{I}$ is the propagation operator, and $S^{m}$ is the vector of source terms influencing the material equations. Slope limiting is performed for each of the material quantities. Although any traditional slope limiter can be used, the Nike code employs the extremum-preserving \cite{colellasekora2008, sekoracolella2009} as well as the traditional van Leer limiter (also referred to as the MUSCL limiter). These techniques can be implemented either componentwise or across characteristic fields and are illustrated in Appendix 2. \\

\noindent
After reconstructing the material quantities, an approximate Riemann solver evaluates the passage of material across each cell interface using left/right states $U_{L/R,i+1/2}^{m,n+1/2}$ \cite{roe1981}. The Nike code employs three such flux functions (Lax-Friedrichs, Lax-Wendroff, and HLLE), which are presented in Appendix 3. It is important to emphasize that these flux functions do not directly account for the influence of radiation on the material quantities. Rather, the radiation effects are included via the source terms, propagation operator, and effective material Jacobian.


\section{Modified Godunov Corrector Scheme}
\noindent
The time discretization for the source term is a single-step, second-order accurate scheme based on the ideas from Dutt et al 2000, Minion 2003, and Miniati \& Colella 2007 \cite{mc2007, dutt2000, minion2003}. Given the material system of balance laws, one aims for a scheme that has an explicit approach for the flux divergence term $\nabla \cdot F^{m}$ and an implicit approach for the stiff source term $S^{m}(U)$. \\

\noindent
At each grid point, one solves the following collection of ordinary differential equations:
\begin{equation}
\frac{dU^{m}}{dt} = S^{m}(U) - ( \nabla \cdot F^{m} )^{n+1/2},
\end{equation}

\noindent
where the time-centered flux divergence term is inputted from the predictor step and taken to be constant valued. Using Picard iteration and the method of deferred corrections, an initial guess for the solution to the collection of ordinary differential equations is:
\begin{equation}
\hat{U} = U^{m,n} + \Delta t (I - \Delta t \nabla_{U^{m}} S^{m}(U) |_{U^{m,n},U^{r,n+1}})^{-1} (S^{m}(U^{m,n},U^{r,n+1}) - ( \nabla \cdot F^{m} )^{n+1/2}) ,
\end{equation}

\noindent
where $\nabla_{U^{m}} S^{m}(U)$ has the same functional form as that which was used to define the propagation operator $\mathcal{I}$ in a previous section. Therefore:
\begin{equation}
\left( I - \Delta t \nabla_{U^{m}} S^{m}(U) \right) = \left( \begin{array}{ccc} 
1 & 0 & 0 \\
0 & 1 & 0 \\
\Delta t \mathbb{P} \mathbb{C} S^{E}_{\rho} & \Delta t \mathbb{P} \mathbb{C} S^{E}_{m} & 1 + \Delta t \mathbb{P} \mathbb{C} S^{E}_{E} 
\end{array} \right) .
\end{equation}

\noindent
By inverting the above matrix, one finds:
\begin{equation}
\left( I - \Delta t \nabla_{U^{m}} S(U) \right)^{-1} = \left( \begin{array}{ccc} 
1 & 0 & 0 \\
0 & 1 & 0 \\
\frac{- \Delta t \mathbb{P} \mathbb{C} S^{E}_{\rho}}{1 + \Delta t \mathbb{P} \mathbb{C} S^{E}_{E}} & \frac{- \Delta t \mathbb{P} \mathbb{C} S^{E}_{m}}{1 + \Delta t \mathbb{P} \mathbb{C} S^{E}_{E}} & \frac{1}{1 + \Delta t \mathbb{P} \mathbb{C} S^{E}_{E}} \\
\end{array} \right) .
\end{equation}

\noindent
The error estimate $\epsilon$ is the difference between the solution obtained from one iteration of the Picard technique where $\hat{U}$ is used as the starting value and the initial guess $\hat{U}$:
\begin{equation}
\epsilon(\Delta t) = U^{m,n} + \frac{\Delta t}{2} \left( S^{m}(\hat{U},U^{r,n+1}) + S^{m}(U^{m,n},U^{r,n+1}) \right) - \Delta t ( \nabla \cdot F^{m} )^{n+1/2} - \hat{U} .
\end{equation}

\noindent
Following Miniati \& Colella 2007, the correction to the initial guess is given by:
\begin{equation}
\delta(\Delta t) = \left( I - \Delta t \nabla_{U^{m}} S^{m}(U) |_{\hat{U},U^{r,n+1}} \right)^{-1} \epsilon(\Delta t) .
\end{equation}

\noindent
Therefore, the material quantities at time $t_{n+1}$ are:
\begin{equation}
U^{m,n+1} = \hat{U} + \delta(\Delta t) .
\end{equation}

\noindent
Appendix 4 explains how to apply various types of boundary conditions in the modified Godunov corrector scheme.


\section{Numerical Tests of Nike}
\noindent
The hybrid Godunov method was implemented in a new radiation hydrodynamical code called \textbf{Nike}. A suite of ten numerical tests were conducted to gauge the temporal and spatial accuracy of the hybrid Godunov method. Tests were carried out using the same code with no special adjustments made for any test and span a wide range of mathematical and physical behavior. These tests are grouped into three categories (Hydrodynamics, Radiation Subsystem, and Full Radiation Hydrodynamics) which define the dominant physical system. The numerical solution is compared with the analytic solution where possible. Otherwise, a self-similar comparison is made. \\

\noindent
The following definitions for the $n$-norms and convergence rates are used throughout this paper. Given the numerical solution $q^{r}$ at resolution $r$ and the analytic solution $u$, the error at a given point $i$ is $\epsilon^{r}_{i} = q^{r}_{i} - u$. Likewise, given the numerical solution $q^{r}$ at resolution $r$ and the numerical solution $q^{r+1}$ at the next finer resolution $r+1$ (spatially averaged onto the coarser grid), the self-similar error at a given point $i$ is $\epsilon^{r}_{i} = q^{r}_{i} - q^{r+1}_{i}$. The 1-norm and max-norm are:
\begin{equation}
L_1 = \sum_i | \epsilon^{r}_{i} | \Delta x^{r} , ~~~~ L_{\max} = \max_i | \epsilon^{r}_{i} | .
\end{equation}

\noindent
The convergence rate is measured using Richardson extrapolation:
\begin{equation}
R_n = \frac{ \textrm{ln} \left( L_n(\epsilon^r)/L_n(\epsilon^{r+1}) \right) }{ \textrm{ln} \left( \Delta x^{r}/\Delta x^{r+1} \right) } . 
\end{equation}

\noindent
For all of the tests presented in this paper, the material fluxes were computed with a componentwise van Leer limiter and an HLLE flux function using second-order (piecewise linear) spatial reconstruction. Moreover, $\gamma = 5/3$ and the CFL number is $\nu = 0.5$. This value for the CFL number was chosen for convenience but any $\nu < 1$ is valid.


\section{Hydrodynamics}
\noindent 
In order to determine how Nike's modified Godunov scheme resolves purely hydrodynamical phenomena, the following set of parameters were chosen to force the radiation hydrodynamical system towards the free streaming limit. Therefore, Equations \ref{eq:cons_law_1d}-\ref{eq:cons_law_source_1d} transform into Equations \ref{eq:streamingD}-\ref{eq:streamingE}. The following tests examine processes associated with the Euler Equations. \\

\noindent
\textbf{Parameters:}
\begin{equation}
\mathbb{C} = 10^{5} , ~ \mathbb{P} = 10^{-20} , ~ \sigma_{a}, \sigma_{t} = 10^{-20} , ~ f = 1 , ~ E_{r}^{0}, F_{r}^{0} = 10^{-20} , \nonumber 
\end{equation}
\begin{equation}
\Delta t = \frac{\nu \Delta x}{\max_{i} ( |u_i| + a_{\textrm{eff},i} )} , ~ x_{\min} = 0 , ~ x_{\max} = 1 , ~ N_{\textrm{cell}} = [32, ~ 64, ~ 128, ~ 256] . \nonumber 
\end{equation}

\noindent
When conducting purely hydrodynamical tests on the modified Godunov scheme, one expects $R = 2.0$ for smooth initial data (e.g., Gaussian pulse) and $R \simeq 0.67$ for discontinuous initial data (e.g., square pulse). This claim is true for all second-order spatially accurate numerical methods when applied to an advection-type problem $(u_t + a u_x = 0)$ \cite{leveque1}. It is important to present these hydrodynamical tests because some radiation hydrodynamical codes perform poorly in this free streaming limit. The results herein demonstrate the hybrid Godunov method to be robust and applicable across different parameter regimes.

\subsection{Linear Advection}
\noindent
For the linear advection tests, the initial values of the conserved quantities are given by:
\begin{equation}
U(x,0) =
\left( \begin{array}{c} 
       \rho^{0} \\
			 m^{0} \\
			 E^{0} \\
			 E_{r}^{0} \\
			 F_{r}^{0} \end{array} \right) =
\left( \begin{array}{c} 
       \rho^{0} \\
			 \rho^{0} u^{0} \\
			 \frac{1}{2} \rho^{0} \left( u^{0} \right)^{2} + \frac{p^{0}}{(\gamma-1)} \\
			 E_{r}^{0} \\
			 F_{r}^{0} \end{array} \right) ,
\end{equation}

\noindent
\textbf{Parameters:}
\begin{equation}
\textrm{IC for Gaussian Pulse:} ~ 
\left\{ \begin{array}{l}
\rho^{0} = 1 + \exp \left( -(v (x - \mu) )^2 \right) , ~ v = 20 , ~ \mu = 0.5 , \\
u^{0} = 1 , ~ p^{0} = 1 , ~ t_{\textrm{end}} = 1 \textrm{ (one crossing time)}
\end{array} \right . \nonumber
\end{equation}
\begin{equation}
\textrm{IC for Square Pulse:} ~
\left\{ \begin{array}{l}
\rho^{0}(0.4<x<0.6) = 1 , ~  \rho^{0}(x<0.4, 0.6<x) = 0.1 , \\
u^{0} = 1 , ~ p^{0} = 1 , ~ t_{\textrm{end}} = 1 \textrm{ (one crossing time)}
\end{array} \right . \nonumber
\end{equation}
\begin{equation}
\textrm{periodic BC} . \nonumber
\end{equation}

\begin{figure}
\begin{center}
\begin{minipage}{3in}
\includegraphics[width=3in,angle=0]{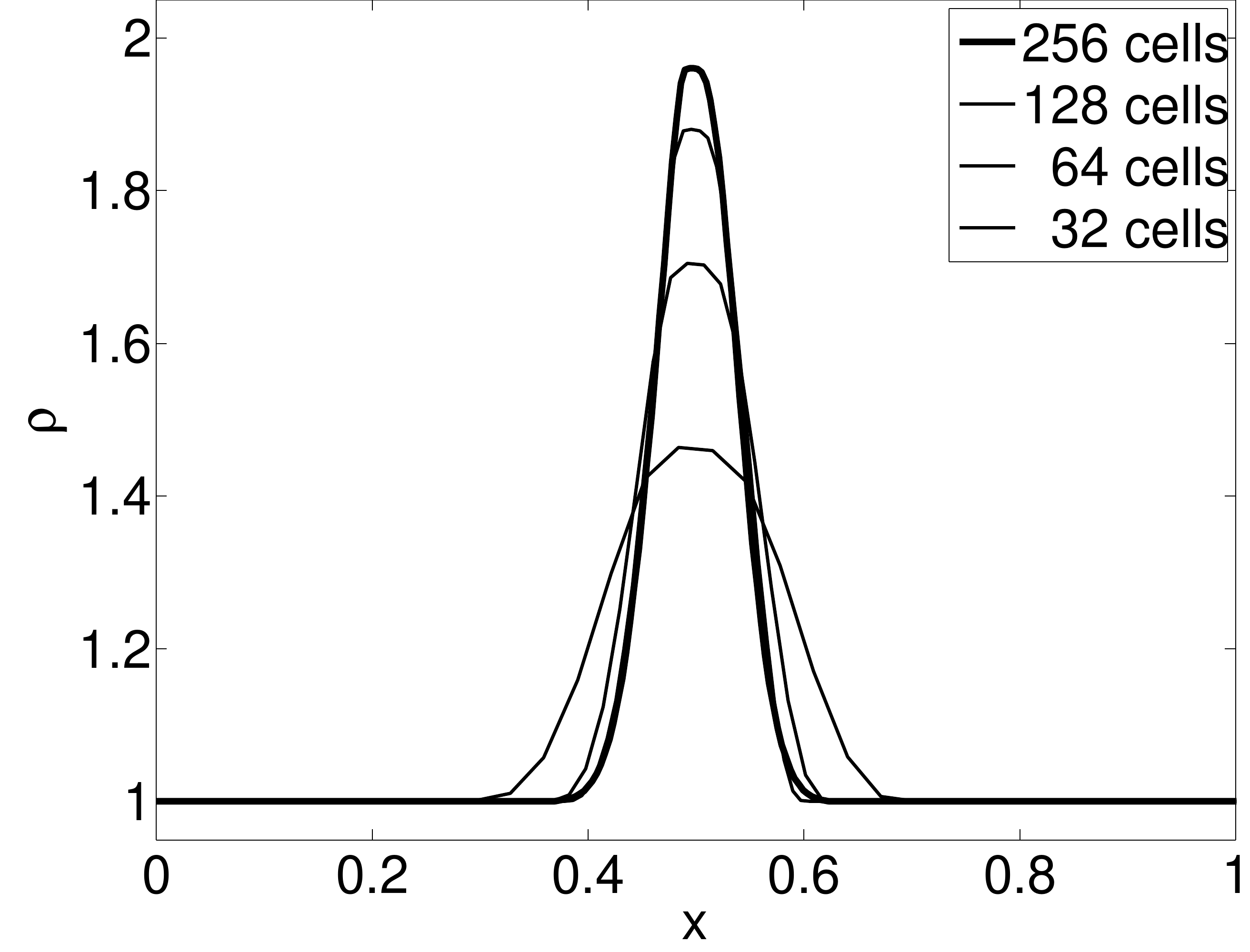}
\caption{\label{fig:hydro_stream_g} Gaussian pulse in the free streaming limit (hydro variables). $t = 1$.}
\end{minipage} 
\hspace{0.4in}
\begin{minipage}{3in}
\includegraphics[width=3in,angle=0]{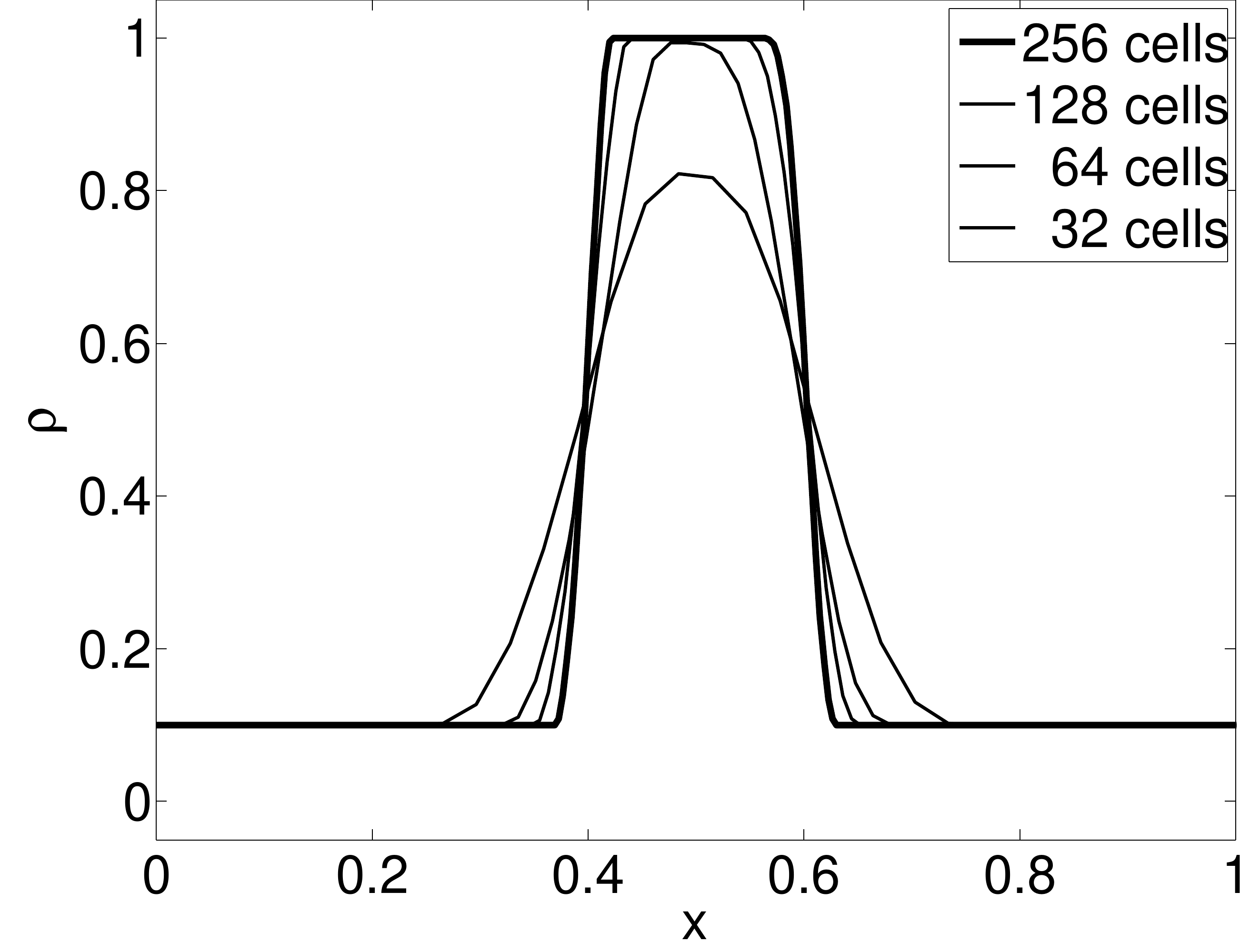}
\caption{\label{fig:hydro_stream_s} Square pulse in the free streaming limit (hydro variables). $t = 1$.}
\end{minipage}
\end{center}
\end{figure} 

\begin{table}
\begin{center}
\begin{tabular*}{1.0\textwidth}{@{\extracolsep{\fill}}rcccccccccccc}
\hline
$N_{\textrm{cell}}$ & $L_1(\rho)$ & $R$ & $L_{\infty}(\rho)$ & $R$ & $L_1(m)$ & $R$ & $L_{\infty}(m)$ & $R$ & $L_1(E)$ & $R$ & $L_{\infty}(E)$ & $R$ \\
\hline
 32 & 5.6E-2 & -   & 4.5E-1 & -   & 5.6E-2 & -   & 4.5E-1 & -   & 2.8E-2 & -   & 2.2E-1 & -   \\
 64 & 2.5E-2 & 1.2 & 2.7E-1 & 0.7 & 2.5E-2 & 1.2 & 2.7E-1 & 0.7 & 1.2E-2 & 1.2 & 1.4E-1 & 0.7 \\
128 & 8.0E-3 & 1.6 & 1.2E-1 & 1.2 & 8.0E-3 & 1.6 & 1.2E-1 & 1.2 & 4.0E-3 & 1.6 & 5.8E-2 & 1.3 \\
256 & 1.9E-3 & 2.1 & 3.8E-2 & 1.7 & 1.9E-4 & 2.1 & 3.8E-2 & 1.7 & 9.6E-4 & 2.1 & 1.8E-2 & 1.7 \\
\hline
\label{tbl:stream_gauss}
\end{tabular*}
Table 1: Convergence analysis for a Gaussian pulse in the free streaming limit (hydro variables). Errors were determined by comparing the numerical/analytic solutions. $t = 1$.
\end{center} 
\end{table}

\begin{table}
\begin{center}
\begin{tabular*}{1.0\textwidth}{@{\extracolsep{\fill}}rcccccc}
\hline
$N_{\textrm{cell}}$ & $L_1(\rho)$ & $R$ & $L_1(m)$ & $R$ & $L_1(E)$ & $R$ \\
\hline
 32 & 9.6E-2 & -   & 9.6E-2 & -   & 4.8E-2 & -   \\
 64 & 5.2E-2 & 0.9 & 5.2E-1 & 0.9 & 2.6E-2 & 0.9 \\
128 & 3.1E-3 & 0.7 & 3.1E-1 & 0.7 & 1.6E-2 & 0.7 \\
256 & 1.9E-3 & 0.7 & 1.9E-2 & 0.7 & 9.6E-3 & 0.7 \\
\hline
\label{tbl:stream_sq}
\end{tabular*}
Table 2: Convergence analysis for a square pulse in the free streaming limit (hydro variables). Errors were determined by comparing the numerical/analytic solutions. $t = 1$.
\end{center} 
\end{table}

\noindent
As the spatial resolution increases, the numerical solutions in Figures \ref{fig:hydro_stream_g} and \ref{fig:hydro_stream_s} converge to Gaussian and square pulses. Furthermore, Tables 1 and 2 show that the convergence rates approach 2.0 and 0.67 for smooth and discontinuous initial data.

\subsection{Propagation of Linear Modes}
\noindent
One of the most discriminating hydrodynamical tests is the propagation of linear modes on a periodic domain. One initializes eigenfunctions of sound and contact waves in a uniform medium by using the right material eigenvectors $R^{m}_{\textrm{eff}}$ and setting the wavelength of each small amplitude perturbation $\mathcal{A}$ equal to the size of the domain $(L = x_{max}-x_{min} = 1)$ \cite{athena}:
\begin{equation}
U(x,0) =
\left( \begin{array}{c} 
       \rho^{0} + \mathcal{A} ~ \sin(k x) ~ R^{m,(1,j)}_{\textrm{eff}}(\rho^{0},u^{0},p^{0}) \\
			 m^{0}    + \mathcal{A} ~ \sin(k x) ~ R^{m,(2,j)}_{\textrm{eff}}(\rho^{0},u^{0},p^{0}) \\
			 E^{0}    + \mathcal{A} ~ \sin(k x) ~ R^{m,(3,j)}_{\textrm{eff}}(\rho^{0},u^{0},p^{0}) \\
			 E_{r}^{0} \\
			 F_{r}^{0} \end{array} \right) .
\end{equation}

\noindent
\textbf{Parameters:}
\begin{equation}
\rho^{0} = 1 , ~ u^{0} = 0 \textrm{ (sound waves: $u \pm a$)} , ~ u^{0} = 1 \textrm{ (contact waves: $u$)} , ~ p^{0} = 1/\gamma , \nonumber
\end{equation}
\begin{equation}
\mathcal{A} = 10^{-6} , ~ k = 2 \pi / L , ~ j = \{ 1, 2, 3 \} , ~ t_{\textrm{end}} = 1 \textrm{ (one crossing time)} , ~ \textrm{periodic BC} . \nonumber
\end{equation}

\noindent
This choice of parameters causes the material sound speed to take the following value $a_{\textrm{eff}}^{0} = 1$. As seen in Tables 3 and 5, the errors associated with the sound waves $(u \pm a)$ are identical because the initialization is the same except that the waves propagate in opposite directions. As evidenced in Tables 3-5, all of the convergence rates approach 2.0.

\begin{table}
\begin{center}
\begin{tabular*}{1.0\textwidth}{@{\extracolsep{\fill}}rcccccccccccc}
\hline
$N_{\textrm{cell}}$ & $L_1(\rho)$ & $R$ & $L_{\infty}(\rho)$ & $R$ & $L_1(m)$ & $R$ & $L_{\infty}(m)$ & $R$ & $L_1(E)$ & $R$ & $L_{\infty}(E)$ & $R$ \\
\hline
 32 & 8.9E-9  & -   & 3.3E-8  & -   & 8.9E-9  & -   & 3.3E-8  & -   & 1.3E-8  & -   & 5.0E-8  & -   \\
 64 & 2.1E-9  & 2.1 & 1.2E-8  & 1.5 & 2.1E-9  & 2.1 & 1.2E-8  & 1.5 & 3.1E-9  & 2.1 & 1.8E-8  & 1.5 \\
128 & 4.2E-10 & 2.3 & 4.2E-9  & 1.5 & 4.2E-10 & 2.3 & 4.2E-9  & 1.5 & 6.3E-10 & 2.3 & 6.3E-9  & 1.5 \\
256 & 8.2E-11 & 2.4 & 1.4E-9  & 1.6 & 8.2E-11 & 2.4 & 1.4E-9  & 1.6 & 1.2E-10 & 2.4 & 2.1E-9  & 1.6 \\
\hline
\label{tbl:hydro_sine1}
\end{tabular*}
Table 3: Convergence analysis for the hydrodynamical linear mode $(u-a)$. Errors were determined by comparing the numerical/analytic solutions. $t = 1$.
\end{center} 
\end{table}

\begin{table}
\begin{center}
\begin{tabular*}{1.0\textwidth}{@{\extracolsep{\fill}}rcccccccccccc}
\hline
$N_{\textrm{cell}}$ & $L_1(\rho)$ & $R$ & $L_{\infty}(\rho)$ & $R$ & $L_1(m)$ & $R$ & $L_{\infty}(m)$ & $R$ & $L_1(E)$ & $R$ & $L_{\infty}(E)$ & $R$ \\
\hline
 32 & 1.3E-8  & -   & 4.0E-8  & -   & 1.3E-8  & -   & 4.0E-8  & -   & 6.7E-9  & -   & 2.0E-8  & -   \\
 64 & 3.6E-9  & 1.9 & 1.4E-8  & 1.5 & 3.6E-9  & 1.9 & 1.4E-8  & 1.5 & 1.8E-9  & 1.9 & 7.1E-9  & 1.5 \\
128 & 8.8E-10 & 2.0 & 5.0E-9  & 1.5 & 8.8E-10 & 2.0 & 5.0E-9  & 1.5 & 4.4E-10 & 2.0 & 2.5E-9  & 1.5 \\
256 & 2.3E-10 & 1.9 & 1.7E-9  & 1.6 & 2.3E-10 & 1.9 & 1.7E-9  & 1.6 & 1.2E-10 & 1.9 & 8.5E-10 & 1.6 \\
\hline
\label{tbl:hydro_sine2}
\end{tabular*}
Table 4: Convergence analysis for the hydrodynamical linear mode $(u)$. Errors were determined by comparing the numerical/analytic solutions. $t = 1$.
\end{center} 
\end{table}

\begin{table}
\begin{center}
\begin{tabular*}{1.0\textwidth}{@{\extracolsep{\fill}}rcccccccccccc}
\hline
$N_{\textrm{cell}}$ & $L_1(\rho)$ & $R$ & $L_{\infty}(\rho)$ & $R$ & $L_1(m)$ & $R$ & $L_{\infty}(m)$ & $R$ & $L_1(E)$ & $R$ & $L_{\infty}(E)$ & $R$ \\
\hline
 32 & 8.9E-9  & -   & 3.3E-8  & -   & 8.9E-9  & -   & 3.3E-8  & -   & 1.3E-8  & -   & 5.0E-8  & -   \\
 64 & 2.1E-9  & 2.1 & 1.2E-8  & 1.5 & 2.1E-9  & 2.1 & 1.2E-8  & 1.5 & 3.1E-9  & 2.1 & 1.8E-8  & 1.5 \\
128 & 4.2E-10 & 2.3 & 4.2E-9  & 1.5 & 4.2E-10 & 2.3 & 4.2E-9  & 1.5 & 6.3E-10 & 2.3 & 6.3E-9  & 1.5 \\
256 & 8.2E-11 & 2.4 & 1.4E-9  & 1.6 & 8.2E-11 & 2.4 & 1.4E-9  & 1.6 & 1.2E-10 & 2.4 & 2.1E-9  & 1.6 \\
\hline
\label{tbl:hydro_sine3}
\end{tabular*}
Table 5: Convergence analysis for the hydrodynamical linear mode $(u+a)$. Errors were determined by comparing the numerical/analytic solutions. $t = 1$.
\end{center} 
\end{table}

\subsection{Sod Shock Tube}
\noindent
The Sod shock tube is an excellent test of an algorithm's ability to resolve nonlinear behavior. This well known test consists of two constant states separated by a discontinuity (i.e., a Riemann problem). \\
 
\noindent
\textbf{Parameters:}
\begin{equation}
U(x<x_{int},0) =
\left( \begin{array}{c} 
       1 \\
			 0 \\
			 \frac{1}{(\gamma-1)} \\
			 E_{r}^{0} \\
			 F_{r}^{0} \end{array} \right) , ~~~~
U(x_{int}<x,0) =
\left( \begin{array}{c} 
       0.125 \\
			 0 \\
			 \frac{0.1}{(\gamma-1)} \\
			 E_{r}^{0} \\
			 F_{r}^{0} \end{array} \right) , \nonumber
\end{equation}
\begin{equation}
x_{int} = 0.5 , ~ t_{\textrm{end}} = 0.2, ~ N_{\textrm{cell}} = 256, ~ \textrm{outflow BC} . \nonumber
\end{equation}

\begin{figure}
\begin{center}
\includegraphics[width=3in,angle=0]{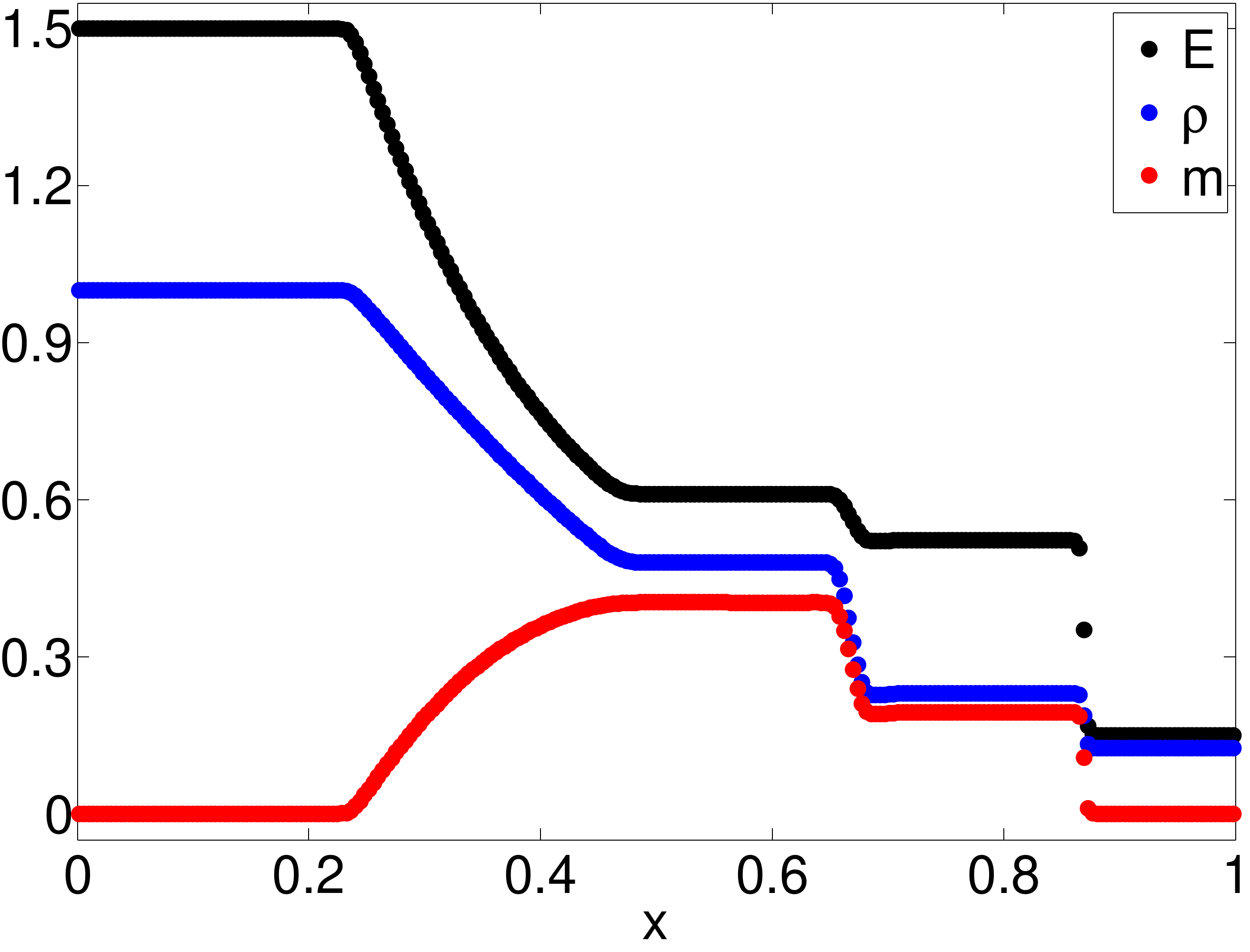}
\caption{\label{fig:hydro_sod} Sod shock tube. $t = 0.2$, $N_{\textrm{cell}} = 256$.}
\end{center}
\end{figure}

\noindent
As the Euler equations evolve in time, a rarefaction fan, contact discontinuity, and shock wave form \cite{sod1978, torobook}. Figure \ref{fig:hydro_sod} shows a higher-order accurate solution to the Sod shock tube problem, where discontinuities are separated by only one grid cell.

\section{Radiation Subsystem}
\noindent 
In order to develop a reliable and robust numerical method for full radiation hydrodynamics, Sekora \& Stone 2009 examined the radiation subsystem, which is a simpler set of equations that minimizes complexity while maintaining the rich hyperbolic-parabolic behavior associated with balance laws that have stiff source terms. If one initially sets the material flow to be stationary such that $u \rightarrow 0$ in Equations \ref{eq:radsub1} and \ref{eq:radsub2}, then the variables, fluxes, and source terms for the radiation subsystem are given by:
\begin{equation}
\partdif{E_r}{t} + \mathbb{C}   \partdif{F_r}{x} =  \mathbb{C} \sigma_a (T^4 - E_r) \label{eq:rad_sub_e} ,
\end{equation}
\begin{equation}
\partdif{F_r}{t} + \mathbb{C} f \partdif{E_r}{x} = -\mathbb{C} \sigma_t F_r \label{eq:rad_sub_f} .
\end{equation}

\noindent
In order to determine how Nike's backward Euler upwinding scheme resolves the radiation subsystem, the following set of parameters were chosen. \\

\noindent
\textbf{Parameters:}
\begin{equation}
\mathbb{C} = 10^{5} , ~ \mathbb{P} = 10^{-20} , ~ \rho^{0}, m^{0}, E^{0} = 10^{-20} , \nonumber 
\end{equation}
\begin{equation}
\Delta t = \frac{\nu \Delta x}{f^{1/2} \mathbb{C}} , ~ N_{\textrm{cell}} = [32, ~ 64, ~ 128, ~ 256] . \nonumber 
\end{equation}

\noindent
When conducting tests on the backward Euler upwinding scheme, one expects $R = 1.0$ for smooth initial data (e.g., Gaussian pulse) and $R \simeq 0.5$ for discontinuous initial data (e.g., square pulse). This claim is true for all first-order spatially accurate numerical methods when applied to an advection-type problem $(u_t + a u_x = 0)$ \cite{leveque1}. By no means was the material flow held stationary $(u=0)$ throughout the evolution of the following numerical tests. All problems were solved with the full dynamical Nike code. The fact that the material quantities remain zero in these tests demonstrates the robustness of the overall method. The following problems were initialized according to:
\begin{equation}
U(x,0) = \left( \rho^{0} , m^{0} , E^{0} , E_{r}^{0} , F_{r}^{0} \right) .
\end{equation}

\subsection{Linear Advection - Free Steaming Limit}
\noindent
In the free streaming limit, the radiation subsystem reduces to Equations \ref{eq:streamingER} and \ref{eq:streamingFR}. If one takes an additional temporal and spatial partial derivative of the radiation subsystem in the free streaming limit and subtracts the resulting equations, then one finds two decoupled wave equations that have the following analytic solutions:
\begin{eqnarray}
E_r(x,t) = E_0(x - f^{1/2} \mathbb{C} t) , \\
F_r(x,t) = F_0(x - f^{1/2} \mathbb{C} t) .
\end{eqnarray}

\noindent
\textbf{Parameters:}
\begin{equation}
\sigma_{a}, \sigma_{t} = 10^{-20} , ~ f = 1 , ~ x_{\min} = 0 , ~ x_{\max} = 1 , ~ \textrm{periodic BC} , \nonumber
\end{equation}
\begin{equation}
\textrm{IC for Gaussian Pulse:} ~ E_r^0, F_r^0 = \exp \left( -(v (x - \mu) )^2 \right) , ~ v = 20 , ~ \mu = 0.5 , \nonumber
\end{equation}
\begin{equation}
\textrm{IC for Square Pulse:} ~ 
E_r^0, F_r^0 = \left\{ \begin{array}{ll}
               1 & 0.4 < x < 0.6 \\
               0 & \rm{otherwise} \end{array} \right . \nonumber
\end{equation}

\begin{table}
\begin{center}
\begin{tabular*}{1.0\textwidth}{@{\extracolsep{\fill}}rcccccccc}
\hline
$N_{\textrm{cell}}$ & $L_1(E_{r})$ & $R$ & $L_{\infty}(E_{r})$ & $R$ & $L_1(F_{r})$ & $R$ & $L_{\infty}(F_{r})$ & $R$ \\
\hline
 32  & 1.2E-1 & -   & 7.2E-1 & -   & 1.2E-1 & -   & 7.2E-1 & -    \\
 64  & 9.9E-2 & 0.3 & 7.1E-1 & 0.0 & 9.9E-2 & 0.3 & 7.1E-1 & 0.0  \\
128  & 7.9E-2 & 0.3 & 6.3E-1 & 0.2 & 7.9E-2 & 0.3 & 6.3E-1 & 0.2  \\
256  & 5.9E-2 & 0.4 & 5.1E-1 & 0.3 & 5.9E-2 & 0.4 & 5.1E-1 & 0.3  \\
512  & 3.9E-2 & 0.6 & 3.7E-1 & 0.5 & 3.9E-2 & 0.6 & 3.7E-1 & 0.5  \\
1024 & 2.4E-2 & 0.7 & 2.5E-1 & 0.6 & 2.4E-2 & 0.7 & 2.5E-1 & 0.6  \\
2048 & 1.3E-2 & 0.9 & 1.4E-1 & 0.8 & 1.3E-2 & 0.9 & 1.4E-1 & 0.8  \\
\hline
\label{tbl:rad_gauss}
\end{tabular*}
Table 6: Convergence analysis for a Gaussian pulse in the free streaming limit (radiation subsystem). Errors were determined by comparing the numerical/analytic solutions. $t = 1$.
\end{center} 
\end{table}

\begin{table}
\begin{center}
\begin{tabular*}{1.0\textwidth}{@{\extracolsep{\fill}}rcccc}
\hline
$N_{\textrm{cell}}$ & $L_1(E_{r})$ & $R$ & $L_1(F_{r})$ & $R$ \\
\hline
 32  & 2.1E-1 & -   & 2.1E-1 & -    \\
 64  & 1.7E-1 & 0.3 & 1.7E-1 & 0.3  \\
128  & 1.3E-1 & 0.4 & 1.3E-1 & 0.4  \\
256  & 9.1E-2 & 0.5 & 9.1E-2 & 0.5  \\
512  & 6.4E-2 & 0.5 & 6.4E-2 & 0.5  \\
1024 & 4.5E-2 & 0.5 & 4.5E-2 & 0.5  \\
2048 & 3.1E-2 & 0.5 & 3.1E-2 & 0.5  \\
\hline
\label{tbl:rad_sq}
\end{tabular*}
Table 7: Convergence analysis for a square pulse in the free streaming limit (radiation subsystem). Errors were determined by comparing the numerical/analytic solutions. $t = 1$.
\end{center} 
\end{table}

\noindent
Graphically, these initial conditions give rise to numerical solutions with profiles that are identical to those shown in Figures \ref{fig:hydro_stream_g} and \ref{fig:hydro_stream_s}. In Tables 6 and 7, one observes first-order accurate convergence rates. What is most interesting is how long it takes (i.e., how high of a spatial resolution is necessary) to see $R \rightarrow 1$ for a Gaussian pulse in the free streaming limit. Yet, this result is not wholly unexpected. In this test, one is examining advective behavior. It is widely known that implicit algorithms perform poorly when trying to capture the propagation of individual wave modes. The fact that the backward Euler upwinding scheme obtains the correct solution in a stable and accurate manner is a testament to the robustness of the overall method. Moreover, resolving individual radiative waves associated with the speed of light is not the primary concern of the hybrid Godunov method. Therefore, it is of little consequence that such high spatial resolution is necessary to see a first-order accurate convergence rate for radiation quantities in the free streaming limit.

\subsection{Exponential Growth/Decay to Thermal Equilibrium}
\noindent
This test examines the temporal accuracy of how radiation variables are updated in the backward Euler upwinding scheme. Given the radiation subsystem and the following initial conditions:
\begin{displaymath}
E_r^0 = \textrm{constant across space} , ~~
F_r^0 = 0, ~~
T     = \textrm{constant across space and time} ,
\end{displaymath}

\noindent
$F_r \rightarrow 0$ for all time. Therefore, the radiation subsystem reduces to the following ordinary differential equation:
\begin{equation}
\frac{d E_r}{dt} = \mathbb{C} \sigma_a (T^4-E_r),
\end{equation}

\noindent 
which has the following analytic solution:
\begin{equation}
E_r = T^4 + (E_r^0 - T^4) \rm{exp}(-\mathbb{C} \sigma_a t) .
\end{equation}

\noindent
For $E_r^0 < T^4$ and $F_r^0 = 0$, one expects exponential growth in $E_r$ until thermal equilibrium $(E_r = T^4)$ is reached. For $E_r^0 > T^4$ and $F_r^0 = 0$, one expects exponential decay in $E_r$ until thermal equilibrium is reached. Since $E_r$ is spatially uniform for this system, this numerical test allows one to examine the temporal order of accuracy of the backward Euler upwinding scheme as a stiff ODE integrator. \\

\noindent
\textbf{Parameters:}
\begin{equation}
\sigma_{a}, \sigma_{t} = 1 , ~ f = 1 , ~ x_{\min} = 0 , ~ x_{\max} = 1 , \nonumber
\end{equation}
\begin{equation}
\textrm{IC for Growth:} ~ E_r^0 = 1, ~ F_r^0 = 0 , ~ T = 10 , \nonumber
\end{equation}
\begin{equation}
\textrm{IC for Decay:} ~ E_r^0 = 10^4, ~ F_r^0 = 0 , ~ T = 1 . \nonumber
\end{equation}

\begin{figure}
\begin{center}
\includegraphics[width=3in,angle=0]{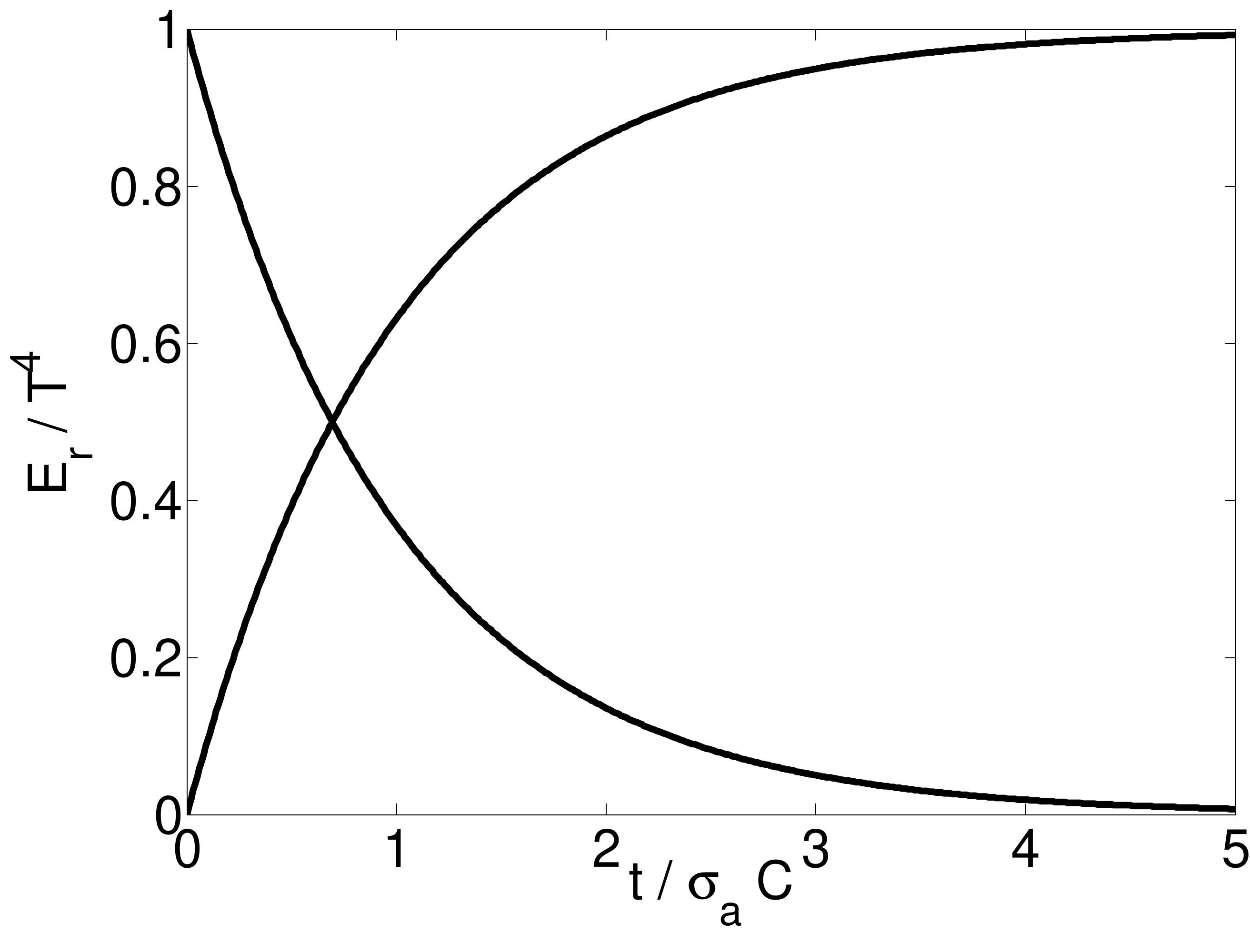}
\caption{\label{fig:rad_exp}Exponential growth/decay of $E_{r}$ to thermal equilibrium. $N_{\textrm{cell}}=256$.}
\end{center}
\end{figure} 

\begin{table}
\begin{center}
\begin{tabular*}{1.0\textwidth}{@{\extracolsep{\fill}}rcccccccc}
\hline
$N_{\textrm{cell}}$ & $L_1(E^{g}_{r})$ & $R$ & $L_{\infty}(E^{g}_{r})$ & $R$ & $L_1(E^{d}_{r})$ & $R$ & $L_{\infty}(E^{g}_{r})$ & $R$ \\
\hline
 32 & 2.9E-1 & -   & 2.9E-1 & -   & 2.9E-1 & -   & 2.9E-1 & -   \\
 64 & 1.4E-1 & 1.0 & 1.4E-1 & 1.0 & 1.4E-2 & 1.0 & 1.4E-1 & 1.0 \\
128 & 7.8E-2 & 0.8 & 7.8E-2 & 0.8 & 6.2E-3 & 1.2 & 6.2E-2 & 1.2 \\
256 & 4.5E-2 & 0.8 & 4.5E-2 & 0.8 & 1.7E-3 & 1.9 & 1.7E-2 & 1.9 \\
\hline
\label{tbl:rad_exp}
\end{tabular*}
Table 8: Convergence analysis for exponential growth/decay in $E_r$ to thermal equilibrium. Errors were determined by comparing the numerical/analytic solutions. $t = 10^{-5} = 1/\sigma_a \mathbb{C}$.
\end{center} 
\end{table}

\noindent
From Figure \ref{fig:rad_exp}, one sees that the numerical solution corresponds with the analytic solution. In Table 8, ones sees that at lower resolution the errors and convergence rates are identical for growth and decay. However, as the resolution increases one sees disparity between these values, such that the growth tests exhibit convergence rates that are less than first-order while the decay tests exhibit convergence rates that are greater than first-order. This asymmetry is expected for this situation because it is associated with the unconditional stability of a backward Euler-type integrator. Whenever a problem is defined by dynamical exponential growth, an algorithm works hard to maintain numerical stability. However, when a problem is defined by dynamical exponential decay, the nature of the problem already ensures stability and the algorithm doesn't have to work as hard. Nonetheless, one finds that the method is well behaved and obtains the correct solution with at least first-order accuracy for even stiff values of the $e$ folding time $(\frac{\Delta t}{1 / \sigma_{a} \mathbb{C}} \geq 1)$. This result credits the flexibility of the backward Euler upwinding scheme as being a robust temporal integrator.

\subsection{Weak Equilibrium Diffusion Limit}
\noindent
In the weak equilibrium diffusion limit, the radiation subsystem reduces to the following set of equations: 
\begin{equation}
\partdif{E_r}{t} = \frac{\mathbb{C}}{3 \sigma_t} \partdif{^2 E_r}{x^2} \label{eq:rad_sub_weak_e} ,
\end{equation}
\begin{equation}
F_r = -\frac{1}{3 \sigma_t} \partdif{E_r}{x} \label{eq:rad_sub_weak_f} .
\end{equation}

\noindent
The optical depth suggests the range of opacities for which diffusion is observed for this subsystem. If $\mathcal{L} = \sigma_t ~ \ell_{\textrm{diff}} > 1$, then one expects diffusive behavior for $\sigma_t > 1 / \ell_{\textrm{diff}}$. Additionally, Equations \ref{eq:rad_sub_weak_e} and \ref{eq:rad_sub_weak_f} set the time scale $t_{\textrm{diff}}$ and length scale $\ell_{\textrm{diff}}$ for diffusion, where $t_{\textrm{diff}} \sim \ell_{\textrm{diff}}^{~2}/D$ and $D = f \mathbb{C} / \sigma_{t}$. Given a diffusion problem with a Gaussian pulse defined over the entire real line $(u_t - D u_{xx} = 0)$, the analytic solution is given by the method of Green's functions:
\begin{equation}
u(x,t) = \int_{-\infty}^{\infty} f(\bar{x}) G(x,t;\bar{x},0) d\bar{x} = \frac{1}{(4 D t v^2 + 1)^{1/2}} \textrm{exp} \left( \frac{-(v (x - \mu) )^2}{4 D t v^2 + 1} \right) . \label{rad_sub_diff_anal}
\end{equation}

\noindent
\textbf{Parameters:}
\begin{equation}
\sigma_{a}, \sigma_{t} = 40 , ~ f = 1/3 , ~ x_{\min} = -5 , ~ x_{\max} = 5 , \nonumber 
\end{equation}
\begin{equation}
N_{\textrm{cell}} = [320, ~ 640, ~ 1280, ~ 2560] , ~ \textrm{Outflow BC} , \nonumber
\end{equation}
\begin{equation}
\textrm{IC for Gaussian Pulse:} ~ \left\{ \begin{array}{l}
E_r^0 = \exp \left( -(v (x - \mu) )^2 \right) , ~ v = 20 , ~ \mu = 0 , \\
F_r^0 = -\frac{f}{\sigma_t} \partdif{E_r^0}{x} = \frac{2 f v^2 (x-\mu)}{\sigma_t} E_r^0 ,  \\
T^4 = E_r \end{array} \right . \nonumber
\end{equation}
\begin{equation}
t_{\textrm{end}} = [0, ~ 1, ~ 4, ~ 16] \times 10^{-6} . \nonumber
\end{equation}

\begin{figure}
\begin{center}
\begin{minipage}{3in}
\includegraphics[width=3in,angle=0]{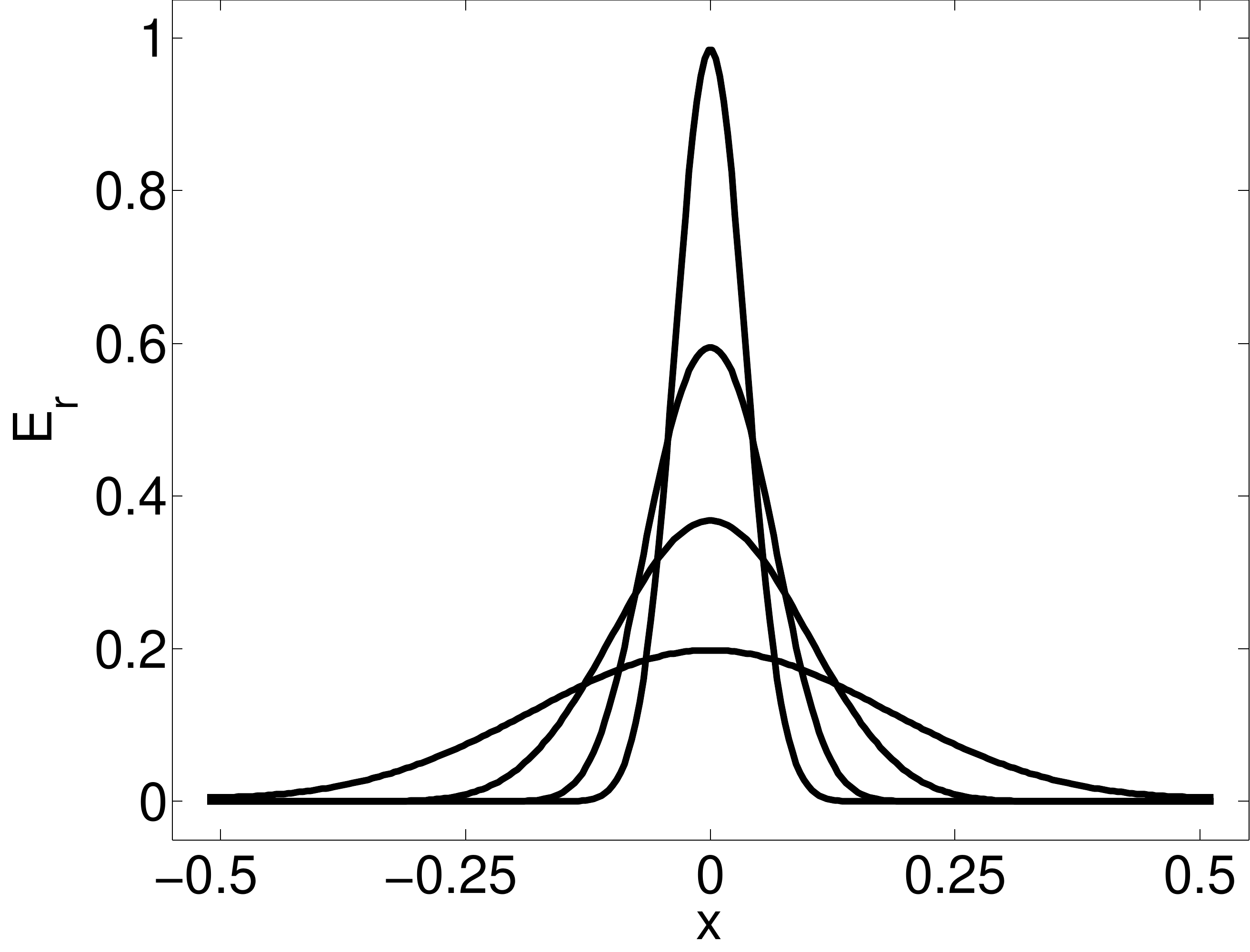}
\caption{\label{fig:rad_weak_diff_e} $E_{r}$ in weak diff limit (rad subsystem). $t = [0,1,4,16] \times 10^{-6}$, $N_{\textrm{cell}} = 2560$.}
\end{minipage} 
\hspace{0.4in}
\begin{minipage}{3in}
\includegraphics[width=3in,angle=0]{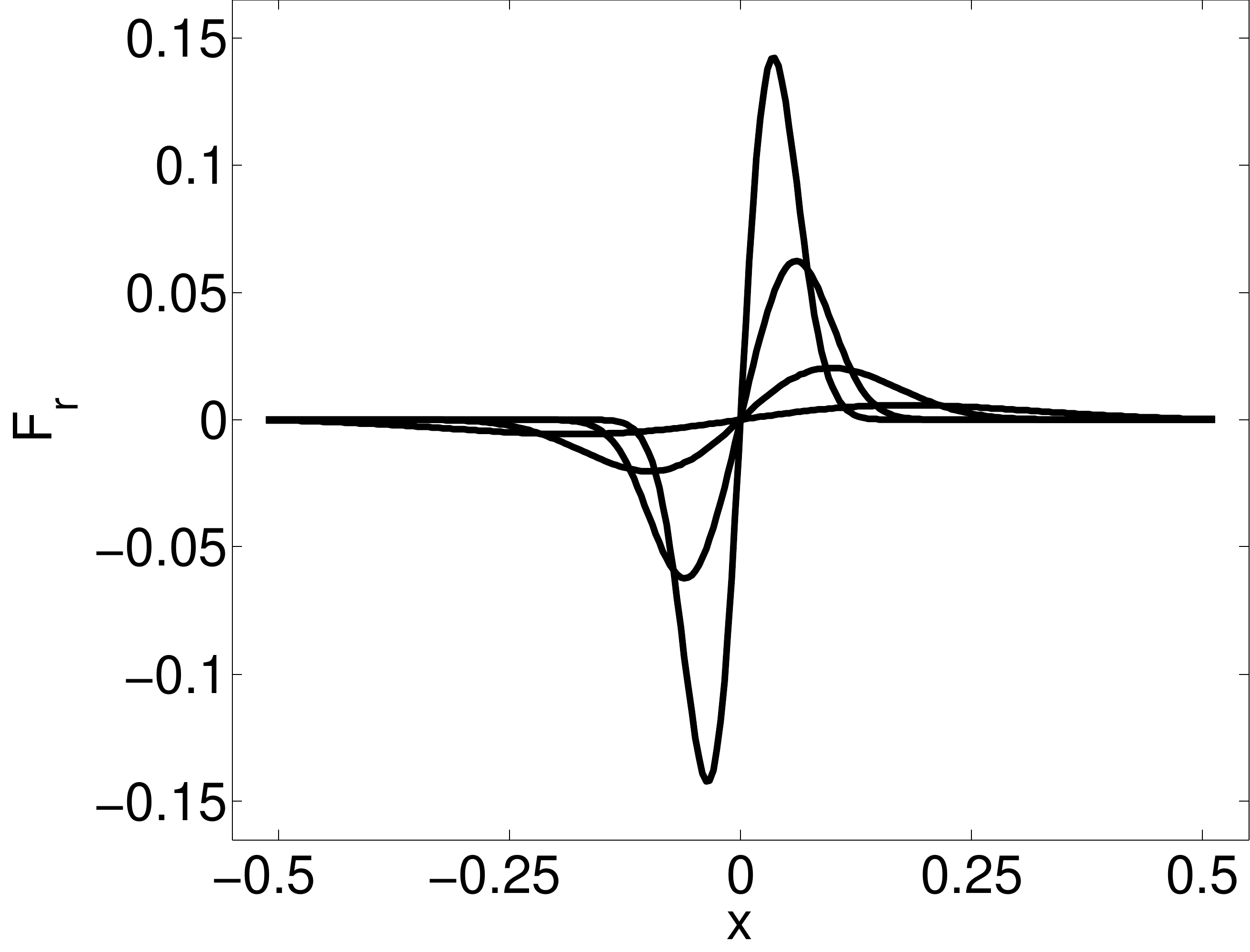}
\caption{\label{fig:rad_weak_diff_f} $F_{r}$ in weak diff limit (rad subsystem). $t = [0,1,4,16] \times 10^{-6}$, $N_{\textrm{cell}} = 2560$.}
\end{minipage}
\end{center}
\end{figure}

\begin{table}
\begin{center}
\begin{tabular*}{1.0\textwidth}{@{\extracolsep{\fill}}rcccccccc}
\hline
$N_{\textrm{cell}}$ & $L_1(E_{r})$ & $R$ & $L_{\infty}(E_{r})$ & $R$ & $L_1(F_{r})$ & $R$ & $L_{\infty}(F_{r})$ & $R$ \\
\hline
 320 & 3.9E-2 & -   & 1.1E-1 & -   & 3.6E-3 & -   & 1.3E-2 & -   \\
 640 & 2.2E-2 & 0.8 & 6.7E-2 & 0.7 & 2.3E-3 & 0.6 & 9.0E-3 & 0.5 \\
1280 & 1.2E-2 & 0.9 & 3.7E-2 & 0.9 & 1.3E-3 & 0.8 & 5.4E-3 & 0.7 \\
2560 & 6.0E-3 & 1.0 & 1.9E-2 & 1.0 & 7.0E-4 & 0.9 & 2.9E-3 & 0.9 \\
\hline
\label{tbl:rad_weak_diff_self_sim}
\end{tabular*}
Table 9: Convergence analysis for $E_r$, $F_r$ in the weak equilibrium diffusion limit. Errors were determined by a self-similar comparison of the numerical solutions. $t = 4 \times 10^{-6}$.
\end{center} 
\end{table}

\noindent
One's intuition about diffusive processes is based on considering an infinite domain. So to minimize boundary effects in the numerical calculation, the computational domain and number of grid cells were expanded by a factor of 10. In Figures \ref{fig:rad_weak_diff_e} and \ref{fig:rad_weak_diff_f}, one observes the diffusive behavior that is expected for this parameter regime. Additionally, the numerical solution compares well with the analytic solution for a diffusion process defined over the entire real line (Equation \ref{rad_sub_diff_anal}). However, diffusive behavior is only a first-order approximation to more complicated hyperbolic-parabolic dynamics taking place in radiation hydrodynamics as well as the radiation subsystem. Therefore, one compares the numerical solution self-similarly and sees the expected first-order convergence in Table 9. \\

\noindent
In order to examine self-similar convergence in the weak equilibrium diffusion limit for the above tests, the backward Euler upwinding scheme uses parameter values and a fine spatial resolution $\Delta x$ to resolve the photon mean free path $\lambda_{t}$. Therefore, the cell-optical depth $\mathcal{L}_{\Delta}$ is always fairly small, such that $\mathcal{L}_{\Delta} = \Delta x / \lambda_{t} = \Delta x ~ \sigma_{t} = [1.25, 0.625, 0.3125, 0.15625]$ for $\sigma_{t} = 40$ and $N_{\textrm{cell}} = [320, 640, 1280, 2560]$, respectively. Consequently, the photon mean free path is moderately resolved which suggests that the backward Euler upwinding scheme is numerically consistent but not necessarily asymptotic preserving. This idea was first developed in the context of radiation hydrodynamics by Larsen et al 1987 although these authors did not use the term asymptotic preserving. In order to explore this idea, consider the following hypothetical question - can one say that a conventional Godunov method is asymptotic preserving for the the system defined by Equations \ref{eq:cons_law_1d}-\ref{eq:cons_law_source_1d}? If one evolves the system according to $\Delta t \sim 1 / \mathbb{C}$, then one expects diffusion behavior for certain parameter regimes because for high spatial and temporal resolution consistency implies asymptotic preservation. However, because one has to use such fine spatial and temporal resolution, a conventional Godunov method is not asymptotic preserving. \\

\noindent
To demonstrate that the backward Euler upwinding scheme has some asymptotic preserving properties, one can use the problem settings that were used earlier but rescale the spatial domain, initial conditions, and time for how long the computation runs so that $\mathcal{L}_{\Delta}$ increases by some significant amount, say 1000. According to this scaling, which is defined by the relation $t_{\textrm{diff}} \sim \ell_{\textrm{diff}}^{~2}/D$, one expects the profiles of $E_{r}$ and $F_{r}$ to match the profiles shown in Figures \ref{fig:rad_weak_diff_e} and \ref{fig:rad_weak_diff_f} except that the spatial scale will have increased by a factor of 1000. Additionally, one expects the magnitude of $F_{r}$ to decrease by a factor of 1000 because in the weak equilibrium diffusion limit, $E_{r}$ behaves according to Equation \ref{rad_sub_diff_anal} and $F_{r}$ behaves according to the following relation:
\begin{equation}
F_{r}(x,t) = - \frac{f}{\sigma_{t}} \partdif{E_{r}}{x} = \frac{f}{\sigma_{t}} \frac{2 v^2 (x-\mu)}{4 D t v^2 + 1} E_{r} .
\end{equation}

\noindent
\textbf{Parameters:}
\begin{equation}
\sigma_{a}, \sigma_{t} = 40 , ~ f = 1/3 , ~ x_{\min} = -5000 , ~ x_{\max} = 5000 , \nonumber 
\end{equation}
\begin{equation}
N_{\textrm{cell}} = [320, ~ 640, ~ 1280, ~ 2560] , ~ \textrm{Outflow BC} , \nonumber
\end{equation}
\begin{equation}
\textrm{IC for Gaussian Pulse:} ~ \left\{ \begin{array}{l}
E_r^0 = \exp \left( -(v (x - \mu) )^2 \right) , ~ v = 0.02 , ~ \mu = 0 , \\
F_r^0 = -\frac{f}{\sigma_t} \partdif{E_r^0}{x} = \frac{2 f v^2 (x-\mu)}{\sigma_t} E_r^0 ,  \\
T^4 = E_r \end{array} \right . \nonumber
\end{equation}
\begin{equation}
t_{\textrm{end}} = [0, ~ 1, ~ 4, ~ 16] . \nonumber
\end{equation}

\begin{figure}
\begin{center}
\begin{minipage}{3in}
\includegraphics[width=3in,angle=0]{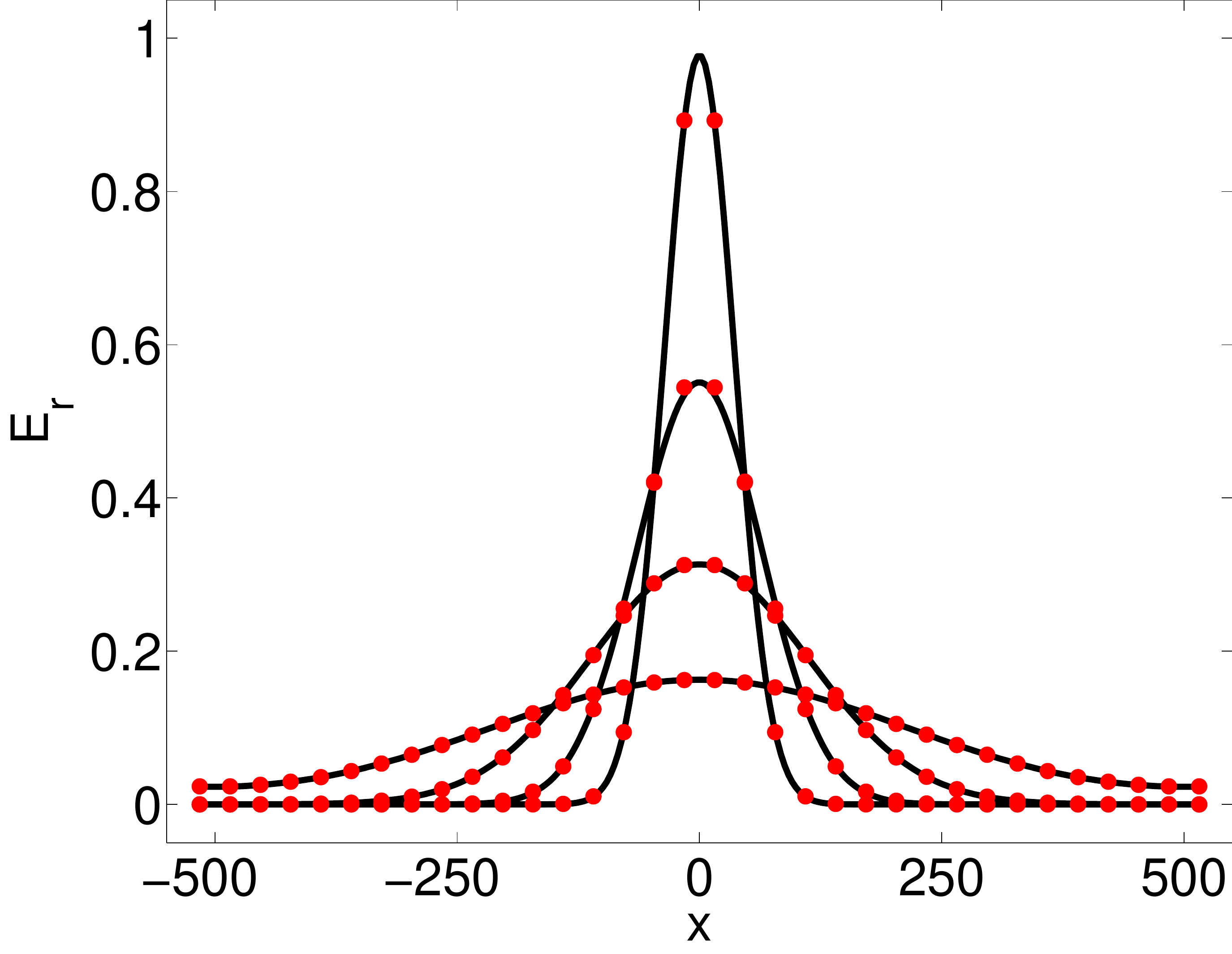}
\caption{\label{fig:rad_weak_diff_e_large_od} $E_{r}$ in weak diff limit (rad subsystem) defined by large cell-optical depths. Red circles designate $N_{\textrm{cell}} = 320$ $(\mathcal{L}_{\Delta} = 1250)$ and black lines designate $N_{\textrm{cell}} = 2560$ $(\mathcal{L}_{\Delta} = 156.25)$. $t = [0,1,4,16]$.}
\end{minipage} 
\hspace{0.4in}
\begin{minipage}{3in}
\includegraphics[width=3in,angle=0]{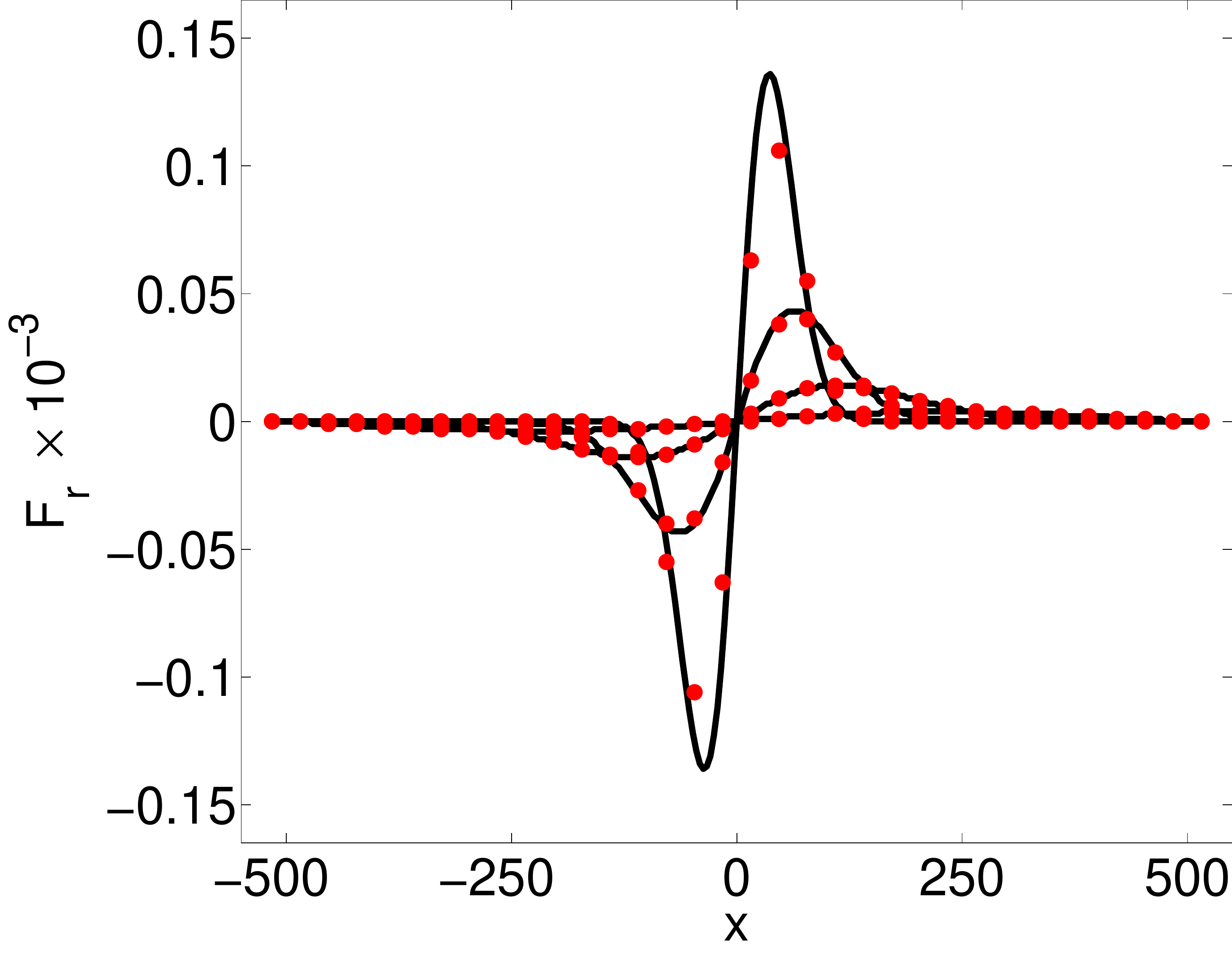}
\caption{\label{fig:rad_weak_diff_f_large_od} $F_{r}$ in weak diff limit (rad subsystem) defined by large cell-optical depths. Red circles designate $N_{\textrm{cell}} = 320$ $(\mathcal{L}_{\Delta} = 1250)$ and black lines designate $N_{\textrm{cell}} = 2560$ $(\mathcal{L}_{\Delta} = 156.25)$. $t = [0,1,4,16]$.}
\end{minipage}
\end{center}
\end{figure}

\begin{table}
\begin{center}
\begin{tabular*}{1.0\textwidth}{@{\extracolsep{\fill}}rcccc}
\hline
$N_{\textrm{cell}}$ & $L_1(E_{r})$ & $L_{\infty}(E_{r})$ & $L_1(F_{r})$ & $L_{\infty}(F_{r})$ \\
\hline
 320 & 3.9E-2 & 1.0E-1 & 3.4E-6 & 1.3E-5 \\
 640 & 2.0E-2 & 5.2E-2 & 1.8E-6 & 6.5E-6 \\
1280 & 8.8E-2 & 6.6E-1 & 1.7E-5 & 1.3E-4 \\
2560 & 8.8E-2 & 6.6E-1 & 1.7E-5 & 1.3E-4 \\
\hline
\label{tbl:rad_weak_diff_self_sim}
\end{tabular*}
Table 10: Convergence analysis for $E_r$, $F_r$ in the weak equilibrium diffusion limit defined by large cell-optical depths. Errors were determined by a self-similar comparison of the numerical solutions. $t = 4$.
\end{center} 
\end{table}

\noindent
Results for these parameters are shown in Figures \ref{fig:rad_weak_diff_e_large_od} and \ref{fig:rad_weak_diff_f_large_od}. Again, one compares the numerical solutions self-similarly across different spatial resolutions at $t = 4$ in Table 10 to examine the convergence properties of the backward Euler upwinding scheme. From the values seen in Table 10, one initially sees first-order convergence. However, as the spatial resolution is refined the error norms stop decreasing. This behavior suggests that for large cell-optical depths, the numerical solution has converged on some asymptotic profile where higher spatial resolution does not enable further convergence because of a competing process associated with the implicit discretization of the radiation subsystem in the backward Euler upwinding scheme. Furthermore, if one examines the discretization of the backward Euler HLLE scheme, one finds the following difference equation for the radiation energy:
\begin{equation}
- \left( d_{1} f^{1/2} \right) E_{r, i-1}^{n+1} + \left( 1 + 2 d_{1} f^{1/2} + d_{2} \sigma_{a} \right) E_{r, i}^{n+1} - \left( d_{1} f^{1/2} \right) E_{r, i+1}^{n+1} + \ldots = E_{r, i}^{n} + \ldots 
\end{equation}

\noindent
This difference equation is important because it has the same functional discretization as a backward-time centered-space scheme for a parabolic/diffusion equation. Therefore, one sees consistency associated with the truncation error and discretization operator. All of the above insights suggest that the backward Euler upwinding scheme can preserve certain asymptotic limits. However, this does not suggest that the overall algorithm (i.e., hybrid Godunov method) is globally asymptotic preserving. Further discussion of this topic is reserved for the conclusions at the end of the paper.

\subsection{Strong Equilibrium Diffusion Limit}
\noindent
In the strong equilibrium diffusion limit, the radiation subsystem reduces to Equations \ref{eq:rad_sub_strong_e} and \ref{eq:rad_sub_strong_f}, which suggest that $F_r \rightarrow 0$ for all time and space while $E_r = E_r^0$.
\begin{equation}
\partdif{E_r}{t} = 0 \label{eq:rad_sub_strong_e} ,
\end{equation}
\begin{equation}
F_r = 0 \label{eq:rad_sub_strong_f} .
\end{equation}

\noindent
\textbf{Parameters:}
\begin{equation}
\sigma_{a}, \sigma_{t} = 10^{6} , ~ f = 1/3 , ~ x_{\min} = -5 , ~ x_{\max} = 5 , \nonumber 
\end{equation}
\begin{equation}
N_{\textrm{cell}} = [320, ~ 640, ~ 1280, ~ 2560] , ~ \textrm{Outflow BC} \nonumber
\end{equation}
\begin{equation}
\textrm{IC for Gaussian Pulse:} ~ \left\{ \begin{array}{l}
E_r^0 = \exp \left( -(v (x - \mu) )^2 \right) , ~ v = 20 , ~ \mu = 0 , \\
F_r^0 = -\frac{f}{\sigma_t} \partdif{E_r^0}{x} = \frac{2 f v^2 (x-\mu)}{\sigma_t} E_r^0 ,  \\
T^4 = E_r \end{array} \right . \nonumber
\end{equation}

\begin{table}
\begin{center}
\begin{tabular*}{1.0\textwidth}{@{\extracolsep{\fill}}rcccccccc}
\hline
$N_{\textrm{cell}}$ & $L_1(E_{r})$ & $R$ & $L_{\infty}(E_{r})$ & $R$ & $L_1(F_{r})$ & $R$ & $L_{\infty}(F_{r})$ & $R$ \\
\hline
 320 & 1.1E-1 & -   & 8.1E-1 & -   & 9.4E-7 & -   & 7.5E-6 & -   \\
 640 & 6.1E-2 & 0.9 & 5.0E-1 & 0.7 & 6.3E-7 & 0.6 & 7.0E-6 & 0.1 \\
1280 & 3.1E-2 & 1.0 & 2.6E-1 & 0.9 & 3.5E-7 & 0.8 & 4.5E-6 & 0.6 \\
2560 & 1.6E-2 & 1.0 & 1.3E-1 & 1.0 & 1.9E-7 & 0.9 & 2.4E-6 & 0.9 \\
\hline
\label{tbl:rad_weak_diff_self_sim}
\end{tabular*}
Table 11: Convergence analysis for $E_r$, $F_r$ in the strong equilibrium diffusion limit. Errors were determined by a self-similar comparison of the numerical solutions. $t = 4 \times 10^{-6}$.
\end{center} 
\end{table}

\noindent
In this test, the numerical solution remains fixed at the initial distribution because $\sigma_a$ and $\sigma_t$ are so large. However, if one fixed $\ell_{\textrm{diff}}$ and scaled time such that $t_{\textrm{diff}} \approx \ell_{\textrm{diff}}^{~2}/D = \ell_{\textrm{diff}}^{~2} \sigma_t / f \mathbb{C}$, then one would observe behavior similar to Figures \ref{fig:rad_weak_diff_e} and \ref{fig:rad_weak_diff_f}. This test illustrates the robustness of the backward Euler upwinding scheme to handle very stiff source terms.

\subsection{Su-Olson Non-Equilibrium Diffusion}
\noindent
Due to the complexity of the equations for radiation hydrodynamics, reference problems with analytic solutions are rare outside of the free streaming limit. However, Marshak 1958 presented a time-dependent radiative transfer problem. The test which is described below is a variation of the problem presented by Marshak 1958 and was analyzed by Pomraning 1979 but more extensively by Su \& Olson 1996. Therefore, this test is referred to as the Su-Olson problem. \\

\noindent
The following test investigates non-equilibrium radiation diffusion as well as the physics of matter-radiation coupling and consists of a purely absorbing, homogeneous medium that occupies the one-dimensional, semi-infinite space $0 \leq x < \infty$. Initially, the medium is defined by zero radiation energy $E_{r}$ and material temperature $T$. Upon beginning the calculation, a time independent radiative flux $F_{r}^{\textrm{inc}}$ impinges upon the left boundary at $x = 0$ \cite{suolson1996}. If one ignores hydrodynamic motion as well as heat conduction and further assumes a diffusion-type approximation, then one can write the following simplified expressions \cite{suolson1996, reynolds2009}: 
\begin{equation}
\partdif{E_{r}(x,t)}{t} - \partdif{}{x} \left( \frac{\mathbb{C} f}{\sigma_{t}} \partdif{E_{r}(x,t)}{x} \right) = \mathbb{C} \sigma_{a} \left( T^{4}(x,t) - E_{r}(x,t) \right) , \label{eq:marshak_e}
\end{equation}
\begin{equation}
c_{v}(T) \partdif{T(x,t)}{t} = - \mathbb{C} \sigma_{a} \left( T^{4}(x,t) - E_{r}(x,t) \right) , \label{eq:marshak_t}
\end{equation}

\noindent
where $c_{v}$ is the specific heat capacity of the material medium and $T$ is related to the material internal energy $e = \rho c_{v} T$. If one chooses the following functional form for the specific heat $c_{v} = \zeta T^{3}$, then Equations \ref{eq:marshak_e} and \ref{eq:marshak_t} can be written as linear ODEs in $E_{r}$ and $T^4$. Here, $\zeta$ is an arbitrary constant that if chosen appropriately simplifies how material and radiation quantities evolve \cite{suolson1996, reynolds2009}. Lastly, the boundary and initial conditions are defined by:
\begin{equation}
\textrm{BC:} ~~
E_{r}(0,t) - \frac{2}{\sigma_{t}} \partdif{E_{r}(0,t)}{x} = 4 F_{r}^{\textrm{inc}} , ~~
E_{r}(\infty,t) = 0  ,  \label{eq:marshak_bc}
\end{equation}
\begin{equation}
\textrm{IC:} ~~ E_{r}(x,0) = T(x,0) = 0 . \label{eq:marshak_ic}
\end{equation}

\noindent
If one sets $\zeta = 4$ and defines the radiation flux as a dependent variable, then the above system can be cast into a form that resembles the radiation subsystem where the material temperature temporally evolves with the radiation quantities:
\begin{equation}
\partdif{E_{r}}{t} + \partdif{F_{r}}{x} = \mathbb{C} \sigma_{a} \left( T^{4} - E_{r} \right) , \label{eq:suolson_e}
\end{equation}
\begin{equation}
F_{r} = - \frac{\mathbb{C} f}{\sigma_{t}} \partdif{E_{r}}{x} , \label{eq:suolson_f}
\end{equation}
\begin{equation}
\partdif{T^{4}}{t} = - \epsilon \mathbb{C} \sigma_{a} \left( T^{4} - E_{r} \right) , \label{eq:suolson_t}
\end{equation}
\begin{equation}
\textrm{IC:} ~~ E_{r}(0,t) + 2 F_{r}(0,t) = 4 F_{r}^{\textrm{inc}} . \label{eq:suolson_bc}
\end{equation}

\noindent
This problem setup along with specific initial and boundary conditions as well as a unique choice of parameters causes the radiation subsystem to exhibit well defined dynamics characterized by the diffusion limit and non-equilibrium behavior. Su \& Olson 1996 arrived at a semi-analytic solution to the above system by defining dimensionless independent variables $(\mathcal{X}, \mathcal{T})$ and dependent variables $(\mathcal{U}, \mathcal{V})$ such that:
\begin{equation}
\mathcal{X} \equiv \sigma_{a} \sqrt{3} x , ~~~~
\mathcal{T} \equiv \epsilon \mathbb{C} \sigma_{a} t ,
\end{equation}
\begin{equation}
\mathcal{U}(\mathcal{X},\mathcal{T}) \equiv E_{r}(x,t) / 4 F_{r}^{\textrm{inc}} , ~~~~
\mathcal{V}(\mathcal{X},\mathcal{T}) \equiv T^{4}(x,t) / 4 F_{r}^{\textrm{inc}} .
\end{equation}

\noindent
Using these dimensionless variables, Equations \ref{eq:marshak_e}-\ref{eq:marshak_ic} become \cite{suolson1996, reynolds2009}:
\begin{equation}
\epsilon \partdif{\mathcal{U}}{\mathcal{T}} - \partdif{^2 \mathcal{U}}{\mathcal{X}^2} = \mathcal{V} - \mathcal{U} , \label{eq:marshak_nd_e}
\end{equation}
\begin{equation}
\partdif{\mathcal{V}}{\mathcal{T}} = \mathcal{U} - \mathcal{V} , \label{eq:marshak_nd_t}
\end{equation}
\begin{equation}
\textrm{BC:} ~~
\mathcal{U}(0,\mathcal{T}) - \frac{2}{\sqrt{3}} \partdif{U(0,\mathcal{T})}{\mathcal{X}} = 1 , ~~
\mathcal{U}(\infty,\mathcal{T}) = 0 , \label{eq:marshak_nd_bc}
\end{equation}
\begin{equation}
\textrm{IC:} ~~ \mathcal{U}(\mathcal{X},0) = \mathcal{V}(\mathcal{X},0) = 0 . \label{eq:marshak_nd_ic}
\end{equation}

\noindent
The first boundary condition in Equation \ref{eq:marshak_nd_bc} enforces the constraint of constant flux on the left side of the computational domain. This special boundary condition can be implemented by imposing the time-varying Dirichlet condition $\mathcal{U}(0,\mathcal{T})$ \cite{reynolds2009}. This quantity, which is evaluated at $\mathcal{X}=0$, is given by \cite{suolson1996}:
\begin{eqnarray}
\mathcal{U}(\mathcal{X},\mathcal{T}) = 1 &-& \frac{2 \sqrt{3}}{\pi} \int_{0}^{1} d \xi \exp( - \mathcal{T} \xi^{2} ) \left( \frac{ \sin( \mathcal{X} \Upsilon_{1}(\xi) + \Theta_{1}(\xi) ) }{ \xi \sqrt{3 + 4 \Upsilon_{1}^{2}(\xi)} } \right)  \label{eq:marshak_bc_int}  \\ 
~  &-& \frac{\sqrt{3}}{\pi} \exp( - \mathcal{T} ) \int_{0}^{1} d \xi \exp( - \mathcal{T} / \epsilon \xi ) \left( \frac{ \sin( \mathcal{X} \Upsilon_{2}(\xi) + \Theta_{2}(\xi) ) }{ \xi (1 + \epsilon \xi) \sqrt{3 + 4 \Upsilon_{2}^{2}(\xi) } } \right) ,  \nonumber
\end{eqnarray}
\begin{equation}
\Upsilon_{1}(\xi) = \xi \sqrt{ \epsilon + \frac{1}{1-\xi^{2}} } , ~~~~
\Upsilon_{2}(\xi) = \sqrt{ (1 - \xi) \left( \epsilon + \frac{1}{\xi} \right) } ,
\end{equation}
\begin{equation}
\Theta_{n}(\xi) = \cos^{-1} \sqrt{ \frac{3}{3 + 4 \Upsilon_{n}^{2}(\xi)}} , ~~~~ n = 1,2 .
\end{equation}

\noindent
Since the backward Euler upwinding scheme is a first-order accurate method, it suffices to use the Trapezoidal Rule $T_{\textrm{int}}(f) = (f(b)+f(a))(b-a)/2$ to numerically evaluate the integrals in Equation \ref{eq:marshak_bc_int}. Higher-order integration techniques were also attempted, but these methods had little effect on the accuracy of the solution computed by the backward Euler upwinding scheme. One additional difficulty associated with evaluating \ref{eq:marshak_bc_int} is that the integrands are highly oscillatory for $\mathcal{T} \ll 1$. Therefore, one uses the following asymptotic expression whenever $\mathcal{T} < 10^{-3}$ \cite{suolson1996, reynolds2009}:
\begin{equation}
\mathcal{U}(0,\mathcal{T}) = \sqrt{ \frac{3 \mathcal{T}}{\pi \epsilon} } - \frac{3 \mathcal{T}}{4 \epsilon} + \frac{\mathcal{T}}{2 \epsilon} \sqrt{ \frac{\mathcal{T}}{3 \pi \epsilon} } + \mathcal{O}(\mathcal{T}^{2})  .
\end{equation}

\noindent
The last element in setting up this problem is updating the material temperature $T$. In previous tests, $T$ was computed by $(i)$ using other material quantities: $T = p / \rho$, $(ii)$ defining a fixed background temperature: $T(x,t)=T^{0}(x,0)$, or $(iii)$ assuming the system to be in thermal equilibrium: $T^{4}=E_{r}$. Due to the non-equilibrium nature of this problem, one implicitly advances $T$ according to the following relation prior to solving the radiation variables in the backward Euler upwinding scheme \cite{suolson1996}:
\begin{equation}
\left( T^{4}_{i} \right)^{n+1} = \left( \frac{1}{1 + \Delta t \epsilon \mathbb{C} \sigma_{a}} \right) \left( \left( T^{4}_{i} \right)^{n} + \Delta t \epsilon \mathbb{C} \sigma_{a} E_{r,i}^{n} \right) .  \label{eq:imp_mat_temp}
\end{equation}

\noindent
The following parameters were chosen because of earlier work by by Sekora \& Stone 2009, Su \& Olson 1996, and Reynolds et al 2009. By setting $F_{r}^{\textrm{inc}} = 1/4$, one normalizes the problem such that $E_{r}(x,t)$ and $T^{4}(x,t)$ directly correspond to $\mathcal{U}(\mathcal{X},\mathcal{T})$ and $\mathcal{V}(\mathcal{X},\mathcal{T})$ under the appropriate change of spatial and temporal coordinates \cite{suolson1996}. Since the Su-Olson problem assumes a diffusion approximation, one chooses $\sigma_{a}, \sigma_{t} = 40$ like in the test problem related to the weak equilibrium diffusion limit. Reynolds et al 2009 examine the spatial range $0 \leq \mathcal{X} \leq 35$ with a resolution of 2048 grid cells. These  values and the relation $\mathcal{X} \equiv \sigma_{a} \sqrt{3} x$ define the spatial aspect of the computational domain. Furthermore, Su \& Olson 1996 tabulate semi-analytic solutions for the dimensionless time $\mathcal{T} = [1, 3, 10, 30, 100]$. Therefore, these values and the relation $\mathcal{T} \equiv \epsilon \mathbb{C} \sigma_{a} t$ define the temporal aspect of the computational domain. Again, this problem was solved using the full dynamical Nike node as calculations were performed using $\epsilon = 0.1, 1.0$.  \\

\noindent
\textbf{Parameters:}
\begin{equation}
\sigma_{a}, \sigma_{t} = 40 , ~ f = 1/3 , ~ F_{r}^{\textrm{inc}} = 1/4 , \nonumber
\end{equation}
\begin{equation}
x_{\min} = 0 , ~ x_{\max} = 0.5 , ~ N_{\textrm{cell}} = [32, ~ 64, ~ 128, ~ 256, ~ 512] , \nonumber
\end{equation}
\begin{equation}
\mathcal{T} = [1, 3, 10, 30, 100]: ~ \left \{ \begin{array}{c}
\epsilon = 0.1  \rightarrow  t = [2.5, 7.5, 25, 75, 250] \times 10^{-6} \\
\epsilon = 1.0  \rightarrow  t = [2.5, 7.5, 25, 75, 250] \times 10^{-7} 
\end{array} \right .
\nonumber
\end{equation}

\begin{figure}
\begin{center}
\begin{minipage}{3in}
\includegraphics[width=3in,angle=0]{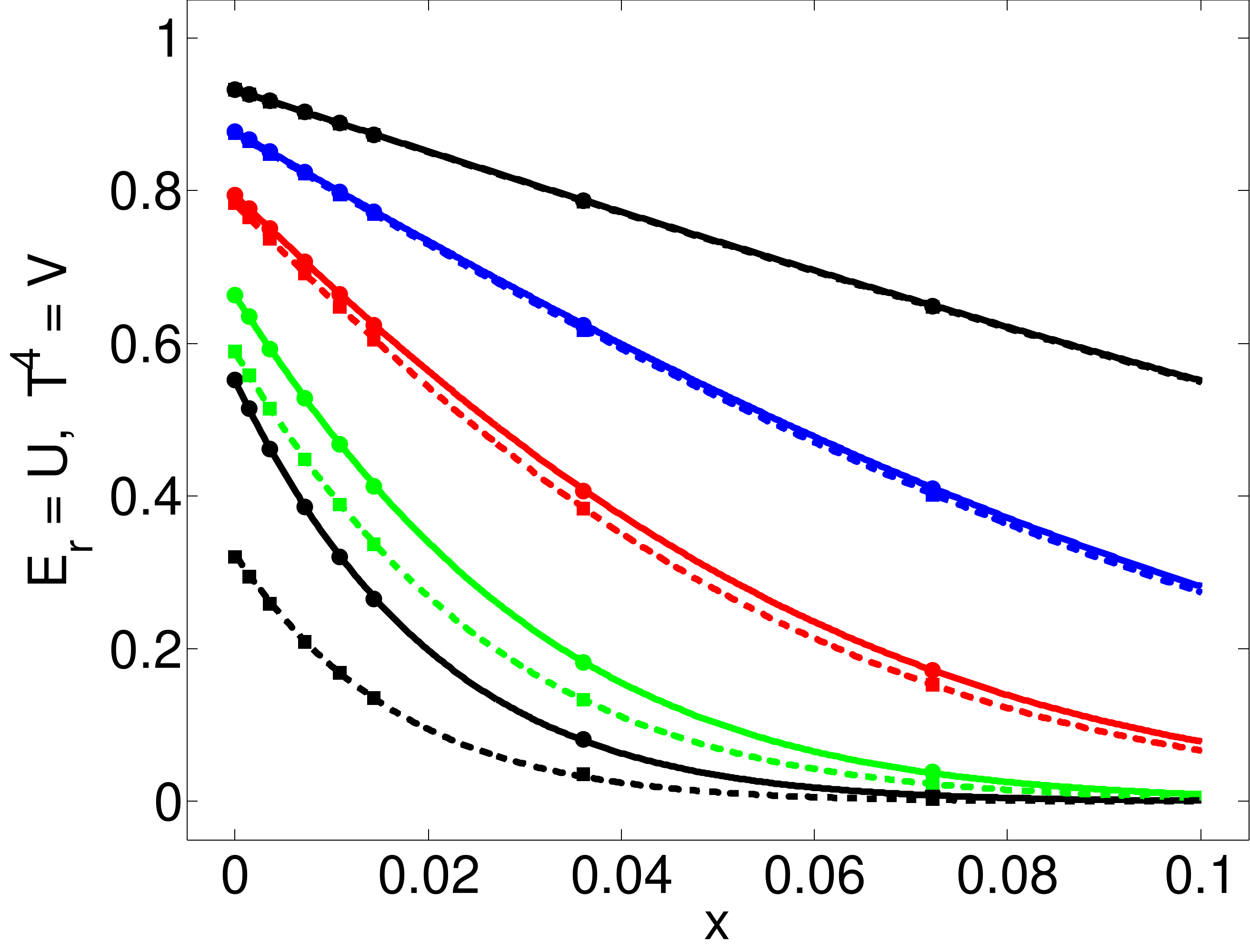}
\caption{\label{fig:rad_marshak_ep01} Su-Olson problem with $\epsilon = 0.1$, where the sets of curves from bottom to top correspond to $t = [2.5, 7.5, 25, 75, 250] \times 10^{-6}$, respectively. $N_{\textrm{cell}} = 512$.}
\end{minipage} 
\hspace{0.4in}
\begin{minipage}{3in}
\includegraphics[width=3in,angle=0]{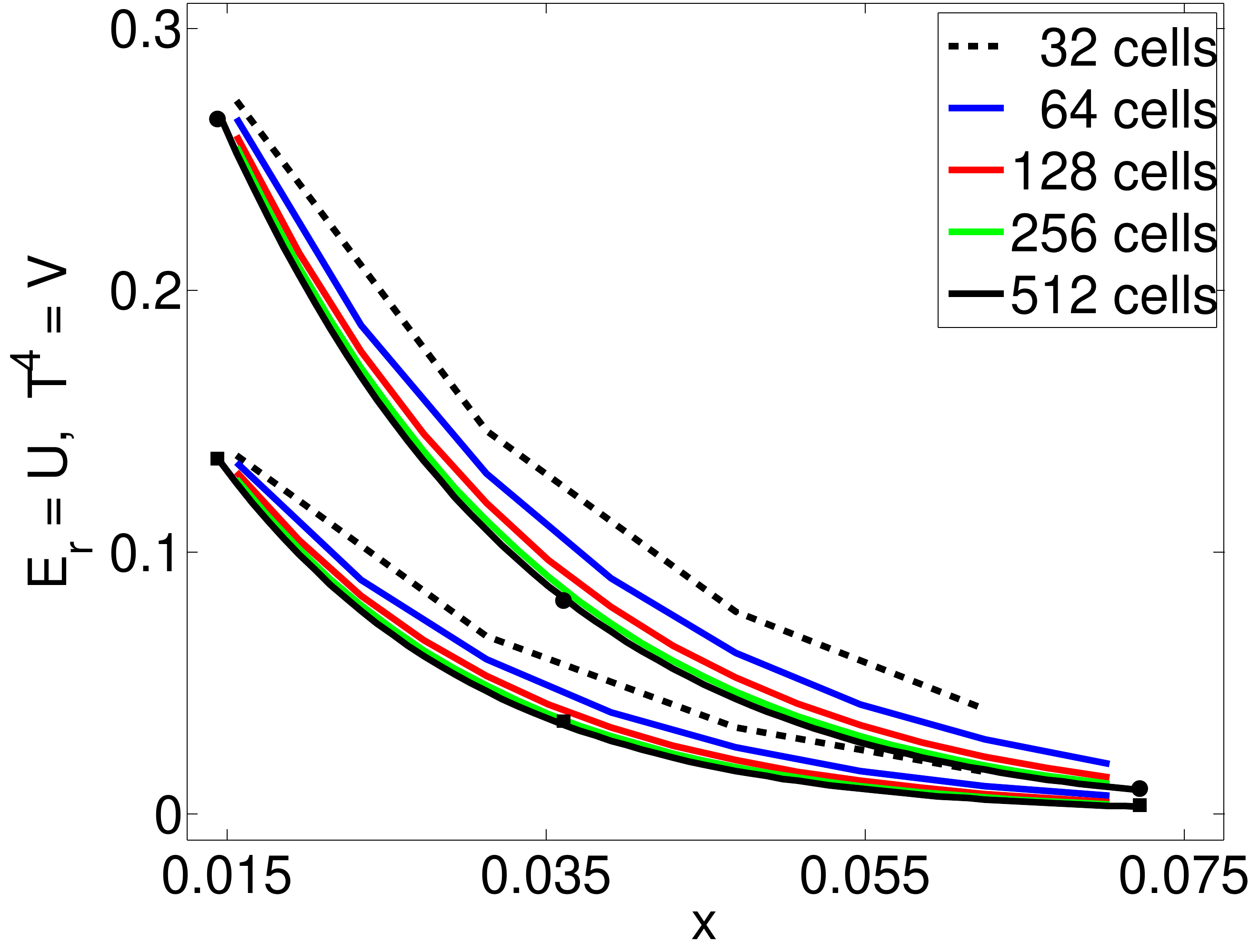}
\caption{\label{fig:rad_marshak_ep01_comp} Convergence results for the Su-Olson problem with $\epsilon = 0.1$ and $t = 2.5 \times 10^{-6}$. The bottom and top sets of curves correspond to $T^{4}$ and $E_r$, respectively.}
\end{minipage}
\end{center}
\end{figure}

\noindent
Figure \ref{fig:rad_marshak_ep01} shows how the Su-Olson problem evolves for $\epsilon = 0.1$. Here, one sees how $T^{4}$ (dashed lines) and $E_{r}$ (solid lines) approach the same value at $t = 250 \times 10^{-6}$ and the problem reaches thermal equilibrium. Also included in Figure \ref{fig:rad_marshak_ep01} are values for the semi-analytic solution that were tabulated by Su \& Olson 1996. Circles and squares mark semi-analytic values for $E_{r}$ and $T^{4}$, respectively. \\

\noindent
Figure \ref{fig:rad_marshak_ep01_comp} demonstrates how the numerical solution for the Su-Olson problem converges to the semi-analytic values. By visually inspecting this plot, one observes two important results. First, one notices that the magnitude of the error at a given point decreases by a factor of 2 as the grid spacing $\Delta x$ decreases by a factor of 2. This behavior characterizes a first-order convergence rate, which one expects for the backward Euler upwinding scheme. Second, this convergence test validates that the hybrid Godunov method can capture diffusion behavior arising from a system of balance laws with stiff relaxation source terms. When one compares the results in Figure \ref{fig:rad_marshak_ep01_comp} to the results in Figure 5 of Reynolds et al 2009, one sees that the magnitude of the error is smaller in Figure \ref{fig:rad_marshak_ep01_comp}. Calculations were also performed for $\epsilon = 1$. These solutions also passed through the semi-analytic solutions of Su \& Olson 1996.


\section{Full Radiation Hydrodynamics}
\noindent
One example of full radiation hydrodynamics that has been studied analytically and numerically is the radiating shock. For sufficiently strong shocks, the post shock material emits radiation which penetrates upstream and preheats the system \cite{ensman1994, hayes2003}. Here, the radiative transport of energy plays a significant role in defining the shock structure and evolution \cite{reynolds2009}. Furthermore, if $P_r > p$ then radiation also transports momentum. \\

\noindent
Lowrie \& Rauenzahn 2007 and Lowrie \& Edwards 2008 give an excellent summary of the solution phenomenology associated with radiating shock waves which are classified into two main types: subcritical and supercritical. As the strength of a shock increases, the temperature behind the shock $T_{1}$ rises and produces a flux that penetrates the upstream material and heats the region from a reference temperature $T_{0}$ to temperature $T_{p} > T_{0}$. If $T_{p} < T_{1}$, then the shock is termed \textbf{subcritical}. Another characteristic of these shock waves is that at higher flow velocities, $T_{p}$ increases relative to $T_{1}$. At some critical velocity $u_{\textrm{crit}}$, $T_{p}=T_{1}$ and such a shock is termed \textbf{critical}. However, if $T_{p} = T_{1}$ and $u > u_{\textrm{crit}}$, then the shock is termed \textbf{supercritical} \cite{turner2001, ensman1994, hayes2003, lowrieRauenzahn2007}. If the material density $\rho_{p}$ associated with temperature $T_{p}$ is less than the material density associated with the temperature behind the shock $T_{1}$, then the material temperature profile has the following features $(i)$ precursor: material is preheated ahead of the shock front by radiation, $(ii)$ Zel'dovich spike: overshoot at the shock front that reaches a non-equilibrium temperature $T_{s} > T_{1}$, and $(iii)$ relaxation region: as extra energy is radiated away, the material temperature declines from $T_{s}$ to $T_{1}$ where the width of the region is proportional to the post-shock optical depth \cite{reynolds2009, lowrieRauenzahn2007}. \\

\noindent
The test problems in this section are based on Ensman 1994, which considered the case of a piston moving through static media and followed the evolution with a Lagrangian code. Since Nike evolves the dynamical quantities in the Eulerian frame, the problem can be recast such that a moving medium impacts a stationary reflecting boundary. The material velocity is initialized to match the piston speed specified in the Ensman paper. To produce figures that are directly comparable to results obtained from a Lagrangian code, one transforms the coordinates of the Eulerian frame calculation according to $x_{\textrm{Eulerian}} = x_{\textrm{lab}} - u_{\textrm{inflow}} t$, where $x_{\textrm{lab}}$ is the lab frame coordinate, $x_{\textrm{Eulerian}}$ is the Eulerian frame coordinate, and $u_{\textrm{inflow}}$ is the inflow velocity \cite{hayes2003}. The following physical parameters were adapted from Ensman 1994 and Hayes \& Norman 2003: \\

\noindent
\textbf{Physical Parameters:}
\begin{equation}
c = 3 \times 10^{10} ~ cm ~ s^{-1} , ~ a_{\infty} = 3 \times 10^{4} ~ cm ~ s^{-1} , ~ f = 1/3 \nonumber 
\end{equation}
\begin{equation}
L = 7 \times 10^{10} ~ cm , ~ \mathcal{L} = L / \lambda_{t} = 22 \Rightarrow \lambda_{t} = 3.2 \times 10^{9} ~ cm , ~ N_{\textrm{cell}} = 300 , \nonumber 
\end{equation}
\begin{equation}
\rho^{0} = 7.78 \times 10^{-10} ~ g ~ cm^{-3} , ~ T_{\infty} = T_{r,\infty} = \left( E_{r,\infty} \right)^{1/4} = 10 ~ K , ~ \sigma_{a} = 3.1 \times 10^{-10} ~ cm^{-1} , \nonumber
\end{equation}
\begin{equation}
\textrm{Subcritical: } u_{\textrm{inflow}} = -6 \times 10^{5} ~ cm ~ s^{-1} , ~ t = [1.7, 2.8, 3.8] \times 10^{4} ~ s , \nonumber
\end{equation}
\begin{equation}
\textrm{Supercritical: } u_{\textrm{inflow}} = -20 \times 10^{5} ~ cm ~ s^{-1} , ~ t = [4, 7.5, 13] \times 10^{3} ~ s , \nonumber
\end{equation}
\begin{equation}
T(x,0) = T_{\infty} + 75 (x/L) ~ K , ~ \textrm{Left BC: Reflecting} , ~ \textrm{Right BC: Inflow} .  \nonumber
\end{equation}

\noindent
Motivated by these physical parameters and the non-dimensionalization adopted throughout this paper \cite{lowrie2001, lowrie1999, lowrieRauenzahn2007}, one arrives at the following set of values: 
\begin{equation}
U(x,0) =
\left( \begin{array}{c} 
       \rho^{0} \\
			 m^{0} \\
			 E^{0} \\
			 E_{r}^{0} \\
			 F_{r}^{0} \end{array} \right) =
\left( \begin{array}{c} 
       \rho^{0} \\
			 \rho^{0} u_{\textrm{inflow}} \\
			 \frac{1}{2} \rho^{0} u^{2}_{\textrm{inflow}} + \frac{\rho^{0} T(x,0)}{(\gamma-1)} \\
			 T^{4}(x,0) \\
			 \frac{-f}{\sigma_{t}} \partdif{E_{r}}{x} = \frac{-f}{\sigma_{t}} \left( 4 T^{3}(x,0) \frac{d T}{dx} \right) = \frac{-30 f T^{3}(x,0)}{\sigma_{t}} \end{array} \right) .
\end{equation}

\noindent
\textbf{Non-dimensionalized Parameters:}
\begin{equation}
\mathbb{C} = 10^{6} , ~ \mathbb{P} = 10^{-11} , ~ \sigma_{a}, \sigma_{t} = 10 , ~ f = 1/3 , \nonumber 
\end{equation}
\begin{equation}
\Delta t = \frac{\nu \Delta x}{\max_{i} ( |u_i| + a_{\textrm{eff},i} )} , ~ x_{\min} = 0 , ~ x_{\max} = 2 , ~ N_{\textrm{cell}} = 512 , \nonumber
\end{equation}
\begin{equation}
\rho^{0} = 1 , ~ T(x,0) = T_{\infty} + 7.5 \left( \frac{x}{x_{\max}-x_{\min}} \right) , ~ T_{\infty} = T_{r,\infty} = \left( E_{r,\infty} \right)^{1/4} = 1 , \nonumber
\end{equation}
\begin{equation}
\textrm{Subcritical: } u_{\textrm{inflow}} = -20 , ~ t = [2, 2.5, 3] \times 10^{-2} , \nonumber
\end{equation}
\begin{equation}
\textrm{Supercritical: } u_{\textrm{inflow}} = -66.6 , ~ t = [3, 6, 9] \times 10^{-3} , \nonumber
\end{equation}
\begin{equation}
\textrm{Left BC: Reflecting} , ~ \textrm{Right BC: Inflow} . \nonumber
\end{equation}

\begin{figure}
\begin{center}
\begin{minipage}{3in}
\includegraphics[width=3in,angle=0]{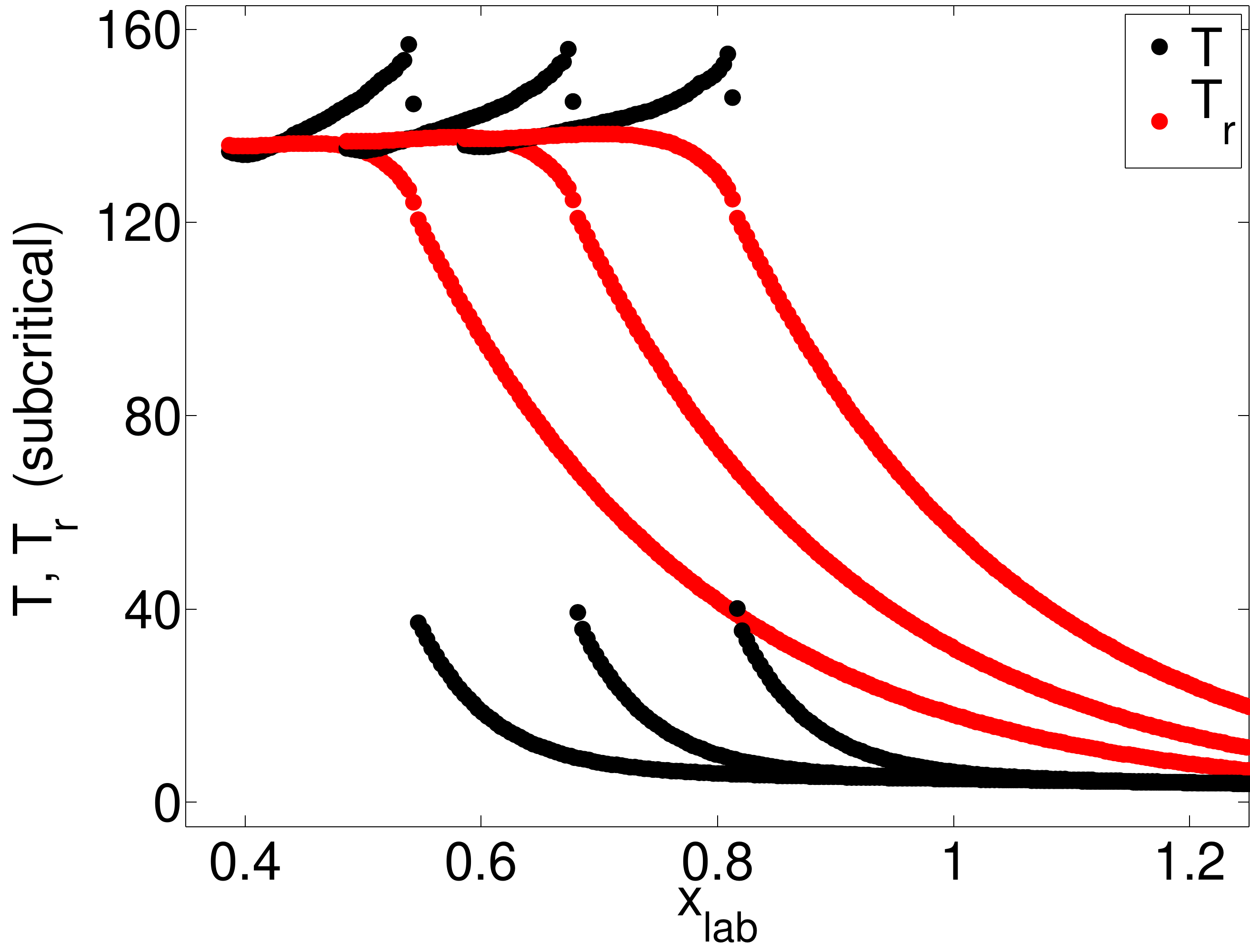}
\caption{\label{fig:rh_sub_t} $T$ and $T_{r}$ for subcritical shock. $t = [2, 2.5, 3] \times 10^{-2}$. $N_{\textrm{cell}} = 512$.}
\end{minipage} 
\hspace{0.4in}
\begin{minipage}{3in}
\includegraphics[height=3in,angle=90]{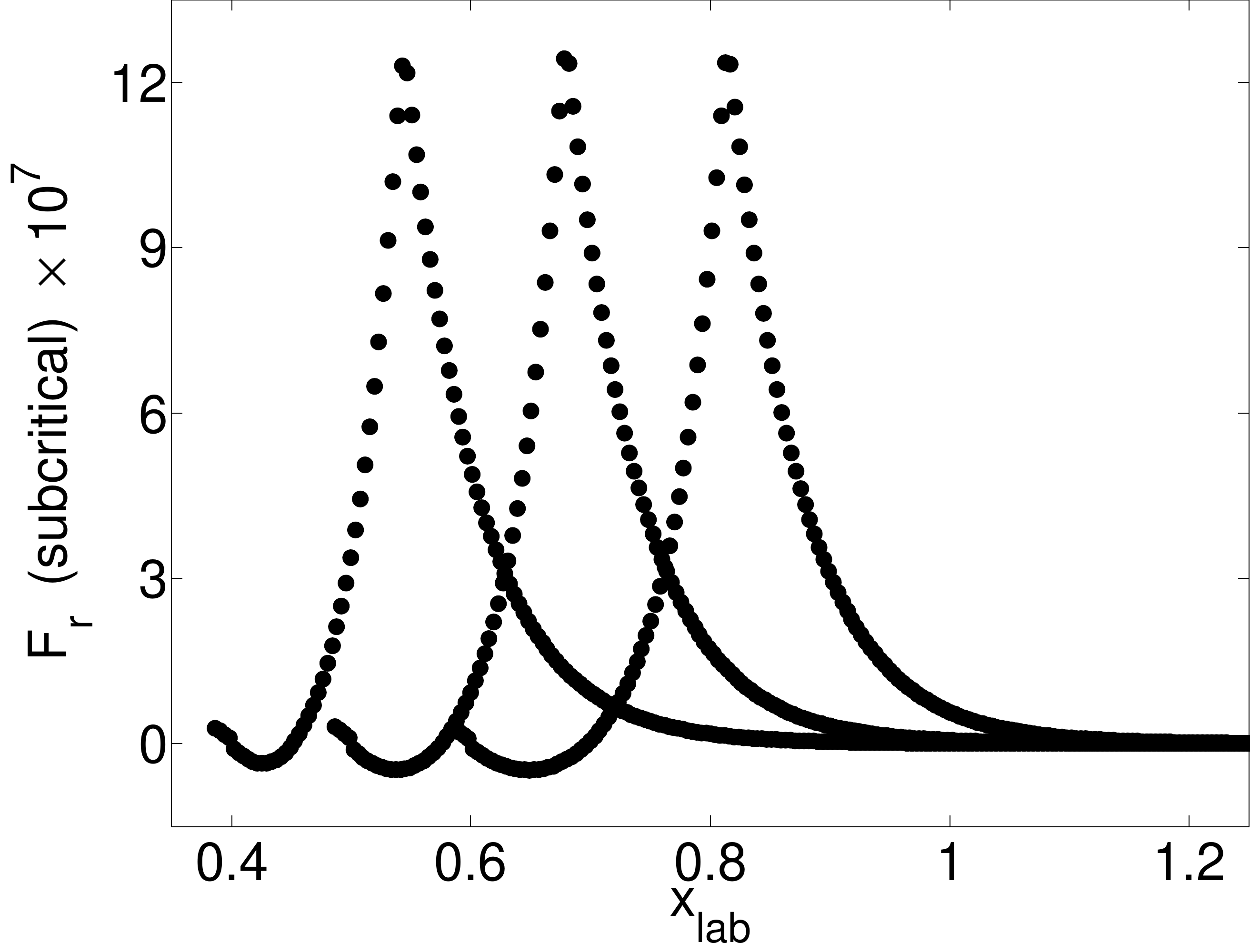}
\caption{\label{fig:rh_sub_fr} $F_{r}$ for subcritical shock. $t = [2, 2.5, 3] \times 10^{-2}$. $N_{\textrm{cell}} = 512$.}
\end{minipage}
\end{center}
\end{figure}

\noindent
Figures \ref{fig:rh_sub_t} and \ref{fig:rh_sub_fr} show the results for the subcritical shock. Here, the the Zel'dovich spike (temperature overshoot) is sharply peaked, lies just behind the shock, and is only seen in the material temperature profile. Additionally, the material and radiation temperatures demonstrate non-equilibrium behavior on both sides of the shock front \cite{hayes2003}. Furthermore, the flux profile is nearly symmetric and matches results discussed by Ensman 1994. As the profile moves from left to right, the shock front is located at the point of maximum flux \cite{ensman1994}. It is important to note how sharply the hybrid Godunov method captures radiative shock fronts, where the discontinuity in $T$ is represented with one computational cell. \\

\noindent
Lowrie \& Edward 2008 examined semi-analytic solutions of planar radiative shock waves with a gray non-equilibrium diffusion model for radiation hydrodynamics and derived an estimate of the maximum material temperature at the Zel'dovich spike:
\begin{equation}
T_{\max}^{\textrm{LE}} = \max \left \{ T_{1} , \frac{ \left( 3 (\gamma \mathcal{M}_{0}^{2} + 1) + \gamma \mathbb{P} (1 - T_{1}^{4}) \right)^{2} }{ 36 \gamma \mathcal{M}_{0}^{2} } \right \} ,
\end{equation}

\noindent
where $\mathcal{M}_{0} = u_{\textrm{inflow}}$ is the Mach number which determines the strength of the shock. The definitions of the other quantities remain unchanged from when they were first presented in this paper. Furthermore, the authors cite a coarser estimate of this maximum temperature which was originally derived by Mihalas \& Mihalas 1984:
\begin{equation}
T_{\max}^{\textrm{MM}} = (3 - \gamma) T_{1} .
\end{equation}

\noindent
In Figure \ref{fig:rh_sub_t}, the material temperature behind the shock $T_{1} \approx 135$ and material temperature at the height of the Zel'dovich spike $T_{s} \approx 156$ for times $t = [2, 2.5, 3] \times 10^{-2}$. This value compares well with $T_{\max}^{\textrm{LE}} = 167$ as well as the coarser estimate $T_{\max}^{\textrm{MM}} = 180$. \\

\begin{figure}
\begin{center}
\begin{minipage}{3in}
\includegraphics[width=3in,angle=0]{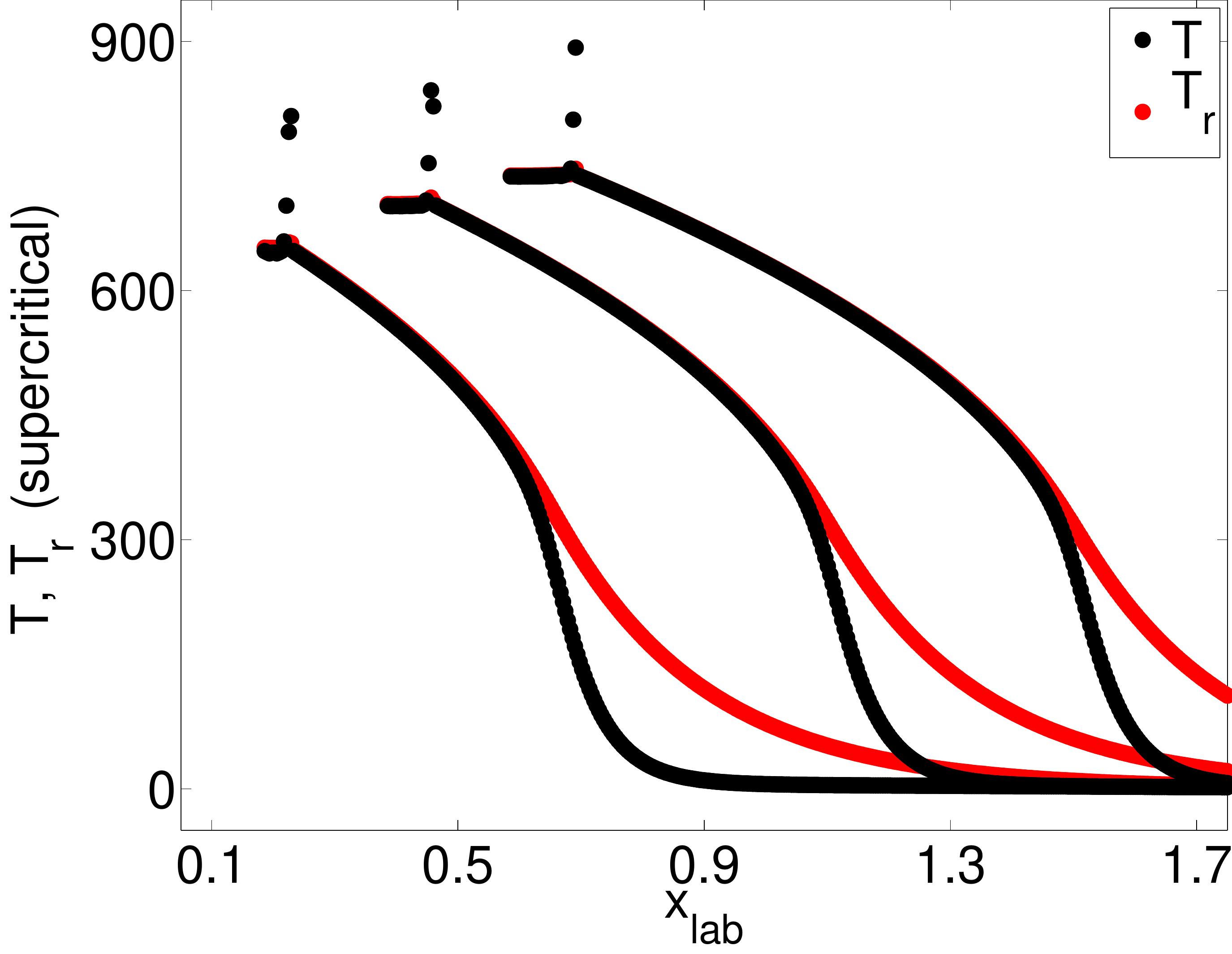}
\caption{\label{fig:rh_super_t} $T$ and $T_{r}$ for supercritical shock. $t = [3, 6, 9] \times 10^{-3}$. $N_{\textrm{cell}} = 512$.}
\end{minipage} 
\hspace{0.4in}
\begin{minipage}{3in}
\includegraphics[width=3in,angle=0]{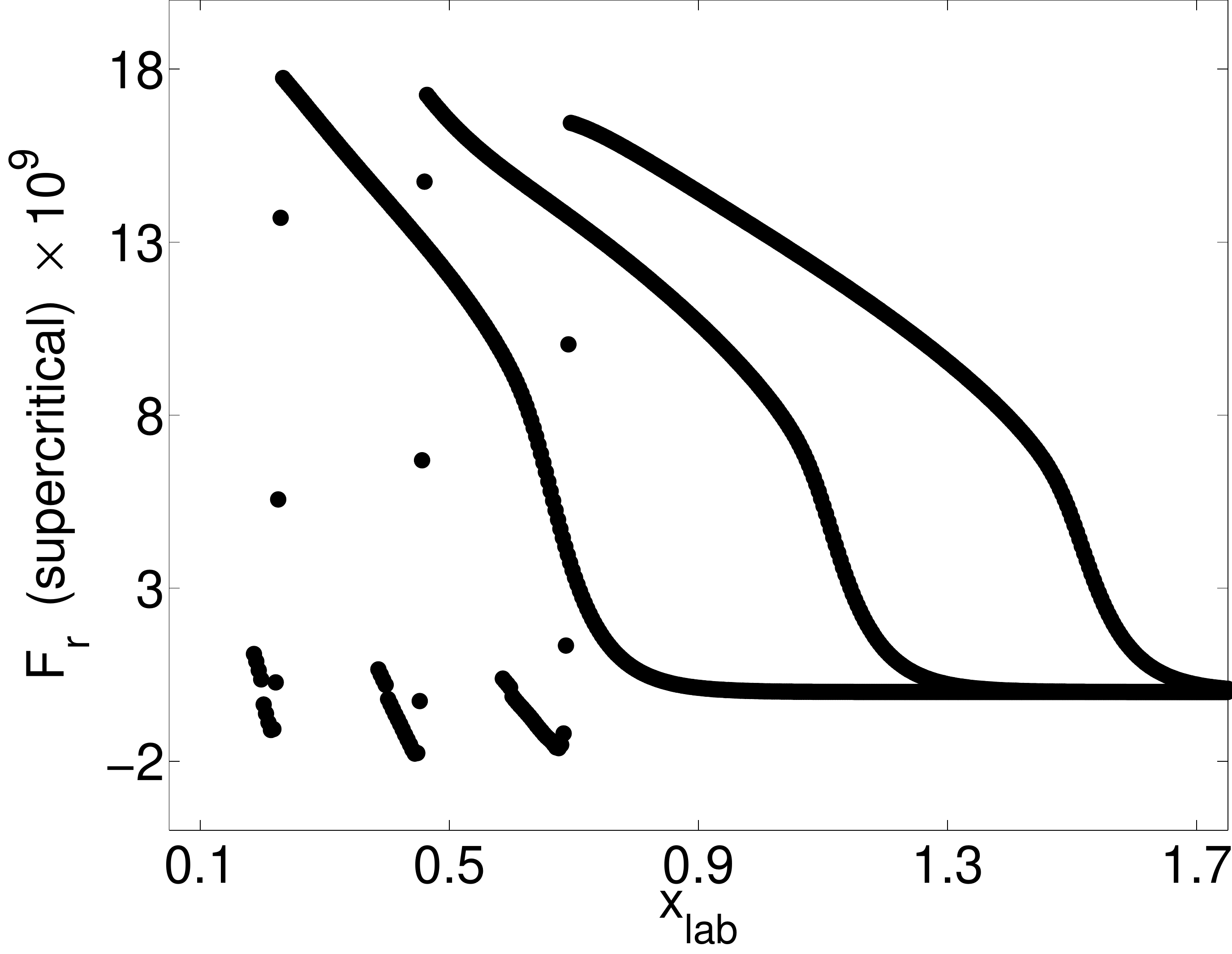}
\caption{\label{fig:rh_super_fr} $F_{r}$ for supercritical shock. $t = [3, 6, 9] \times 10^{-3}$. $N_{\textrm{cell}} = 512$.}
\end{minipage}
\end{center}
\end{figure}

\noindent
Figures \ref{fig:rh_super_t} and \ref{fig:rh_super_fr} show the results for the supercritical shock. Here, the shock front is located at the temperature maximum. In many instances, computer codes have to resort to adaptive mesh refinement to reasonably capture the sharp temperature spike seen in Figure \ref{fig:rh_super_t}. In these plots, one sees the expected results of an extended region where $T_{p} \leq T_{1}$ and the radiation flux $F_{r}$ being asymmetric \cite{ensman1994, lowrie2008, lowrieRauenzahn2007}. In Figure \ref{fig:rh_super_t}, $T_{1} = [646, 702, 738]$ and $T_{s} = [810, 841, 893]$ for $t = [3, 6, 9] \times 10^{-3}$. For this spatial resolution, these values are within the coarser estimate $T_{\max}^{\textrm{MM}} = [861, 936, 984]$. \\

\noindent
The tests herein examine the behavior of full radiation hydrodynamics. As mentioned earlier, radiation is the dominant transport mechanism for energy and momentum when $P_r > p$ which is equivalent to $f E_r > p$. If one considers equilibrium diffusion behavior  and the material equation of state, then the above inequality becomes $f T^4_r > \rho T$. Since $\rho$ is initially set equal to unity and $f = 1/3$, this inequality is satisfied by the profiles in Figures \ref{fig:rh_sub_t} and \ref{fig:rh_super_t} - illustrating that radiation can function as the dominant transport mechanism. \\

\begin{figure}
\begin{center}
\includegraphics[width=3in,angle=0]{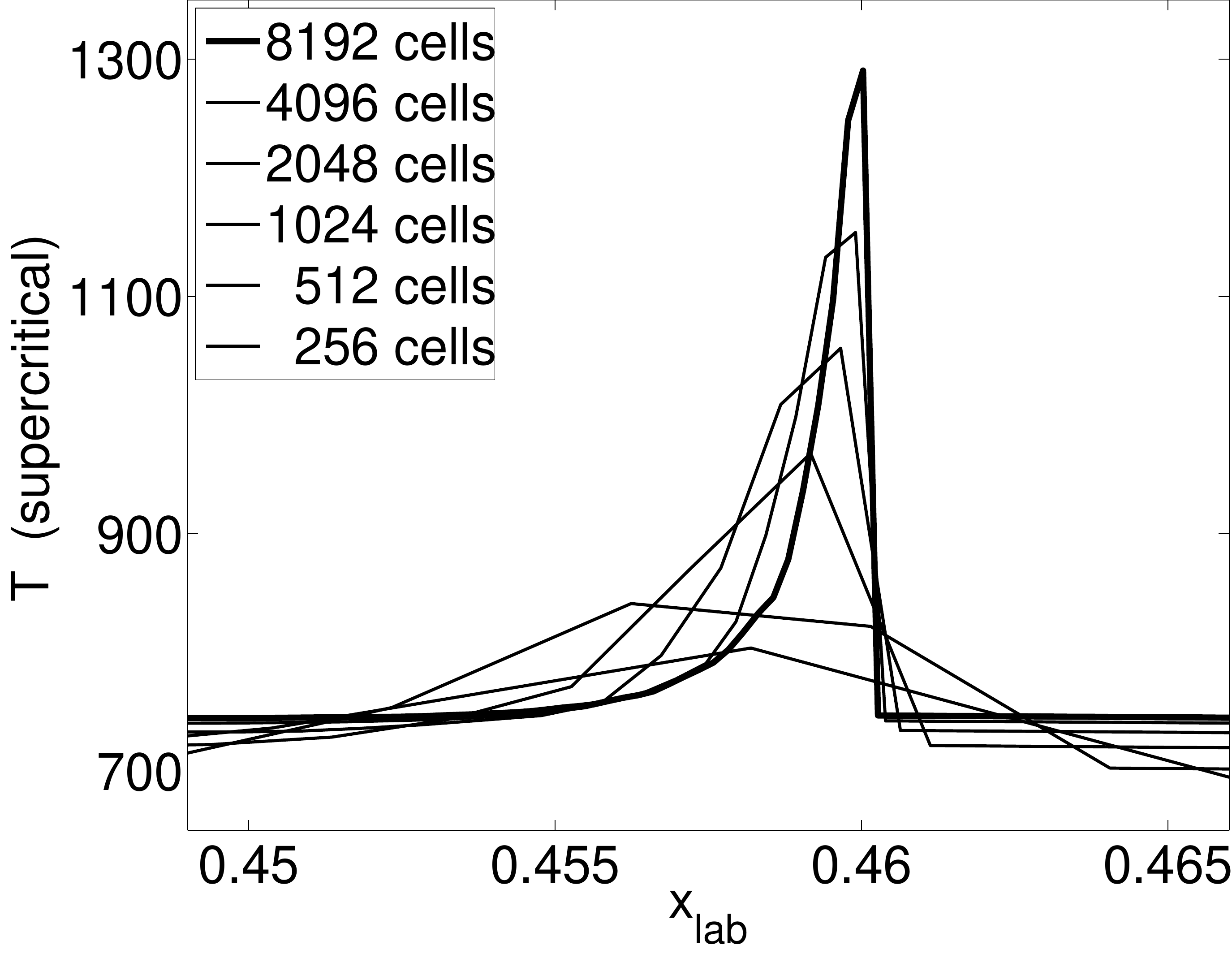}
\caption{\label{fig:rh_super_high_res} A supercritical Zel'dovich spike for various spatial resolutions. $t = 0.006$.}
\end{center}
\end{figure}

\noindent
Figure \ref{fig:rh_super_high_res} demonstrates self-similar convergent behavior in the numerical solution for a Zel'dovich spike in the material temperature $T$ of a supercritical shock wave for a range of resolutions $N_{\textrm{cell}} = [256, 512, 1024, 2048, 4096, 8192]$. These results were initialized with the same conditions as the plot of $t = 6 \times 10^{-3}$ in Figure \ref{fig:rh_super_t}. Figure \ref{fig:rh_super_high_res} clearly shows asymmetry in the supercritical Zel'dovich spike. This observation corresponds well with the discussion in Lowrie \& Rauenzahn 2007 and Lowrie \& Edwards 2008 as well as the numerical results in Sincell et al 1999. However, this physical feature is often missing in many radiation hydrodynamical calculations of supercritical shocks. Lastly, the sharp discontinuity from $T_{s}$ to $T_{p}$ as well as the gradual decay from $T_{s}$ to $T_{1}$ in the relaxation region is a testament to the hybrid Godunov method in the Nike code.


\section{Conclusions and Future Work}
\noindent
This paper presents a hybrid Godunov method for full radiation hydrodynamics. Numerical tests focusing on the material components, radiation components, and fully coupled system show this technique to be uniformly well behaved across various parameter regimes. The advantage of this algorithm is that it is accurate, stable, robust, explicit of the material flow scale, unsplit, and fully couples matter and radiation without invoking a diffusion-type approximation. Two additional numerical tests that should be investigated are the dispersion analysis of the boundary and initial value problems for radiation hydrodynamical linear waves in one spatial dimension. Despite previous computational work being done on the boundary value problem by Mihalas \& Mihalas 1983 and Turner \& Stone 2001, a computational test based upon the initial value problem that resembles the propagation of linear eigenmodes in hydrodynamics has not been carried out. Constructing such a test seems possible given the research of Lowrie et al 1999, Blaes \& Socrates 2003, and Johnson 2009. These tests are important because they will examine the damping rates and propagation speeds of various radiation hydrodynamical waves. \\

\noindent
In this paper, it was shown that the effective eigen-quantities of the modified Godunov scheme for the split material subsystem exhibits adiabatic and isothermal behavior in certain parameter regimes while evolving the overall radiation hydrodynamical system according to an effective CFL condition. Additionally, the modified Godunov scheme resolves the free streaming (hydrodynamical) limit with second-order accuracy for the material quantities and captures the appropriate material temperature profiles for subcritical and supercritical radiating shock waves that are influenced by non-equilibrium diffusion behavior. For these reasons, one concludes that the modified Godunov scheme has some asymptotic preserving properties. Furthermore, this paper showed that the backward Euler upwinding scheme for the split radiation subsystem recovers free streaming, weak equilibrium diffusion, and strong equilibrium diffusion behavior even for large cell-optical depths. Additionally, the backward Euler upwinding scheme can be cast in a form where its discretization resembles a backward-time centered-space differencing of a parabolic/diffusion operator - illustrating consistency in the discretization operator. For these reasons, one concludes that the backward Euler upwinding scheme has some asymptotic preserving properties. Despite the volume of numerical tests that are showcased in this paper, it in not clear that the overall algorithm (i.e., the hybrid Godunov method which combines the modified Godunov and backward Euler upwinding schemes for their respective subsystems) is globally asymptotic preserving. One should carry out a discretization analysis on the full hybrid Godunov method in the spirit of Lowrie \& Morel 2002 which defined in a mathematically rigorous fashion what it means for an algorithm to be asymptotic preserving. With respect to this kind of work, full radiation hydrodynamics may not be the best system to analytically investigate asymptotic preserving properties of the hybrid Godunov method because of the complexity associated with the system. One should conduct such an analysis on a set of equations like the one shown in Appendix 5 which is a simpler system of hyperbolic balance laws with dual multiscale behavior. Additionally, it would be interesting to investigate the algorithmic coupling between the material and radiation subsystems by implicitly updating the material temperature $T$ in a predictor context similar to Equation \ref{eq:imp_mat_temp}. Such an update could be made after the backward Euler upwinding scheme advances the radiation quantities but before the modified Godunov scheme advances the material quantities. \\

\noindent
This numerical method employs familiar algorithmic machinery without a significant increase in computational overhead. Therefore, it seems reasonable that one will be able to extend the ideas in this paper to multiple spatial dimensions via a MUSCL or CTU approach \cite{colella1990}. Such work is currently underway. Additionally, the new algorithmic ideas were cast in such a way that they should be easily implemented into existing codes, particularly ones that carry out MHD calculations. Investigating full radiation MHD in multiple spatial dimensions is an area of research on the forefront of computational science and a subject that the authors want to pursue. Moreover, one would like to combine a method for radiation hydrodynamics with a technique for updating the variable tensor Eddington factor $\mathsf{f}$ which is used in the closure relation $\mathsf{P_r} = \mathsf{f} E_r$ for the radiation moment equations. Short characteristic $S_{N}$ as well as Monte Carlo methods are promising techniques for solving the photon transport equation and updating $\mathsf{f}$ at each temporal iteration \cite{castorbook}. Furthermore, the question remains as to whether the hybrid Godunov method and mixed frame approach are applicable to problems defined by multigroup physics. Lastly, it would be interesting to compare the performance and accuracy of the hybrid Godunov method with an algorithm that is either fully implicit, based upon the $P_{N} P_{M}$ family of schemes \cite{det2008, dbtm2008}, or uses a fully relativistic treatment of radiation and matter \cite{castorTalk, morita1986}.


\section*{Acknowledgment}
\noindent
The authors thank P Colella, R Lowrie, J Morel, J Castor, M Dumbser, E Toro, F Miniati, D Reynolds, and G Hammett for helpful discussions and insightful suggestions. MDS acknowledges support from the DOE CSGF Program which is provided under grant DE-FG02-97ER25308. JMS acknowledges support from grant DE-FG52-06NA26217.



\section*{Appendix 1: Modifications to Matrix Equations}
\noindent
\textbf{Periodic Boundary Conditions}
{\tiny {\begin{equation}
\left( \begin{array}{cccccccccccc}
\theta_{3} & \theta_{4} & \theta_{5} & \theta_{6} & 0 & \ldots & ~ & ~ & \ldots & 0 & \theta_{1} & \theta_{2} \\
  \phi_{3} &   \phi_{4} &   \phi_{5} &   \phi_{6} & 0 & \ldots & ~ & ~ & \ldots & 0 &   \phi_{1} &   \phi_{2} \\
\theta_{1} & \theta_{2} & \theta_{3} & \theta_{4} & \theta_{5} & \theta_{6} & 0 & \ldots & ~ & \ldots & 0 & 0 \\
  \phi_{1} &   \phi_{2} &   \phi_{3} &   \phi_{4} &   \phi_{5} &   \phi_{6} & 0 & \ldots & ~ & \ldots & 0 & 0 \\
         0 &          0 & \theta_{1} & \theta_{2} & \ldots & ~ & ~ & ~ & ~ & ~ & ~ & ~ \\
         0 &          0 &   \phi_{1} &   \phi_{2} & \ldots & ~ & ~ & ~ & ~ & ~ & ~ & ~ \\
    \vdots & ~ & ~ & ~ & ~ & ~ & ~ & ~ & ~ & ~ & ~ & ~ \\
         0 & \ldots & ~ & ~ & \ldots & 0 & \theta_{1} & \theta_{2} & \theta_{3} & \theta_{4} & \theta_{5} & \theta_{6} \\
         0 & \ldots & ~ & ~ & \ldots & 0 &   \phi_{1} &   \phi_{2} &   \phi_{3} &   \phi_{4} &   \phi_{5} &   \phi_{6} \\
\theta_{5} & \theta_{6} & 0 & \ldots & ~ & ~ & \ldots & 0 & \theta_{1} & \theta_{2} & \theta_{3} & \theta_{4} \\
  \phi_{5} &   \phi_{6} & 0 & \ldots & ~ & ~ & \ldots & 0 &   \phi_{1} &   \phi_{2} &   \phi_{3} &   \phi_{4} 
\end{array} \right)
\left( \begin{array}{c}
E_{r,1}^{n+1} + \theta_{7} \\
F_{r,1}^{n+1} + \phi_{7} \\
E_{r,2}^{n+1} + \theta_{7} \\
F_{r,2}^{n+1} + \phi_{7} \\
~ \\
\vdots \\
~ \\
E_{r,N-1}^{n+1} + \theta_{7} \\
F_{r,N-1}^{n+1} + \phi_{7} \\
E_{r,N}^{n+1} + \theta_{7} \\
F_{r,N}^{n+1} + \phi_{7} 
\end{array} \right) =
\left( \begin{array}{c}
E_{r,1}^{n} + \theta_{7} \\
F_{r,1}^{n} + \phi_{7} \\
E_{r,2}^{n} + \theta_{7} \\
F_{r,2}^{n} + \phi_{7} \\
~ \\
\vdots \\
~ \\
E_{r,N-1}^{n} + \theta_{7} \\
F_{r,N-1}^{n} + \phi_{7} \\
E_{r,N}^{n} + \theta_{7} \\
F_{r,N}^{n} + \phi_{7}
\end{array} \right)  \nonumber
\end{equation} }}

\noindent
\textbf{Outflow Boundary Conditions}
{\tiny {\begin{equation}
\left( \begin{array}{cccccccccccc}
\theta_{3} + \theta_{1} & \theta_{4} + \theta_{2} & \theta_{5} & \theta_{6} & 0 & \ldots & ~ & ~ & ~ & ~ & \ldots & 0 \\
  \phi_{3} +   \phi_{1} &   \phi_{4} +   \phi_{2} &   \phi_{5} &   \phi_{6} & 0 & \ldots & ~ & ~ & ~ & ~ & \ldots & 0 \\
\theta_{1} & \theta_{2} & \theta_{3} & \theta_{4} & \theta_{5} & \theta_{6} & 0 & \ldots & ~ & \ldots & 0 & 0 \\
  \phi_{1} &   \phi_{2} &   \phi_{3} &   \phi_{4} &   \phi_{5} &   \phi_{6} & 0 & \ldots & ~ & \ldots & 0 & 0 \\
         0 &          0 & \theta_{1} & \theta_{2} & \ldots & ~ & ~ & ~ & ~ & ~ & ~ & ~ \\
         0 &          0 &   \phi_{1} &   \phi_{2} & \ldots & ~ & ~ & ~ & ~ & ~ & ~ & ~ \\
    \vdots & ~ & ~ & ~ & ~ & ~ & ~ & ~ & ~ & ~ & ~ & ~ \\
         0 & \ldots & ~ & ~ & \ldots & 0 & \theta_{1} & \theta_{2} & \theta_{3} & \theta_{4} & \theta_{5} & \theta_{6} \\
         0 & \ldots & ~ & ~ & \ldots & 0 &   \phi_{1} &   \phi_{2} &   \phi_{3} &   \phi_{4} &   \phi_{5} &   \phi_{6} \\
0 & 0 & 0 & \ldots & ~ & ~ & \ldots & 0 & \theta_{1} & \theta_{2} & \theta_{3} + \theta_{5} & \theta_{4} + \theta_{6} \\
0 & 0 & 0 & \ldots & ~ & ~ & \ldots & 0 &   \phi_{1} &   \phi_{2} &   \phi_{3} +   \phi_{5 }&   \phi_{4} +   \phi_{6}
\end{array} \right)
\left( \begin{array}{c}
E_{r,1}^{n+1} + \theta_{7} \\
F_{r,1}^{n+1} + \phi_{7} \\
E_{r,2}^{n+1} + \theta_{7} \\
F_{r,2}^{n+1} + \phi_{7} \\
~ \\
\vdots \\
~ \\
E_{r,N-1}^{n+1} + \theta_{7} \\
F_{r,N-1}^{n+1} + \phi_{7} \\
E_{r,N}^{n+1} + \theta_{7} \\
F_{r,N}^{n+1} + \phi_{7} 
\end{array} \right) =
\left( \begin{array}{c}
E_{r,1}^{n} + \theta_{7} \\
F_{r,1}^{n} + \phi_{7} \\
E_{r,2}^{n} + \theta_{7} \\
F_{r,2}^{n} + \phi_{7} \\
~ \\
\vdots \\
~ \\
E_{r,N-1}^{n} + \theta_{7} \\
F_{r,N-1}^{n} + \phi_{7} \\
E_{r,N}^{n} + \theta_{7} \\
F_{r,N}^{n} + \phi_{7}
\end{array} \right)  \nonumber
\end{equation} }}

\noindent
\textbf{Reflecting Boundary Conditions}
{\tiny {\begin{equation}
\left( \begin{array}{cccccccccccc}
\theta_{3} + \theta_{1} & \theta_{4} - \theta_{2} & \theta_{5} & \theta_{6} & 0 & \ldots & ~ & ~ & ~ & ~ & \ldots & 0 \\
  \phi_{3} +   \phi_{1} &   \phi_{4} -   \phi_{2} &   \phi_{5} &   \phi_{6} & 0 & \ldots & ~ & ~ & ~ & ~ & \ldots & 0 \\
\theta_{1} & \theta_{2} & \theta_{3} & \theta_{4} & \theta_{5} & \theta_{6} & 0 & \ldots & ~ & \ldots & 0 & 0 \\
  \phi_{1} &   \phi_{2} &   \phi_{3} &   \phi_{4} &   \phi_{5} &   \phi_{6} & 0 & \ldots & ~ & \ldots & 0 & 0 \\
         0 &          0 & \theta_{1} & \theta_{2} & \ldots & ~ & ~ & ~ & ~ & ~ & ~ & ~ \\
         0 &          0 &   \phi_{1} &   \phi_{2} & \ldots & ~ & ~ & ~ & ~ & ~ & ~ & ~ \\
    \vdots & ~ & ~ & ~ & ~ & ~ & ~ & ~ & ~ & ~ & ~ & ~ \\
         0 & \ldots & ~ & ~ & \ldots & 0 & \theta_{1} & \theta_{2} & \theta_{3} & \theta_{4} & \theta_{5} & \theta_{6} \\
         0 & \ldots & ~ & ~ & \ldots & 0 &   \phi_{1} &   \phi_{2} &   \phi_{3} &   \phi_{4} &   \phi_{5} &   \phi_{6} \\
0 & 0 & 0 & \ldots & ~ & ~ & \ldots & 0 & \theta_{1} & \theta_{2} & \theta_{3} + \theta_{5} & \theta_{4} - \theta_{6} \\
0 & 0 & 0 & \ldots & ~ & ~ & \ldots & 0 &   \phi_{1} &   \phi_{2} &   \phi_{3} +   \phi_{5 }&   \phi_{4} -   \phi_{6}
\end{array} \right)
\left( \begin{array}{c}
E_{r,1}^{n+1} + \theta_{7} \\
F_{r,1}^{n+1} + \phi_{7} \\
E_{r,2}^{n+1} + \theta_{7} \\
F_{r,2}^{n+1} + \phi_{7} \\
~ \\
\vdots \\
~ \\
E_{r,N-1}^{n+1} + \theta_{7} \\
F_{r,N-1}^{n+1} + \phi_{7} \\
E_{r,N}^{n+1} + \theta_{7} \\
F_{r,N}^{n+1} + \phi_{7} 
\end{array} \right) =
\left( \begin{array}{c}
E_{r,1}^{n} + \theta_{7} \\
F_{r,1}^{n} + \phi_{7} \\
E_{r,2}^{n} + \theta_{7} \\
F_{r,2}^{n} + \phi_{7} \\
~ \\
\vdots \\
~ \\
E_{r,N-1}^{n} + \theta_{7} \\
F_{r,N-1}^{n} + \phi_{7} \\
E_{r,N}^{n} + \theta_{7} \\
F_{r,N}^{n} + \phi_{7}
\end{array} \right)  \nonumber
\end{equation} }}

\noindent
\textbf{Inflow Boundary Conditions}
{\tiny {\begin{equation}
\left( \begin{array}{cccccccccccc}
\theta_{3} & \theta_{4} & \theta_{5} & \theta_{6} & 0 & \ldots & ~ & ~ & ~ & ~ & \ldots & 0 \\
  \phi_{3} &   \phi_{4} &   \phi_{5} &   \phi_{6} & 0 & \ldots & ~ & ~ & ~ & ~ & \ldots & 0 \\
\theta_{1} & \theta_{2} & \theta_{3} & \theta_{4} & \theta_{5} & \theta_{6} & 0 & \ldots & ~ & \ldots & 0 & 0 \\
  \phi_{1} &   \phi_{2} &   \phi_{3} &   \phi_{4} &   \phi_{5} &   \phi_{6} & 0 & \ldots & ~ & \ldots & 0 & 0 \\
         0 &          0 & \theta_{1} & \theta_{2} & \ldots & ~ & ~ & ~ & ~ & ~ & ~ & ~ \\
         0 &          0 &   \phi_{1} &   \phi_{2} & \ldots & ~ & ~ & ~ & ~ & ~ & ~ & ~ \\
    \vdots & ~ & ~ & ~ & ~ & ~ & ~ & ~ & ~ & ~ & ~ & ~ \\
         0 & \ldots & ~ & ~ & \ldots & 0 & \theta_{1} & \theta_{2} & \theta_{3} & \theta_{4} & \theta_{5} & \theta_{6} \\
         0 & \ldots & ~ & ~ & \ldots & 0 &   \phi_{1} &   \phi_{2} &   \phi_{3} &   \phi_{4} &   \phi_{5} &   \phi_{6} \\
0 & 0 & 0 & \ldots & ~ & ~ & \ldots & 0 & \theta_{1} & \theta_{2} & \theta_{3} & \theta_{4} \\
0 & 0 & 0 & \ldots & ~ & ~ & \ldots & 0 &   \phi_{1} &   \phi_{2} &   \phi_{3} &   \phi_{4}
\end{array} \right)
\left( \begin{array}{c}
E_{r,1}^{n+1} + \theta_{7} \\
F_{r,1}^{n+1} + \phi_{7} \\
E_{r,2}^{n+1} + \theta_{7} \\
F_{r,2}^{n+1} + \phi_{7} \\
~ \\
\vdots \\
~ \\
E_{r,N-1}^{n+1} + \theta_{7} \\
F_{r,N-1}^{n+1} + \phi_{7} \\
E_{r,N}^{n+1} + \theta_{7} \\
F_{r,N}^{n+1} + \phi_{7} 
\end{array} \right) =
\left( \begin{array}{c}
E_{r,1}^{n} + \theta_{7} - \theta_{1} E_{r,1}^{0} - \theta_{2} F_{r,1}^{0} \\
F_{r,1}^{n} + \phi_{7} - \phi_{1} E_{r,1}^{0} - \phi_{2} F_{r,1}^{0} \\
E_{r,2}^{n} + \theta_{7} \\
F_{r,2}^{n} + \phi_{7} \\
~ \\
\vdots \\
~ \\
E_{r,N-1}^{n} + \theta_{7} \\
F_{r,N-1}^{n} + \phi_{7} \\
E_{r,N}^{n} + \theta_{7} - \theta_{5} E_{r,N}^{0} - \theta_{6} F_{r,N}^{0} \\
F_{r,N}^{n} + \phi_{7} - \phi_{5} E_{r,N}^{0} - \phi_{6} F_{r,N}^{0}
\end{array} \right)  \nonumber
\end{equation} }}


\section*{Appendix 2: Piecewise Linear Limiting Techniques}
\noindent
\textbf{Componentwise Limiting}
\begin{enumerate}
\item Compute left, center, and right differences: $\Delta_{-} U_{i} = U_{i} - U_{i-1}$, $\Delta_{c} U_{i} = \frac{1}{2} \left( U_{i+1} - U_{i-1} \right)$, $\Delta_{+} U_{i} = U_{i+1} - U_{i}$
\item Enforce the TVD condition: $\Delta_{\lim} U_{i} = 2 \min \left( |\Delta_{-} U_{i}| , |\Delta_{+} U_{i}| \right)$ 
\item Define limited slope:
\begin{equation}
P_{\Delta}(\Delta U_{i}) = \left\{
\begin{array}{cc}
\min \left( |\Delta_{c} U_{i}|, |\Delta_{\lim} U_{i}| \right) \textrm{sign} \left( \Delta_{c} U_{i} \right) & \Delta_{-} U_{i} \Delta_{+} U_{i} > 0 \\
0 & \Delta_{-} U_{i} \Delta_{+} U_{i} \leq 0
\end{array} \right . \nonumber
\end{equation} 
\end{enumerate}

\noindent
\textbf{Limiting Across Characteristic Fields}
\begin{enumerate}
\item Compute left, center, and right differences like Step 1 above
\item Using the left material eigenvectors $L^{m}_{\textrm{eff}}$ given in Equation \ref{eq:left_eff_evec}, expand the differences in characteristic variables where $k$ defines a specific row vector of $L^{m}_{\textrm{eff}}$: $\Delta_{-} \mathcal{U}^{k}_{i} = L^{m, k}_{\textrm{eff}} \cdot \Delta_{-} U_i$, $\Delta_{c} \mathcal{U}^{k}_{i} = L^{m, k}_{\textrm{eff}} \cdot \Delta_{c} U_i$, $\Delta_{+} \mathcal{U}^{k}_{i} = L^{m, k}_{\textrm{eff}} \cdot \Delta_{+} U_i$
\item Enforce the TVD condition one field at a time: $\Delta_{\lim} \mathcal{U}^{k}_{i} = 2 \min \left( |\Delta_{-} \mathcal{U}^{k}_{i}| , |\Delta_{+} \mathcal{U}^{k}_{i}| \right)$
\item Define limited characteristic variables:
\begin{equation}
\Delta \mathcal{U}^{k}_{i} = \left\{
\begin{array}{cc}
\min \left( |\Delta_{c} \mathcal{U}^{k}_{i}|, |\Delta_{\lim} \mathcal{U}^{k}_{i}| \right) \textrm{sign} \left( \Delta_{c} \mathcal{U}^{k}_{i} \right) & \Delta_{-} \mathcal{U}^{k}_{i} \Delta_{+} \mathcal{U}^{k}_{i} > 0 \\
0 & \Delta_{-} \mathcal{U}^{k}_{i} \Delta_{+} \mathcal{U}^{k}_{i} \leq 0
\end{array} \right . \nonumber
\end{equation}
\item Using the right material eigenvectors $R^{m}_{\textrm{eff}}$ given in Equation \ref{eq:right_eff_evec}, reconstruct the limited slopes for each of the material quantities where $k$ defines a specific column vector of $R^{m}_{\textrm{eff}}$: $\Delta U^{k}_{i} = \sum_{k} \Delta \mathcal{U}^{k}_{i} R^{m, k}_{\textrm{eff}}$
\end{enumerate}


\section*{Appendix 3: Material Flux Evaluations}
\noindent
The \textbf{Lax-Friedrichs} flux is first-order accurate for smooth solutions, where:
\begin{equation}
F^{\textrm{LF}}_{i+1/2} = \frac{1}{2} \left( F(U_{L}) + F(U_{R}) \right) + \frac{1}{2} \frac{\Delta x}{\Delta t} \left( U_{L} - U_{R} \right) . 
\end{equation}  \nonumber

\noindent
The \textbf{Richtmyer} flux is second-order accurate for smooth solutions and is based on an intermediate quantity $U^{R}_{i+1/2}$, where:
\begin{equation}
F^{\textrm{R}}_{i+1/2} = F(U^{\textrm{R}}_{i+1/2}) , ~~~~
U^{\textrm{R}}_{i+1/2} = \frac{1}{2} \left( U_{L} + U_{R} \right) + \frac{1}{2} \frac{\Delta t}{\Delta x} \left( F(U_{L}) - F(U_{R}) \right) .
\end{equation}  \nonumber

\noindent
The \textbf{HLLE} flux is higher-order accurate for smooth solutions and is derived by approximately solving for the Rankine-Hugoniot shock conditions of the Riemann problem, where:
\begin{equation}
F^{\textrm{HLLE}}_{i+1/2} =  \left\{
\begin{array}{cc}
F(U_{L}) & 0 \leq s_{L} \\
F^{\textrm{HLLE}} & s_{L} \leq 0 \leq s_{R} \\
F(U_{R}) & s_{R} \leq 0 
\end{array} \right .  ,
\end{equation}  \nonumber
\begin{equation}
F^{\textrm{HLLE}} = \frac{s_{R} F(U_{L}) - s_{L} F(U_{R}) + s_{L} s_{R} \left( U_{R} - U_{L} \right)}{s_{R} - s_{L}} .
\end{equation}  \nonumber

\noindent
Here, $s_{L,R}$ are the wave speeds to the left and right of a cell interface that can be estimated in the following three ways in the algorithm $(i)$ directly: $s_{L,R} = u_{L,R} \mp a_{\textrm{eff}}(U_{L,R})$, $(ii)$ using Roe-averaged quantities: $s_{L,R} = \bar{u}_{L,R} \mp \bar{a}_{\textrm{eff}}(U_{L,R})$ where $\bar{\kappa} = (\rho_{L}^{1/2} \kappa_{L} + \rho_{R}^{1/2} \kappa_{R}) / (\rho_{L}^{1/2} + \rho_{R}^{1/2})$, and $(iii)$ taking the maximum wave speed over all grid cells: $s_{L,R} = \mp s_{\max}$ where $s_{\max} = \max_{i} ( |u_i| + a_{\textrm{eff},i} )$. It is important to note that when $s_{L,R} = \mp s_{\max}$ the HLLE flux function algebraically becomes the Lax-Friedrichs flux function.


\section*{Appendix 4: Boundary Conditions for Corrector Scheme}
\noindent
If one assumes that there is a computational grid of $N$ cells starting at the index $i_{start}$ and ending at the index $i_{end}$ and if one further assumes that there are $N_{ghost}$ cells on either side of $N$, then one defines boundary conditions in the following ways: \\

\noindent
\textbf{Periodic Boundary Conditions}
{\footnotesize {\begin{equation}
\begin{array}{llcl}
\textrm{Left:}  & U(i_{start}-N_{ghost} , \ldots , i_{start}-1) & \rightarrow & U(i_{end}-N_{ghost}+1 , \ldots , i_{end}) , \\
\textrm{Right:} & U(i_{end}+1 , \ldots , i_{end}+N_{ghost})     & \rightarrow & U(i_{start} , \ldots , i_{start}+N_{ghost}-1) .
\end{array}  \nonumber
\end{equation} }}

\noindent
\textbf{Outflow Boundary Conditions}
{\footnotesize {\begin{equation}
\begin{array}{llcl}
\textrm{Left:}  & U(i_{start}-N_{ghost} , \ldots , i_{start}-1) & \rightarrow & U(i_{start}) , \\
\textrm{Right:} & U(i_{end}+1 , \ldots , i_{end}+N_{ghost})     & \rightarrow & U(i_{end}) .
\end{array}  \nonumber
\end{equation} }}

\noindent
\textbf{Reflecting Boundary Conditions}
{\footnotesize {\begin{equation}
\begin{array}{llcl}
\textrm{Left:}  & \{ \rho,E,E_{r} \}(i_{start}-N_{ghost} , \ldots , i_{start}-1)  & \rightarrow & \{ \rho,E,E_{r} \}(i_{start}+N_{ghost}-1 , \ldots , i_{start} ) , \\
~               & \{ m,F_{r} \}(i_{start}-N_{ghost} , \ldots , i_{start}-1)     & \rightarrow & - \{ m,F_{r} \}(i_{start}+N_{ghost}-1 , \ldots , i_{start} ) , \\
\textrm{Right:} & \{ \rho,E,E_{r} \}(i_{end}+1 , \ldots , i_{end}+N_{ghost})      & \rightarrow &  \{ \rho,E,E_{r} \}(i_{end} , \ldots , i_{end}-N_{ghost}+1 ) , \\
~               & \{ m,F_{r} \}(i_{end}+1 , \ldots , i_{end}+N_{ghost})         & \rightarrow & - \{ m,F_{r} \}(i_{end} , \ldots , i_{end}-N_{ghost}+1 ) .
\end{array}  \nonumber
\end{equation} }}

\noindent
\textbf{Inflow Boundary Conditions}
{\footnotesize {\begin{equation}
\begin{array}{llcl}
\textrm{Left:}  & U(i_{start}-N_{ghost} , \ldots , i_{start}-1) & \rightarrow & U^{0}(i_{start}-N_{ghost} , \ldots , i_{start}-1) , \\
\textrm{Right:} & U(i_{end}+1 , \ldots , i_{end}+N_{ghost})     & \rightarrow & U^{0}((i_{end}+1 , \ldots , i_{end}+N_{ghost}) .
\end{array}  \nonumber
\end{equation} }}


\section*{Appendix 5: Simpler System for Investigation}
\noindent
To illustrate how some systems of hyperbolic balance laws with dual multiscale behavior reduce to non-stiff systems, consider the following set of partial differential equations:
\begin{eqnarray}
\rho_{t} + m_{x} & = & 0 , \label{simple1} \\
m_{t} + \rho_{x} & = & \sigma f , \label{simple2} \\
e_{t} + C f_{x}  & = & C \sigma ( \rho - e ) , \label{simple3} \\
f_{t} + C e_{x}  & = & - C \sigma f , \label{simple4}
\end{eqnarray}

\noindent 
where the terms appearing in the above system do not directly correspond with material and radiation quantities. Equations \ref{simple1} and \ref{simple2} define the slow (i.e., macroscopic) hyperbolic system, which has a wave speed of unity; while Equations \ref{simple3} and \ref{simple4} define the fast (i.e., microscopic) hyperbolic system, which has a wave speed of $C$. Stiffness associated with the different wave speeds defines one multiscale behavior. Additionally, the non-zero source terms in Equations \ref{simple2}-\ref{simple4} are augmented by a magnitude $\sigma$ which may be small or large with respect to unity and defines the other multiscale behavior. In some parameter regime of $C$ and $\sigma$, one expects $e \rightarrow \rho$ because the source terms enable the relaxation of some of the components in the system. If one assumes that there is a small parameter $\epsilon \ll 1$ such that $C, \sigma \rightarrow 1 / \epsilon$ and one employs methods from perturbation theory to construct the following series for $e$ and $f$:
\begin{eqnarray}
e & = & e_{0} + \epsilon e_{1} + \epsilon^{2} e_{2} + \epsilon^{3} e_{3} + \ldots , \\
f & = & f_{0} + \epsilon f_{1} + \epsilon^{2} f_{2} + \epsilon^{3} f_{3} + \ldots ,
\end{eqnarray}

\noindent
after matching terms of like order in $\epsilon$, one arrives at the following reduced system of equations which exhibits no stiffness and has a well-defined solution:
\begin{eqnarray}
\rho_{t} + m_{x}   & = & 0 , \\
m_{t} + 2 \rho_{x} & = & 0 + \mathcal{O}(\epsilon^{2}) , \\
e_{t} - e_{xx}     & = & 0 + \mathcal{O}(\epsilon^{2}) .
\end{eqnarray}


\end{document}